\documentclass[useAMS,usenatbib]{mnras}
\usepackage{times,psfig,epsfig} 
\usepackage{rotating, graphicx, adjustbox}
\usepackage{color}
\usepackage{amssymb, amsmath}
\usepackage[T1]{fontenc} 
\usepackage{aecompl}
\usepackage{gensymb}

\newcommand{\chandra}{{\it Chandra}}
\newcommand{\xmm}{{\it XMM-Newton}}
\newcommand{\ska}{{\it SKA}}

\newcommand{\athena}{{\it ATHENA$+$}}
\newcommand{\planck}{{\it Planck}}

\newcommand{\ccat}{{\it CCAT-prime}}

\newcommand{\ergscm}{erg\,s$^{-1}$cm$^{-2}$}
\newcommand{\ergscmarcmin}{erg\,s$^{-1}$cm$^{-2}$\,arcmin$^{-2}$}
\newcommand{\ergscmdeg}{erg\,s$^{-1}$cm$^{-2}$deg$^{-2}$}

\newcommand{\mincir}{\raise
  -2.truept\hbox{\rlap{\hbox{$\sim$}}\raise5.truept \hbox{$<$}\ }}
\newcommand{\magcir}{\raise
  -2.truept\hbox{\rlap{\hbox{$\sim$}}\raise5.truept \hbox{$>$}\ }}
\newcommand{\siml}{\raise 
  -2.truept\hbox{\rlap{\hbox{$\sim$}}\raise5.truept \hbox{$<$}\ }}
\newcommand{\simg}{\raise
  -2.truept\hbox{\rlap{\hbox{$\sim$}}\raise5.truept \hbox{$>$}\ }}

\providecommand{\abs}[1]{\lvert#1\rvert}
\providecommand{\avg}[1]{\langle#1\rangle}

\title[Multi-wavelength mock observations of the WHIM]
{Multi-wavelength mock observations of the WHIM in a simulated galaxy cluster}

\author[S. Planelles et al.]
{Susana Planelles$^{1}$\thanks{e-mail: susana.planelles@uv.es}, Petar Mimica$^{1}$, Vicent Quilis$^{1,2}$, Carlos Cuesta-Mart{\'i}nez$^{1}$  \\~\\
\footnotesize 
$^1$ Departament d'Astronomia i Astrof{\'i}sica, Universitat de Val\`encia, c/ Dr. Moliner, 50, 46100 - Burjassot (Valencia), Spain\\
$^2$ Observatori Astron\`omic, Universitat de Val\`encia, E-46980 Paterna (Valencia), Spain}

\begin{document}

\maketitle 

\begin{abstract}

About half of the expected total baryon budget in the local Universe is `missing'. Hydrodynamical simulations suggest that most of the missing baryons are located in a mildly overdense, warm-hot intergalactic medium (WHIM), which is difficult to be detected at most wavelengths. In this paper we explore multi-wavelength synthetic observations of a massive galaxy cluster developed in a full Eulerian-AMR cosmological simulation.  A novel numerical procedure is applied on the outputs of the simulation, which are post-processed with a full-radiative transfer code that allows to compute the change of the intensity at any frequency  along the null-geodesic of photons. 
We compare the emission from the whole inter-galactic medium (IGM) and from the WHIM component (defined as the gas with a temperature in the range $10^5-10^7$ K) at three observational bands associated to thermal X-rays, thermal and kinematic Sunyaev-Zel'dovich effect, and radio emission. The synthetic maps produced by this procedure could be directly compared with existing observational maps and could be used as a guide for future observations with forthcoming instruments. The analysis of the different emissions  associated to a high-resolution galaxy cluster is in broad agreement with previous simulated and observational estimates of both gas components.  

\end{abstract} 
 

\section{Introduction}
\label{sec:intro}

Observations of the cosmic microwave background \citep[CMB; e.g.][]{Planck_2016},  primordial nucleosynthesis calculations \citep[e.g.][]{Kirkman_2003} and Ly$\alpha$ absorption systems \citep[][]{Rauch_1998} have reported that about a 5 per cent of the total energy content of the Universe is in the form of baryons. At high redshift, $z\magcir2$, the amount of gas measured from Ly$\alpha$ observations and contained in galaxies accounts for the full cosmic baryon budget. However, in the local Universe, while the Ly$\alpha$ forest can account for about one third of the low-z baryons, most part of them remain ``missing''  \citep[][]{Fukugita_1998, Cen_1999, Bregman_2007, Bertone_2008}.  

Cosmological simulations predict that, according to the hierarchical model of cosmic structure formation, tiny density fluctuations in the primordial density field evolved into the large scale structure we observe today:  an intricate network of voids, sheets and filaments connecting the largest clusters of galaxies \citep[e.g.][for reviews]{Kravtsov_2012, Planelles2015}. As the cosmic evolution proceeds, gravitational shock waves propagate from collapsing and central regions to the surrounding environment, heating in a efficient way the intra-galactic medium (IGM). While in the largest clusters of galaxies the intra-cluster medium (ICM) can be gravitationally heated up to temperatures of $10^7-10^8$ K, mildly overdense regions, such as filaments,  can only reach temperatures of $10^5-10^7$ K \citep[e.g.][]{Cen_1999, Dave_2001,Shull_2012}.  
This warm-hot intergalactic medium (WHIM) is thought to contain about 50 per cent of the total low-z cosmic baryons. The WHIM is expected to be located mainly in filaments but also around rich clusters and between pairs of merging clusters. However, given the typically moderate WHIM densities (from $\sim 4\times 10^{-6}$ to $\sim10^{-4}$ cm$^{-3}$) and temperatures,  its emission in the UV and soft X-ray bands, due to thermal Bremsstrahlung, is difficult to be detected by current observational facilities  \citep[e.g.][]{Paerels_2008}. Moreover, at these temperatures, the IGM is fully ionized and contributes significantly to the X-ray background (XRB) in the form of a soft diffuse emission. In a similar way, absorption features from the WHIM are also too weak to be easily resolved \citep[e.g.][]{Richter_2008}. 

Despite the low surface brightness associated to the WHIM, there have been claims of detection in X-ray analyses, either in absorption or in emission, such as those along the line of sight of distant quasi-stellar radio sources \citep[e.g.][]{Fang_2007, Zappacosta_2010, Nicastro_2013} or in the gaseous filaments connecting merging clusters \citep[e.g.][]{Scharf_2000, Finoguenov_2003, Werner_2008, Nicastro_2010, Eckert_2015}. 
However, since X-ray surface brightness is proportional to  the square of the gas density, X-ray observations are more suited to explore high-density regions, usually associated to the hot ICM within the cluster virial radius.

Besides X-ray observations, galaxy clusters can also be observed at millimeter wavebands through the  Sunyaev-Zel'dovich effect \citep[SZ;][]{SZ_1972, SZ_1980}. There exist a thermal (tSZ) and a kinematic (kSZ) SZ effect. The tSZ effect is the gain in energy of  CMB photons when they are scattered, along the line of sight, by high-energy ICM electrons. Given that this effect is proportional to the gas density,  these observations are crucial to reveal the physics of cluster outskirts \citep[see][for a review]{Reiprich_2013}. In this regard, the diffuse WHIM gas component has  been claimed to be detected  through the tSZ effect  in the microwave band \citep[][]{Planck_2013}.
On the other hand, when galaxy clusters present bulk motions,  a second-order signal generated by the interaction between CMB photons and ICM electrons is the  kSZ effect. In this case,  the CMB photons can gain or lose energy depending on whether the ICM gas is approaching or going away from the observer. Recently, this kinematic Sunyaev-Zel'dovich effect  has been also employed to claim the detection of all the missing baryons within and around the central galaxies  in the Sloan Digital Sky Survey \citep[][]{CHM_2015}.  
 
Although most of the emission from clusters is associated to thermal processes, non-thermal cluster radiation plays also a significant role \citep[e.g.][for a recent review]{Brunetti_2014}. Indeed, observations of radio haloes and radio relics \citep[e.g.][for a review]{Ferrari_2008} are associated to extended synchrotron radio emission from clusters. This radio emission originates as a consequence of structure formation shocks accelerating relativistic particles which, in the presence of a magnetic field, can emit synchrotron radiation.
Therefore, non-thermal phenomena provide information on the pressure of the thermal gas, the distribution of shock waves and the level of ICM magnetic fields \citep[$\sim 0.1-1~\mu$G;][]{Feretti_2008}.
Hence, the detection of accretion shocks around cosmic filaments would be essential to deepen our understanding of the WHIM component. 
Moreover, in addition to radio emission, it has been also claimed the existence of a hard X-ray cluster emission due to inverse Compton scattering of relativistic ICM electrons and CMB photons \citep[e.g.][for a review]{Rephaeli_2008}. However, the origin and detection of this non-thermal emission are still controversial \citep[e.g.][]{Fusco-Femiano_2004, Sanders_2004, Sanders_2005, Eckert_2008, Molendi_2009, PerezTorres_2009}. 

From a theoretical point of view, the first attempts to ``observe'' the WHIM component came from hydronamical simulations \citep[e.g.][]{Cen_1999, Dave_2001}, which   predicted that the WHIM can contribute up to 40 per cent of the total XRB emission \citep[e.g.][]{Croft_2001, Roncarelli_2006}, with most of it coming from $z<0.9$ \citep[see also][]{Kravtsov_2002, Ursino_2006}. Further numerical studies have put a significant effort in analysing the properties of the WHIM component depending on the inclusion and treatment of different physical processes such as radiative cooling, star formation and its associated feedback, chemical enrichment, or magnetic fields \citep[e.g.][]{Cen_2006, Cen_2006b, Bertone_2010a, Bertone_2010b, Tornatore_2010, Roncarelli_2012, Vazza_2015, Gheller_2015}.
Several numerical studies have also exploited the complementarity of X-ray and SZ properties of the WHIM \citep[e.g.][]{CHM_2006, Roncarelli_2007, Ursino_2014}, confirming that such a combination represents a promising tool to study future WHIM observations.  
Simulations have been also employed to explore the non-thermal emission from galaxy clusters \citep[e.g.][]{Pfrommer_2008, Hoeft_2008, Battaglia_2009, Brown_2011, ArayaMelo_2012, Vazza_2015, Vazza_2015b, Vazza_2016, Basu_2016}.

In the near future, a new generation of telescopes will significantly widen our understanding of the main properties and distribution of the WHIM.
In fact, the highly-improved capabilities of these upcoming facilities will certainly unveil a whole new picture of the WHIM in several different wavelength bands like the ones associated with the emissions in X-rays, SZ and radio. 
In this context, our aim is to use  a high-resolution simulated galaxy cluster to produce synthetic detailed maps -- in the aforementioned bands -- directly comparable with existing or future planned and, therefore, improve our knowledge on the WHIM component by getting closer the theoretical and the observational planes. 
To do so, we identify  the  WHIM component (defined as the gas with a temperature in the range $10^5-10^7$ K) at several locations from the cluster centre to its outskirts. Once this is done, we use a novel approach to post-process the data from the simulation by means of a relativistic full-radiative transfer code. This code computes the change in the intensity along each line of sight by integration of the null-geodesic of photons. The code is capable of computing the self-absorption, and we use that capability to verify that the absorption is negligible, so that we are justified in running the code in post-processing and in neglecting any radiation back-reaction on the gas.
In this paper, we focus on the analysis of the thermal X-ray emission of the cluster  (in the soft and hard X-ray bands), the associated tSZ and kSZ effects, and the corresponding radio emission. We wish to emphasize that all bands are treated consistently by the same code,  thus avoiding the usage of different numerical approaches for each separate wavelength. The results derived from this approach are compared with previous existing data, both numerical and observational.

This paper is organised as follows. In Section \ref{sec:simu} we briefly introduce the numerical codes employed in our study, that is, an Eulerian cosmological code to perform the hydrodynamical simulation and a radiative transfer code to process the outputs of the simulation and compute the associated emission in different bands. Moreover, a description of the main properties of our main simulated cluster is also included. In  Section \ref{sec:results} we present our results on the different X-rays, SZ and radio emissions of the WHIM and the IGM. Finally, in  Section \ref{sec:conclu}, we summarise and discuss our main findings.
Besides, while Appendix \ref{app:Cloudy} is devoted to discuss how the contribution from metal lines affects the estimation of the X-ray emission,  Appendix \ref{app:SZeffect} shows instead how relevant is to consider the relativistic corrections when computing the SZ signal.

\section{Numerical details}
\label{sec:simu}

In this section we present the main numerical details of the codes we have employed in the present analysis, namely, the cosmological code MASCLET \citep{Quilis2004}  and the radiative transfer code {\sc SPEV} \citep[][]{Mimica_2009, CMC_2015, Mimica_2016}.
Here, we only provide a brief description of the main features relevant for the present study, while we refer the reader to the corresponding references for further details.

\subsection{The cosmological simulation}
\label{subsec:part1}

In this paper we analyse the outcomes of a  hydrodynamical simulation  performed with the cosmological code MASCLET \citep{Quilis2004}.
MASCLET is an Eulerian code, with an adaptive  mesh refinement  (AMR) scheme, that couples  {\it high-resolution shock capturing}
techniques to resolve the evolution of the gaseous component of the Universe with an  N-body scheme to evolve the dark matter component. 

Our simulation accounts for  a flat $\Lambda CDM$  universe with cosmological  parameters:  $\Omega_{\Lambda}=\Lambda/{3H_o^2}=0.69$, $\Omega_m=0.31$, $\Omega_b=0.048$, $h=H_0/100$ km\, s$^{-1}$\,Mpc$^{-1}=0.678$, $\sigma_8=0.82$ and  $n_s=0.96$. 

The simulation domain, a cubical box with a comoving  side length of $40$ Mpc, has been discretised using $128^3$ cubical cells in the coarse (lowest resolution) level. In this simulation we allow for a maximum of 9 refinement levels \citep[see][for further details on the refinement procedure]{Navarro_2013, Quilis_2017}, providing  a peak physical spatial resolution of $\sim 610$  pc  at $z=0$. 
As for the dark matter particles,  the best  mass resolution  is  $\sim 2\times  10^6\,  M_\odot$, which is equivalent to distributing $1024^3$ particles throughout the computational volume.

Using a CDM transfer function \citep{Eisenstein_1998}, the initial conditions of our simulation were set up at $z=100$.
Moreover, following the ideas presented in  \citet{Hoffman_1991}, we have performed a constrained realization in order to reproduce a big galaxy cluster at the centre of our simulated volume.  

Besides gravity and hydrodynamics, the simulation includes additional physical processes such as inverse Compton and free-free  cooling, atomic and molecular cooling for a primordial gas, and UV heating \citep{Haardt_1996}. With the purpose of computing the abundances of each specie, the gas is considered to be  optically thin  and in ionization  equilibrium, but  not in  thermal equilibrium \citep{Katz_1996, Theuns_1998}. We employ metallicity dependent tabulated cooling rates from  \citet{SD_1993} and we truncate the cooling curve below a temperature of $10^4$ K. In addition, our simulation includes star formation, according to the models by \citet{Yepes_1997} and \citet{Springel_2003}, and feedback from supernova (SN) type II. However,  feedback from stellar winds, SN type Ia, or  active galactic nuclei (AGN) are not accounted for.

\subsection{Computing the emission}
\label{subsec:part2}

Obtaining proper observables from hydrodynamical simulations is not a trivial task. In order to reproduce the pictures obtained by an observer at $z=0$, we need to  consider all the matter contributing to the integrated signal up to a given cosmic time.  There have been several studies trying to overcome this difficulty, for instance, by reconstructing the past light cone as observed at $z=0$ \citep[e.g.][]{Roncarelli_2012}. Despite the improvement of this approach, it is still an approximated way to deal with this issue.
In our study, we use instead a full-radiative transfer code to compute the thermal and non-thermal emission in different X-rays, SZ, and radio bands. We note that our method uses a code developed for the radiative transfer in relativity (see below), so that it follows the emission along the null-geodesics and can provide consistent multi-wavelength and multi-epoch synthetic observations. In this case we use the same null-geodesics to also compute the SZ effect, so that we \emph{consistently} compute \emph{all} the observational signals. Moreover, as we explain below, the method is well-tested and converges with the increasing spatial and temporal simulation resolution.

To compute the emission from MASCLET snapshots we use in post-processing the code {\sc SPEV} \citep{Mimica_2009}. Although it was originally designed to process the outputs of a relativistic hydrodynamics simulation and compute the non-thermal emission, it has since been expanded to include the thermal emission \citep{CMC_2015}. In this work we compute both thermal and non-thermal emission using a two-step procedure. First, the raw output from MASCLET is preprocessed to select the numerical cells that satisfy the criteria for the emission. Then, the preprocessed data is fed to the post-processor that generates a multi-frequency image. {\sc SPEV} first processes each cell to determine which pixels of the image can observe it. It then computes the emission from each zone at the proposed frequencies, and stores it in memory reserved for the lines of sight leading to those pixels, together with the information about the zone position, geometry, and velocity, as well as the electron density and temperature (for the purpose of computing the SZ effect, see Sec.~\ref{sec:SZ}).
We assume that the pixel at the origin of the image (right in its geometrical centre) observes the centre of the simulation box when it is at $z=0.36$ (see Section \ref{subsec:part3} for further details on this choice). This determines the range of redshifts and positions of the cells that can be observed simultaneously with the origin. In order to obtain the continuous spacetime coverage, we interpolate the positions of the cell corners between snapshots assuming a linear motion and taking into account the Hubble flow and their peculiar velocity. This interpolation in spacetime allows {\sc SPEV} to deal with an arbitrary numerical resolution, as well as with the arbitrary frequency with which simulation snapshots are stored. For the numerical tests that demonstrate the convergence of {\sc SPEV} with increasing temporal and spatial numerical resolution see \citet{Mimica_2016}.

After all the zones have been post-processed, {\sc SPEV} solves the radiative transfer equation for each pixel in order to compute the observed intensity at a frequency $\nu$:
\begin{equation}\label{eq:intens}
  {\mathrm d}I_\nu/{\mathrm d}s = j_\nu - \alpha_\nu I_\nu\, ,
\end{equation}
where $I_\nu$, $j_\nu$ and $\alpha_\nu$ are the intensity, emissivity and absorption coefficients, respectively, and $s$ is the path along the line of sight. In practice, the line of sight is composed of finite segments, each corresponding to an intersection with a numerical cell. We assume that $j_\nu$ and $\alpha_\nu$ are constant in each segment and can easily compute the final intensity by sorting the segments in the order of decreasing distance from the observer and then analytically solving Eq.~\ref{eq:intens} in each segment, taking as the initial value for $I_\nu$ the outgoing intensity from the previous segment. The outgoing intensity from the last segment is then the observed intensity of the pixel. The incoming intensity into the first (furthest away from the observer) segment is assumed to be zero (except when computing SZ, where we assume the CMB intensity, see Sec.~\ref{sec:SZ}). We note that the algorithm described here is different from the light-cones scheme used in some previous works \citep[e.g.,][]{Roncarelli_2006} since it does not use stacking at different redshifts, but rather produces a consistent image in the observer frame by creating a continuous spacetime representation of the simulation regardless of its actual spatial and temporal resolution. For more details about the {\sc SPEV} imaging algorithm see \citet{Mimica_2016}.

\subsubsection{Thermal emission}
\label{subsubsec:th}

The free-free thermal Bremsstrahlung is the dominant emission process at the characteristic ICM temperatures. However, as we go to lower temperatures the emission through metal lines also becomes relevant \citep[e.g.][]{Bohringer_2010}. Actually, most of the emission associated to the WHIM in the soft X-ray band is due to line emission \citep[e.g.][]{Bertone_2008, Bertone_2010a}.
Therefore, a proper implementation of metal enrichment is crucial to perform an accurate determination of the observational  IGM and WHIM properties. 

In the current analysis, following the results presented in \cite{Croft_2001}, and in a similar way to previous studies \citep[e.g.][]{Branchini_2009, Takei_2011}, we estimate the gas metallicity in solar units according to $Z=min(0.3, 0.005(1+\delta_g)^{1/2})$, where $\delta_g=\rho_{gas}/\avg{\rho_{gas}}-1$ and $\rho_{gas}$ is the gas density. This metallicity estimate, computed in post-processing, is provided to {\sc SPEV} in order to compute the thermal X-ray emission as contributed by both the free-free Bremsstrahlung and the line emission. As explained in \cite{Takei_2011}, although this metallicity estimate is not totally self-consistent with the thermodynamical state of the gas, it avoids the underestimation of the metal abundance in low-density regions which,  due to the characteristics of our simulation, are worse resolved. 
 
In the default version of {\sc SPEV}, in order to compute the thermal X-ray emission, we assumed that the dominant process is the free-free thermal Bremsstrahlung and use the Kirchhoff's law to compute the absorption coefficient \citep[see Appendix A1 of][for a detailed explanation on how the thermal emission is implemented in {\sc SPEV}. Furthermore, we note that the Gaunt factor tables have been updated with the ones computed by \cite{vanHoof_2014, vanHoof_2015}.]{CMC_2015}\footnote{By comparing the emission maps with and without absorption we verified that it is negligible in the scenario we studied.}. However, this is not an accurate assumption to compute the thermal X-ray emission from galaxy clusters in the energy regime where line emission from metals have a significant contribution. Therefore, in this work, to compute the X-ray images having also into account the contribution from lines, we need to be able to model the plasma emission in a large dynamic range of parameters. In order to do so, we use the publicly available code {\sc Cloudy} \citep[version 17.00;][]{Cloudy_2017} to construct a large interpolation table of the X-ray emission in different energy bands. In Appendix~\ref{app:Cloudy} we provide a detailed explanation about the computation procedure, as well as a comparison between this and the default {\sc SPEV} thermal emission method. According to our tests, the method based on  {\sc Cloudy} tables allows us to reliably predict the X-ray emission from low-density and low-temperature media. 

In practice, for each model generated with {\sc Cloudy} (see Appendix~\ref{app:Cloudy} for details) we assume thermal equilibrium, use the \verb+CMB+ and \verb+HM05+ {\sc Cloudy} commands to generate the CMB and UV background radiation fields \citep[][]{Haardt_1996}, and save the diffuse continuum. We then integrate the continuum in the soft ($0.5-2$ keV) and hard ($2-10$ keV) X-ray energy bands. 
In Section \ref{subsec:xrays} we will analyse the results obtained using this method when we compute synthetic X-ray maps at the soft  and at the hard  X-ray energy bands for the IGM and the WHIM.

\subsubsection{Non-thermal emission}
\label{subsubsec:nth}

The nature, strength and distribution of cosmic magnetic fields is still highly uncertain \citep[e.g. see][for a review]{Dolag_2008}. It is beyond the scope of the present work to perform a detailed analysis of different magnetisation models in order to constrain current uncertainties \citep[e.g. see][for recent works on different ICM magnetic field models]{Vazza_2017a, Vazza_2017b}. We pretend instead to show the capabilities of {\sc SPEV} in generating radio images from cosmological simulations. Therefore, in order to estimate the non-thermal emission, we need to adopt some simplifications. In particular, following the model employed in some recent works \citep[e.g.][]{Hoeft_2008, ArayaMelo_2012}, we assume that the magnetic field, $B$, satisfies flux conservation and is given by 
\begin{equation}\label{eq:Bfield}
\frac{B}{B_{ref}}=\left(\frac{n}{10^{-4}cm^{-3}}\right)^{2/3}\, ,
 \end{equation}
where $B_{ref}=0.1\, \mu$G and $n$ stands for the gas number density in the post-shock region. We only take into account the non-thermal emission when $B<1\, \mu$G \citep[e.g.][]{Feretti_2008}\footnote{We performed tests with different values of the upper limit (up to $5\, \mu$G), but found no significant changes in the final convolved images shown in Section~\ref{subsec:radio}.}. Moreover, this emission is computed only from those zones where the Mach number, $\mathcal{M}$, is greater than $1.01$ \citep[see][for details on the shock-finding algorithm employed to compute the strength of shocks]{Planelles2013, Sergio_2017}.

In this work we do not implement any transport scheme for the non-thermal electrons. We use the formalism of \citet{Miniati_2001} and \citet{EPSJ_2007} to compute the power-law index ($\alpha_{inj}$) and the lower energy cutoff ($\gamma_{min}$) for the non-thermal electron distribution:
\begin{eqnarray}\label{eq:pwl}
  \alpha_{inj} = 2 (\mathcal{M}^2 + 1) / (\mathcal{M}^2 - 1)\\ 
  \gamma_{min} = x_{inj}\sqrt{2kT / (m_p c^2)}\, , 
\end{eqnarray}
where $T$ is the temperature of the shocked gas and $m_p$ and $c$ are the proton mass and the speed of light. We set $x_{inj}$ to $3.5$ \citep{EPSJ_2007}. The upper cut-off is determined by balancing the synchrotron cooling and the acceleration timescales:
\begin{equation}
  \gamma_{max} = \left(\frac{3m_e^2 c^4}{4\pi e^3 B}\right)^{1/2}\, ,
\end{equation}
where $m_e$ and $e$ are the electron mass and charge, respectively. The synchrotron emission is computed using the standard formulae \citep[e.g.,][]{Rybicki:1979} and employing an efficient numerical method (as explained in Sec.~4 of \citealt{Mimica_2009}).

In Section~\ref{subsec:radio} we will discuss and compare the results obtained with our reference model (also labelled as `model B1'), given by Eq.~\ref{eq:Bfield}, with those derived when we  assume the magnetic field to be in a fraction of equipartition with the gas internal energy density, i.e.
\begin{equation}\label{eq:Bfield2}
  B^2 / 8\pi = \epsilon_B P\, ,
\end{equation}
where $P$ is the thermal pressure and $\epsilon_B$ is a proportionality constant (we assume $\epsilon_B = 10^{-3}$). We will labelled this model as `model B2' and, as with our reference model, we will only take into account the non-thermal emission when $B<1\, \mu$G. 

\subsection{Computing the SZ effect}
\label{sec:SZ}

In an intermediate step before producing the final image, {\sc SPEV} needs to organize the data along the lines of sight, so that each line of sight contains zones that are simultaneously observed in the center of a particular pixel. We use these line of sight data structures to also compute the thermal and kinematic SZ effects in the non-relativistic approximation \citep[see][for reviews]{Birkinshaw_1999, Carlstrom_2002, Colafrancesco_2007}. 

As mentioned in the previous section, for each pixel we store the electron density, the electron temperature, and the fluid velocity in all the contributing numerical cells. To compute the SZ effect we have modified the final phase of the {\sc SPEV} algorithm, when the pixel intensities are computed, as follows. First, we initialize the pixel intensities to the CMB intensity, that is, $I_\nu = 2 h \nu^3 / c^2 / (e^x - 1)$, where $h$ is the Planck constant,  $x:=h\nu / k T_{CMB}$ is a dimensionless frequency, $k$ is the Boltzmann constant, and $T_{CMB} = 2.72548$ K \citep[e.g.][]{Fixsen_2009}. Therefore, the change in the intensity due to the thermal SZ effect is given by 
\begin{equation}
\label{eq:tSZ}
  \Delta I_{\nu,th} = \frac{x e^x}{e^x - 1}\left[x\frac{e^x + 1}{e^x - 1} - 4 \right] I_\nu\int \mathrm{d}s \frac{k T_e}{m_e c^2} n_e \sigma_T\, ,
\end{equation}
where $T_e$ and $n_e$ are the electron temperature and number density, and $\sigma_T$ is the Thomson cross-section. The integral in Eq.~\ref{eq:tSZ} is computed using the same numerical method that is used for solving Eq.~\ref{eq:intens}. 

On the other hand, the kinematic SZ effect, contributed by the ICM bulk motion, is computed according to the following equation:
\begin{equation}
\label{eq:kSZ}
  \Delta I_{\nu,k} = \frac{xe^x}{e^x - 1} I_\nu \int \mathrm{d}s n_e \sigma_T \frac{v_\parallel}{c}\, ,
\end{equation}
where $v_\parallel$, the component of the gas velocity parallel to the line of sight, can be either positive or negative depending on whether the gas element is moving away or approaching to the observer. 

In the following, we use Eqs.~\ref{eq:tSZ} and \ref{eq:kSZ} to compute separately the thermal and kinematic SZ contributions, as well as the total SZ effect (see Sec.~\ref{subsec:sz}):
\begin{equation}
\label{eq:totalSZ}
\Delta I_\nu / I_\nu = \Delta I_{\nu,th} / I_\nu + \Delta I_{\nu,k} / I_\nu\, .
\end{equation}
For a value of $T_{CMB} = 2.72548$ K, the tSZ signal given by Eq.~\ref{eq:tSZ} shows its minimum and maximum intensities at frequencies $\nu \sim 128$ GHz and $369$ GHz, respectively.
On the contrary, at $\nu\sim 217$ GHz the tSZ effect vanishes and the total SZ signal is dominated by the kinematic contribution. Therefore, the analysis of the SZ effect at these three characteristic frequencies is very useful to disentangle the thermal and kinematic contributions to the total SZ signal and to derive the cluster thermal distribution \citep[e.g.][]{Kay_2008, Prokhorov_2010, Prokhorov_2011}.  

We note that equations presented above are valid to estimate the SZ effect in the non-relativistic approximation and, therefore, in this work we will compute the different SZ signals without taking into account the relativistic correction. In a future work, we plan to include these corrections in {\sc SPEV} and to perform a detailed analysis on the accuracy of the results for a larger sample of galaxy clusters. For the moment, in order to have an idea of the significance of the approximation and the performance of {\sc SPEV}, we present in Appendix \ref{app:SZeffect} a comparison between the results obtained with {\sc SPEV} for the total SZ signal and those obtained with the {\sc SZpack}\footnote{http://www.Chluba.de/SZpack} code \citep{SZPack_2012, SZPack_2013} at different frequencies. {\sc SZpack} is a publicly available library that, taking into account relativistic corrections, provides a precise estimation of the SZ effects  in massive galaxy clusters for a wide range of electron temperatures and cluster peculiar velocities. A deeper analysis of the effects of the relativistic corrections on the SZ effect of galaxy clusters is out of the scope of the present paper and, therefore, we defer the reader to more suitable  works on this purpose \citep[e.g.][]{Itoh_1998, Nozawa_2005}.

\subsection{The simulated cluster}
\label{subsec:part3}

In order to identify the sample of dark matter haloes in our simulation, we apply the spherical overdensity halo finder ASOHF
\citep[see][for further details]{Planelles2010, Knebe_2011}. Within our simulated volume, we find a sample of $\sim44$ haloes with virial masses\footnote{The viral mass, $M_{vir}$, is the mass contained within a sphere of radius $R_{vir}$ enclosing  an average density equal to $\Delta_{vir}$ \citep{Bryan1998} times the corresponding critical cosmic density.} larger than $10^{13} M_{\odot}$. 
However, for the purpose of the present study, we are only interested in the analysis of the main central halo developed in our simulation. The excellent numerical resolution in this central region allows us to describe in detail the distribution of matter and its associated emission from the centre of the main halo out to its outskirts.  At $z=0$, the central halo developed in the simulated volume, which is also the largest one, has a virial mass of $M_{vir}\sim 4.3\times 10^{14} M_\odot$ and a  virial radius of $R_{vir}\sim 1.98$ Mpc. However, motivated by recent observations of massive galaxy clusters \citep[see e.g.][]{Eckert_2015}, in the present study we will take as a target the main progenitor of this simulated cluster but at $z=0.33$, when it has $M_{vir}\sim 3.2\times 10^{14} M_\odot$ and  $R_{vir}\sim 1.7$ Mpc and it is still the central and largest halo in the simulation. 

Our cosmological simulation, which follows the cosmic evolution from $z=100$ down to $z=0$, generates a total of 80 snapshots non-uniformly distributed in redshift.  These outputs from the cosmological code become the inputs for {\sc SPEV} in order to compute the emission associated to the different gaseous components.  However, since we are mainly interested in the WHIM emission and the contribution from high-z gas is expected to be negligible, we will only take into account the redshift interval  $0.65< z < 0.33$, which corresponds to 11 discrete snapshots. We verified that no contribution to the emission comes from the snapshots with $z \geq 0.65$ or $z \leq 0.33$ and, therefore, to speed up the computation, we do not use the data from those snapshots as inputs for {\sc SPEV}.

Therefore, unless otherwise stated, in the following we will analyse the X-ray, SZ or radio signals coming from inside and from an extended region around our main central  halo as observed at $z=0.33$. In order to do so, we will take into account all simulation snapshots that contain volumes that can be observed simultaneously with the central halo at $z=0.33$. Moreover, given the wide distribution of the WHIM and the typical extensions of recent WHIM studies \citep[e.g.][]{Eckert_2015}, we will consider a region with an amplitude of  $5\times R_{vir}$ from the halo centre, which corresponds to an angular extension of around $0.6$$^{\circ}$ along each coordinate direction. Given the features of our simulation, this choice also ensures an excellent numerical resolution of the selected region throughout the redshift evolution. 

Within our volume of interest we will consider the distribution and emission from two types of gas: the IGM, which includes all the gas cells within the selected volume, and the WHIM component, formed by those cells with a gas temperature in the range $10^5 K < T < 10^7 K$. We note that the definition of the WHIM used in numerical studies is still somehow arbitrary: while some analyses define the WHIM according only to the gas temperature cut, as we do, \citep[e.g.][]{Roncarelli_2006, Tornatore_2010}, some others include as well a condition on the gas density \citep[e.g.][]{Roncarelli_2012, Ursino_2014}.

Given the lack of an efficient central source of feedback in our simulation, such as the one associated to a central AGN \citep[see, e.g.][and references therein]{Planelles_2013b}, our cluster is characterized by an excessive X-ray emission in its central region. Therefore, in order to overcome this issue and avoid an spurious estimate of the  corresponding emission, we have computed the cooling time  associated to each gas element in our simulation and removed those gas cells with  $t_{cool}<10^{10}$ years \citep{Lufkin_2000}. This cut in cooling time slightly compensates for the strong central overcooling in the centre of our cluster.

We note that, when comparing with previous numerical analyses of the WHIM component,  our study may suffer from different caveats such as  the relatively small considered volume or the lack of an efficient  AGN feedback source. However, we would like to point out that, on the one hand, the moderate volume is compensated by an excellent resolution throughout the analysed region and, in any case, it is in agreement with the real extension of some recent cluster observations \citep[e.g.][]{Eckert_2015}. On the other hand, although we do not include any AGN feedback model in our simulation, we compute the emission of each  gas element along the line of sight with a full-radiative transfer code, obtaining a complete and fully consistent analysis of the WHIM at different wavebands.

\begin{figure}
\begin{center}
{\includegraphics[width=7cm]{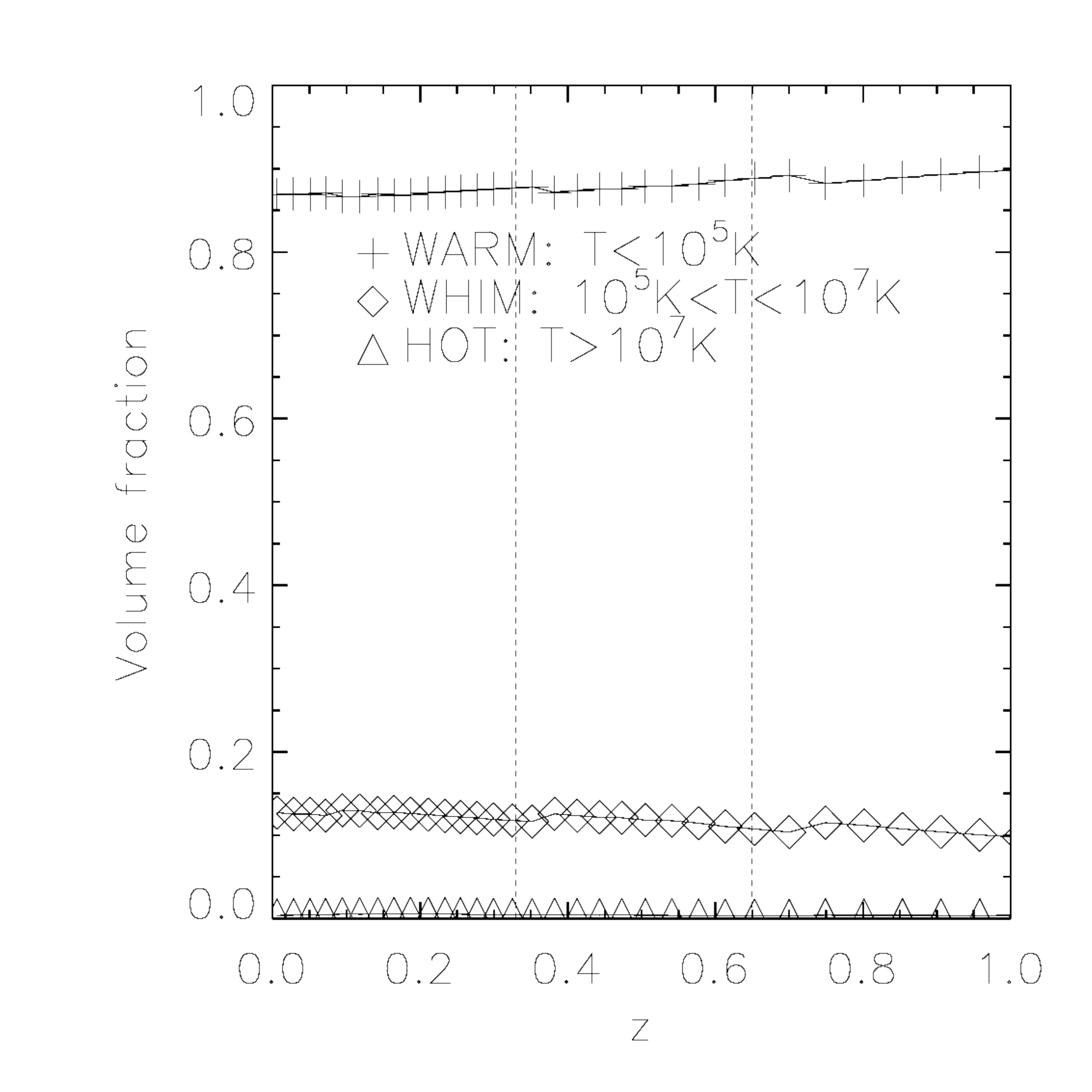}}\\
\vspace{-0.5cm}
{\includegraphics[width=7cm]{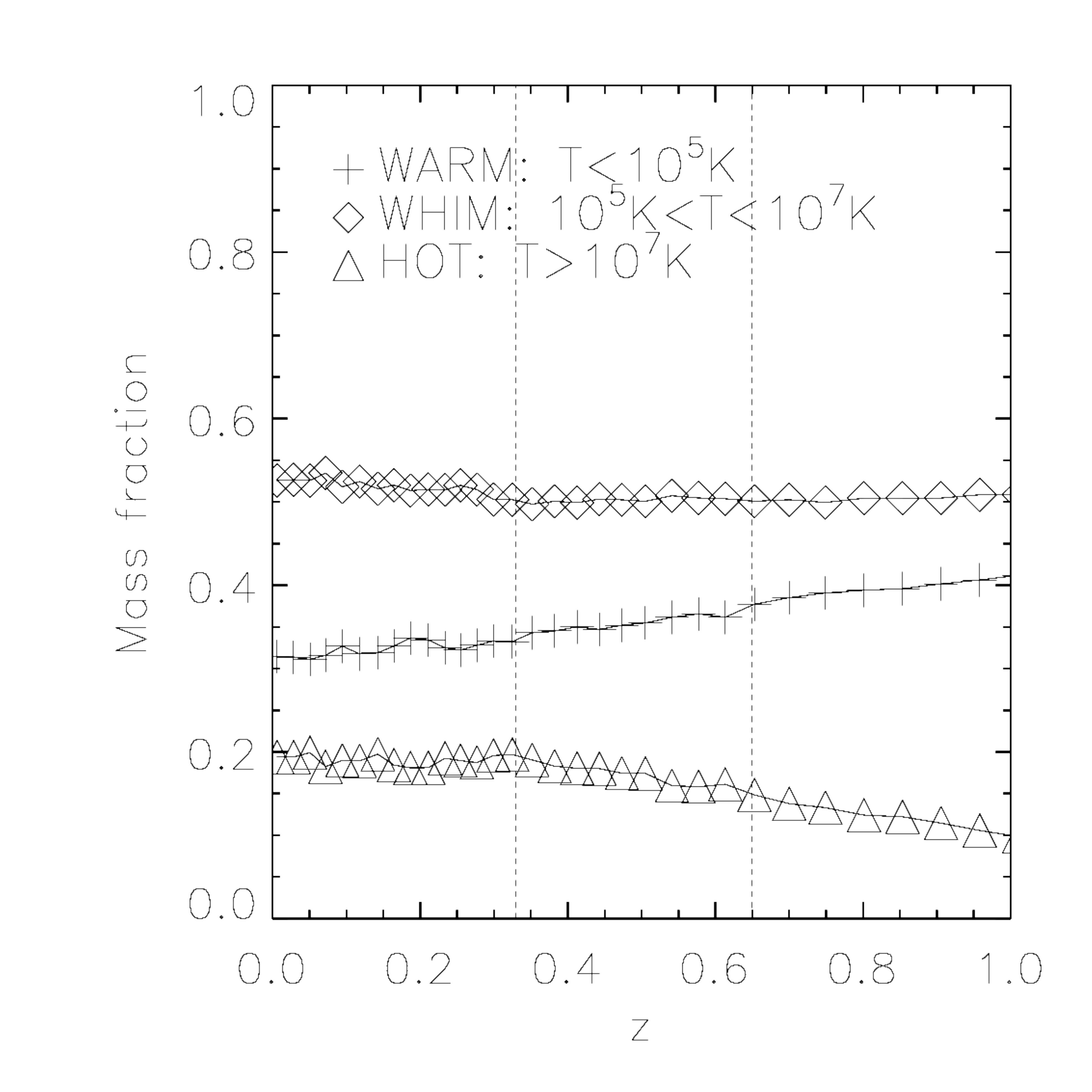}}
\caption{Redshift evolution of the fraction in volume (top panel) and  in mass (bottom panel) occupied by different IGM gas components within the whole simulated volume. Vertical lines in both panels enclose the temporal region within which we will compute the cluster emission.}
\label{fig:zevol}
\end{center}
\end{figure}

\begin{figure*}
\begin{center}
{\includegraphics[width=17.0cm]{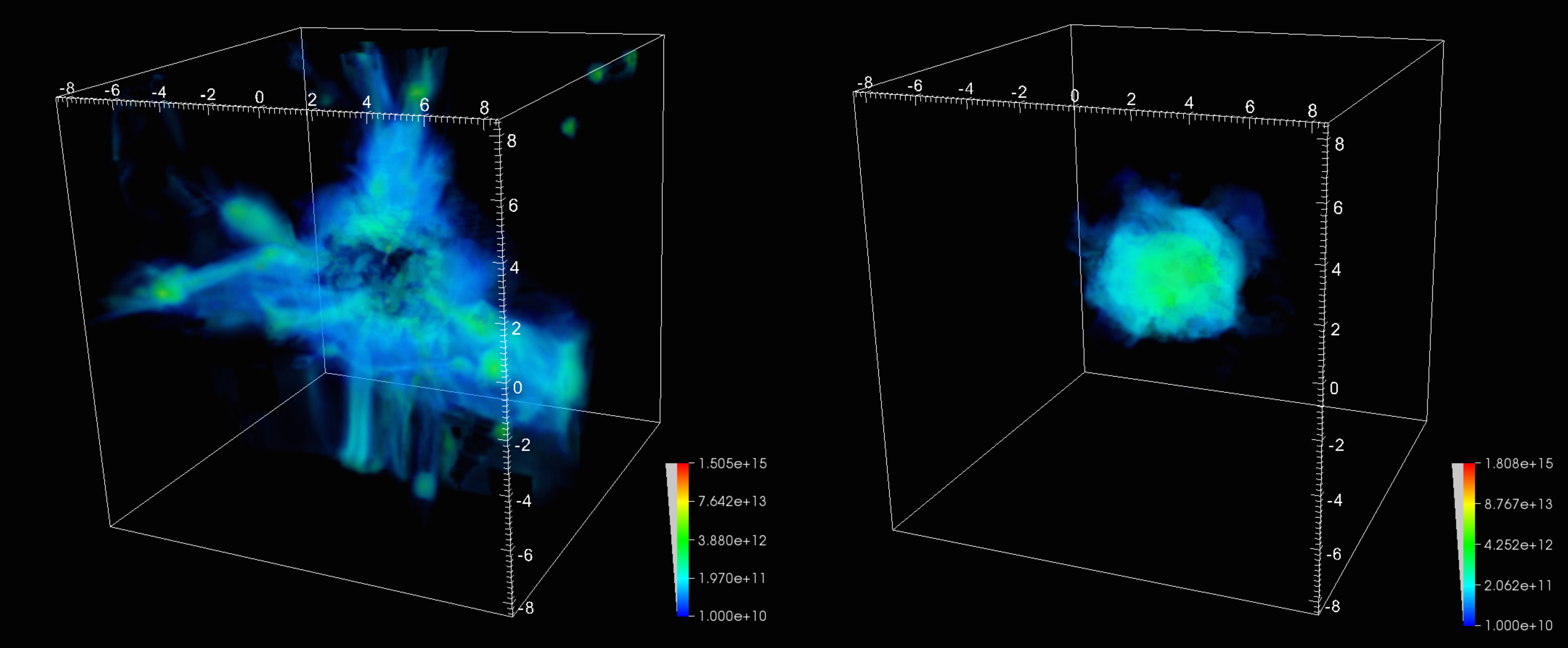}}
\caption{3D distribution of the WHIM ($10^5\, K<T<10^7\, K$; left panel) and ICM ($T>10^7\, K$; right panel) gas components within a simulated volume of $\sim$ 17 Mpc$^3$ centered at the position of the most massive galaxy cluster at $z=0$. The color code stands for the gas density in units of M$_{\odot}/$Mpc$^3$.}
\label{fig:temp}
\end{center}
\end{figure*}

\begin{figure}
\centering
{\includegraphics[width=7cm]{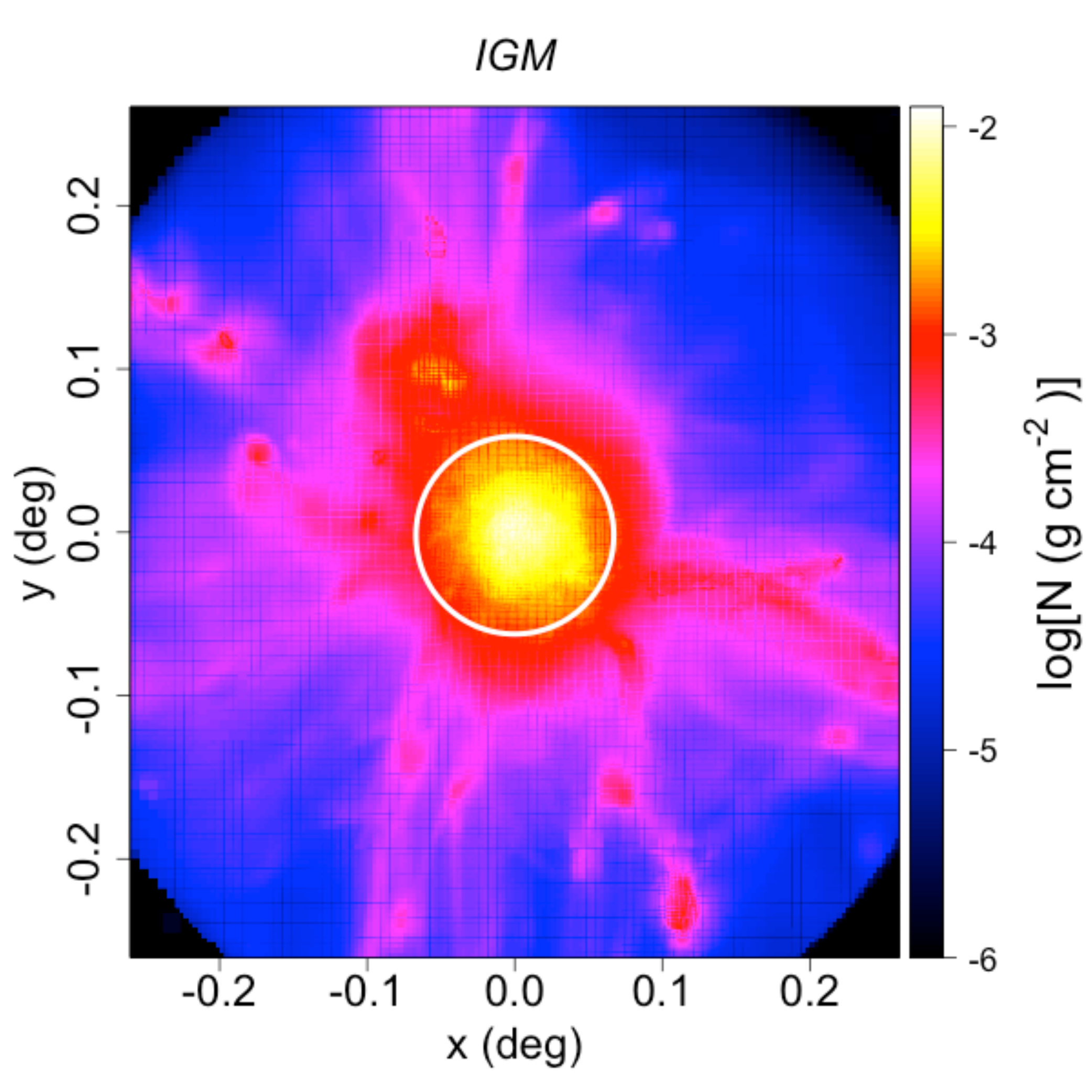}}\\
{\includegraphics[width=7cm]{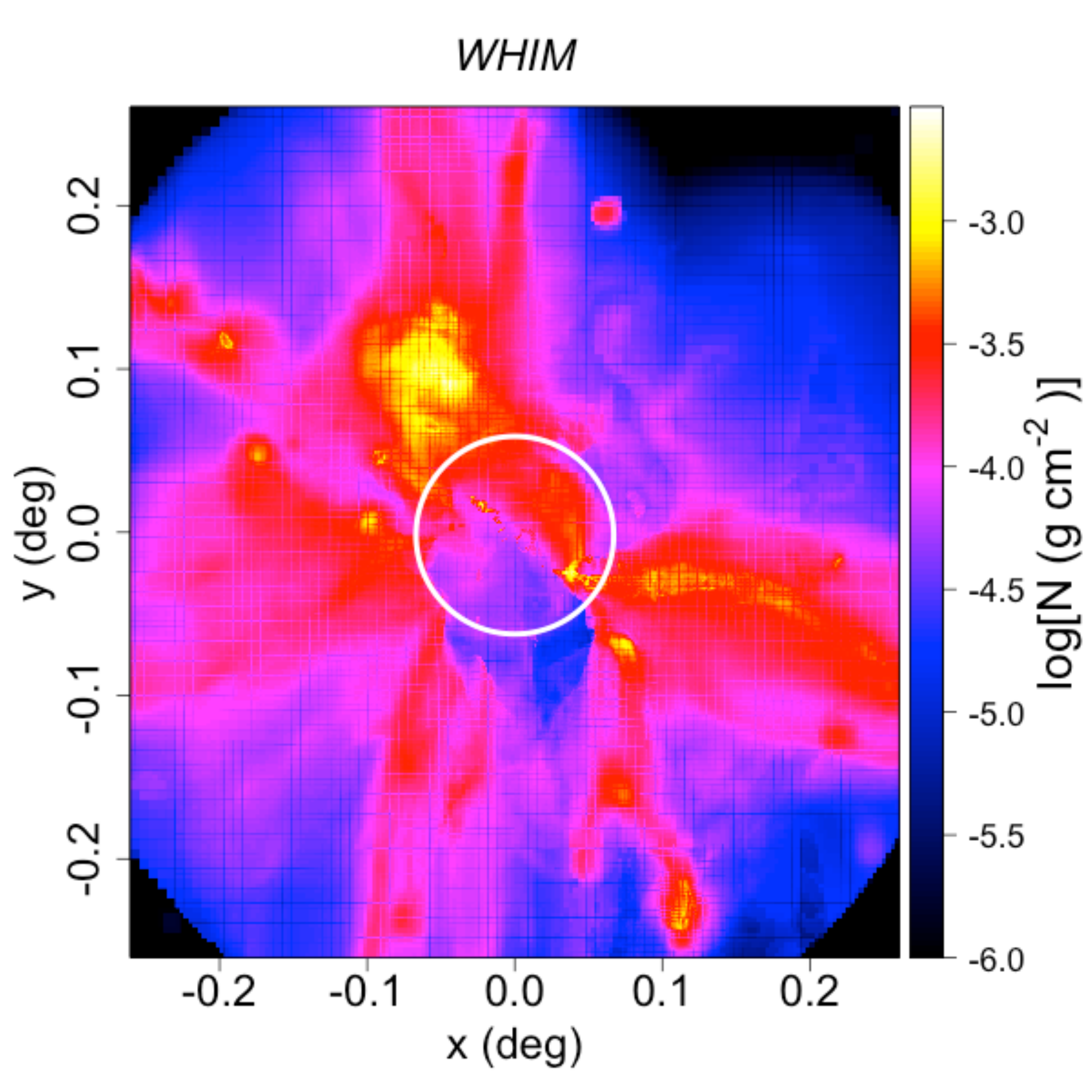}}
\caption{Surface density maps for the IGM and the WHIM gas components (top and bottom panels, respectively) within 
a region of $\sim0.6$$^{\circ}$ around the main and central cluster in the simulation. The white circle represents the virial radius of the central cluster. The value in each pixel is obtained by summing up the contribution of each gas element along the line of sight in the redshift range $0.65< z < 0.33$. The color bar stands for the gas column density in g cm$^{-2}$ in log scale.} 
\label{fig:rho-Lcut}
\end{figure}

\section{Results}
\label{sec:results}

\subsection{Distribution and evolution of different gas components}
\label{subsec:general}

Before analyzing in detail the characteristic WHIM emission through different physical processes, we study the global redshift and spatial distribution of different IGM gas components in our simulated volume.  

Figure \ref{fig:zevol} shows the evolution since $z=1$ of the fractions in volume (top panel) and in mass (bottom panel) occupied by different IGM gas components within the whole  ($40\,  Mpc$) simulated box. In particular, we show separately the evolution of three different IGM regimes classified according to the gas temperature: the warm ($T< 10^5 K$), the WHIM ($10^5 K < T < 10^7 K$), and the hot ($T> 10^7 K$) gas components. Several interesting trends can be drawn from this figure. As expected,  most of the volume is dominated by the warm  ($\sim90\%$) and the WHIM ($\sim10\%$) gas phases, which are the most spread components throughout the redshift evolution. On the contrary, the hot gas component, mainly associated to collapsed regions such as galaxy clusters, represents only a few per cent of the simulated volume. In terms of mass we note that, as evolution proceeds and the largest cosmic structures begin to be formed, the amount of gas in the hot phase increases from $~10\%$ at $z=1$ up to $20\%$ at $z=0$. Simultaneously, while the amount of gas in the warm component slightly decreases down to $~30\%$ at $z=0$, the amount of WHIM mildly augments from $50\%$ at  $z=1$ up to a value of $\sim55\%$ at the present epoch.  
These values are in broad agreement with the results obtained from previous numerical analyses \citep[e.g.][]{Cen_1999}. Nevertheless, we point out that the inclusion  in simulations of different physical processes or the assumption of  slightly different cosmologies may affect the obtained evolutions \citep[e.g.][]{Dave_2001, Cen_2006, Tornatore_2010, Ursino_2014}. On the other hand, although employing larger cosmological boxes allows for a larger number of group- and cluster-size haloes, the simulated box size does not significantly affect the baryon budget \citep[e.g.][]{Tornatore_2010}.

From now on, unless otherwise specified, we will focus on two types of gas: the whole IGM, formed by all the gas within our simulated volume, and the WHIM component, formed by the gas with $10^5 K < T < 10^7 K$. We will analyse the emission from these two gas components in different wavebands and through different physical mechanisms. Moreover, as explained in Section \ref{subsec:part3}, we will only consider a subvolume of radius  $\sim 8.5$ comoving Mpc around the centre of the main halo. 
Within this volume, Fig.~\ref{fig:temp} shows the 3D distribution of the WHIM ($10^5 K < T < 10^7 K$) and the hot ICM  ($T> 10^7 K$) gas components. Both gas components have been selected according only to the gas temperature, whereas the color code stands for their corresponding densities in units of M$_{\odot}/$Mpc$^3$. From this figure, it is clear that the WHIM is distributed around the main cluster with a filamentary structure throughout the considered volume, whereas the ICM is confined to the more central cluster region.

For the sake of completeness, Fig.~\ref{fig:rho-Lcut} shows the projection along the line of sight\footnote{Throughout the paper we will consider the line of sight as parallel to the z-axis of the simulation box. We have verified that considering different projections does not have important implications on our conclusions.}  of the density maps associated to the IGM and the WHIM gas components  within our volume of interest. In the IGM map (top panel), high-density regions (yellowish colors) are clearly correlated with the main haloes and subhaloes in our simulation. As expected, due to the characteristics of our simulation  and to the volume selection we have  performed, the central region is dominated by the largest cluster. If we only take into account the WHIM component (bottom panel),  the most massive (and hottest) virialized systems are almost completely removed and what remains is the distribution of the warm-hot gas, mainly associated  to the smallest objects. This component  is distributed throughout the whole volume with a tendency to reside around the location of the main halo, showing as well  a more filamentary distribution that extends out to the limits of the image. However, we note that, due to projection effects, part of the filamentary structure is diluted with the diffuse background, making more difficult its detection \citep[e.g.][]{Croft_2001, Roncarelli_2006}.

\subsection{Thermal X-ray emission}
\label{subsec:xrays}

\begin{figure*}
{\includegraphics[width=5.cm]{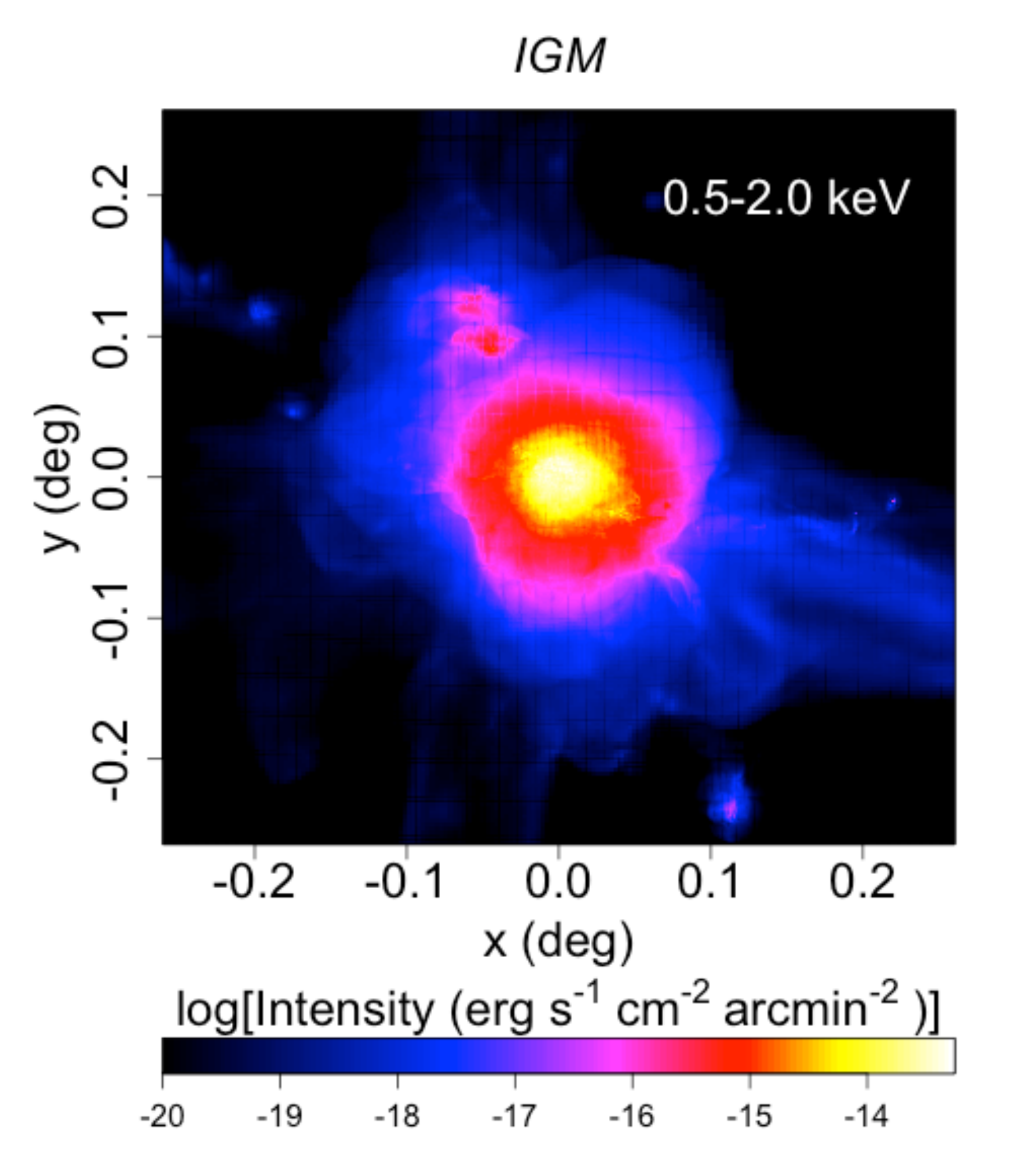}}
{\includegraphics[width=5.cm]{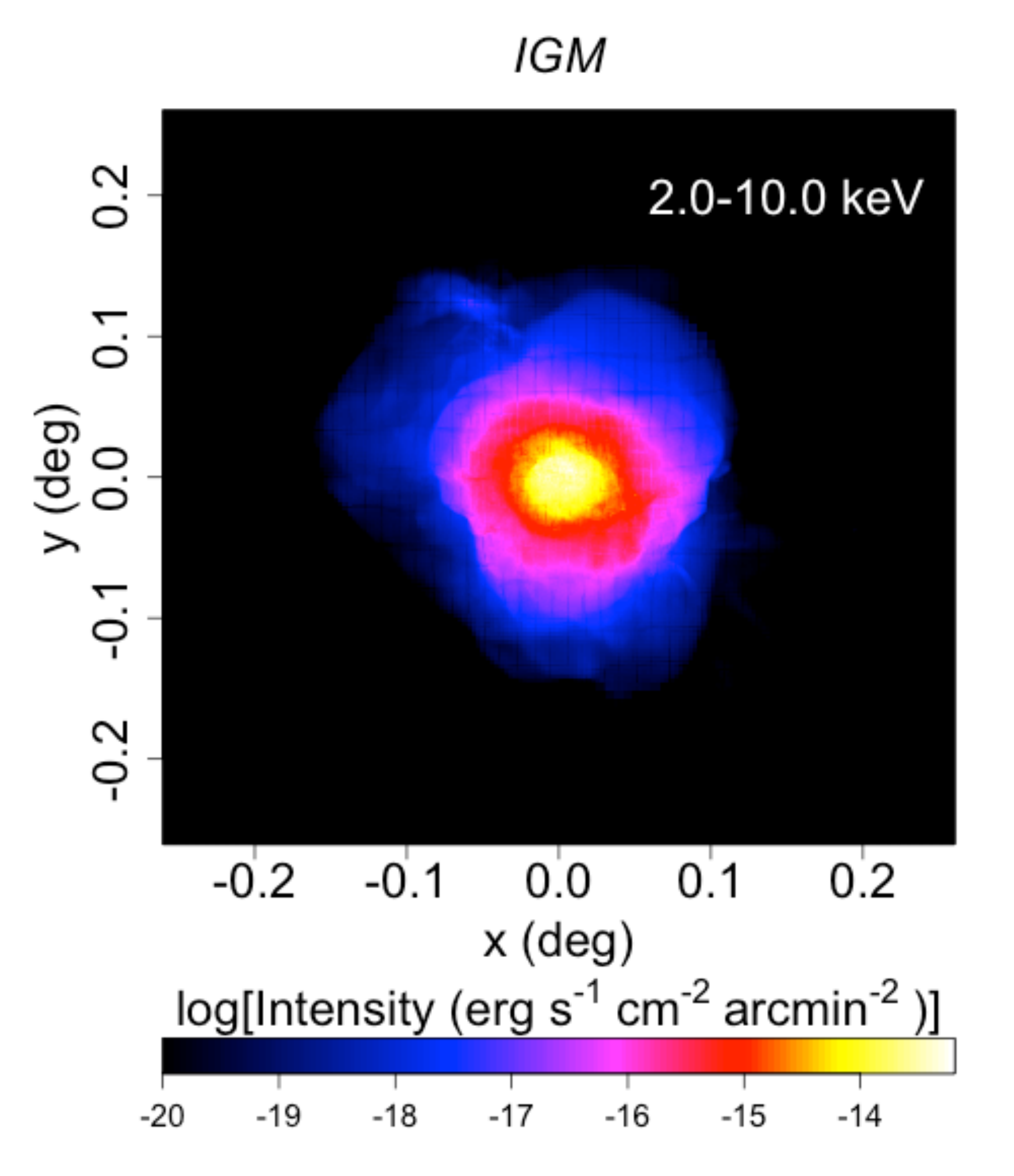}}
{\includegraphics[width=5.cm]{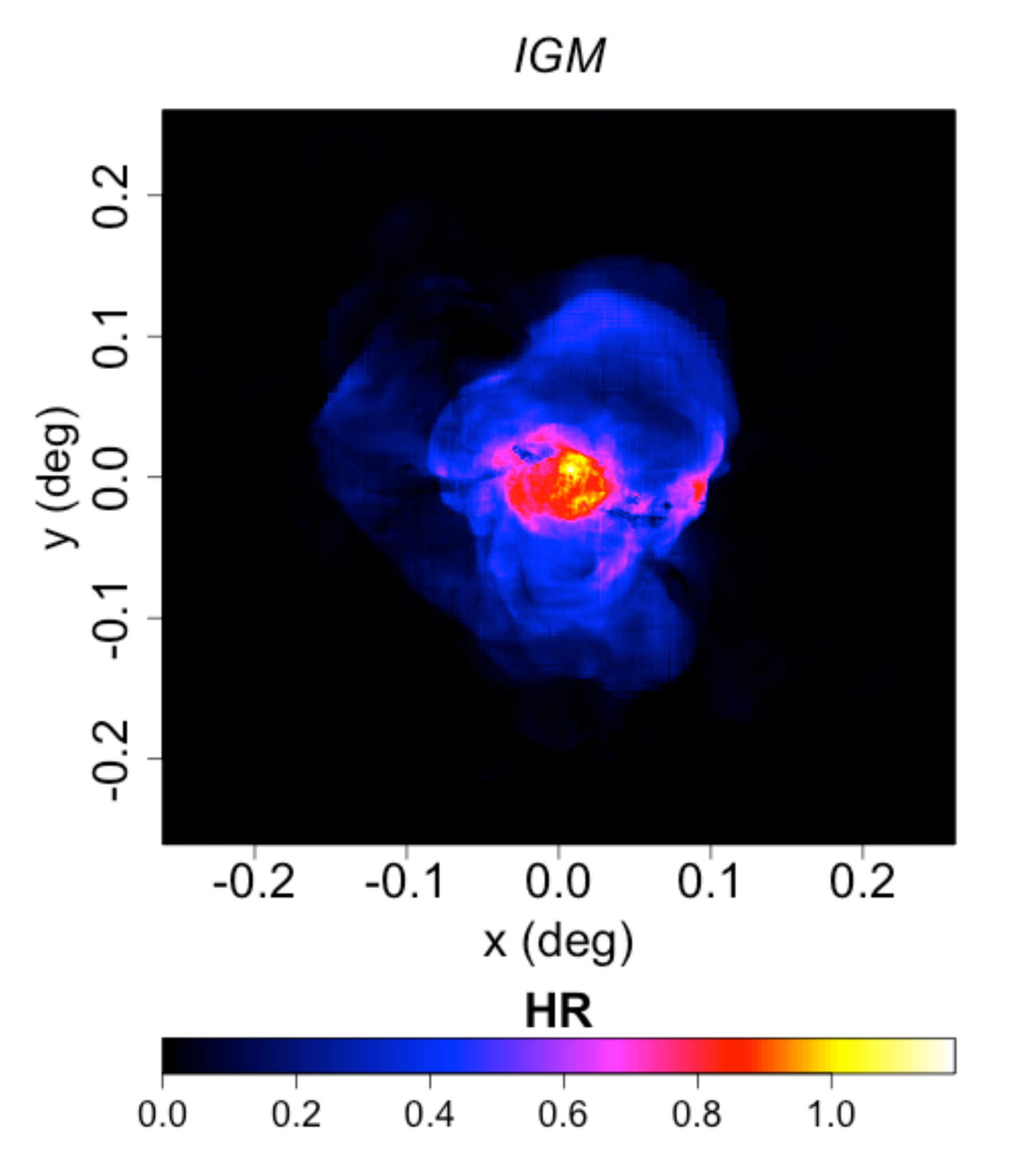}}\\
{\includegraphics[width=5.cm]{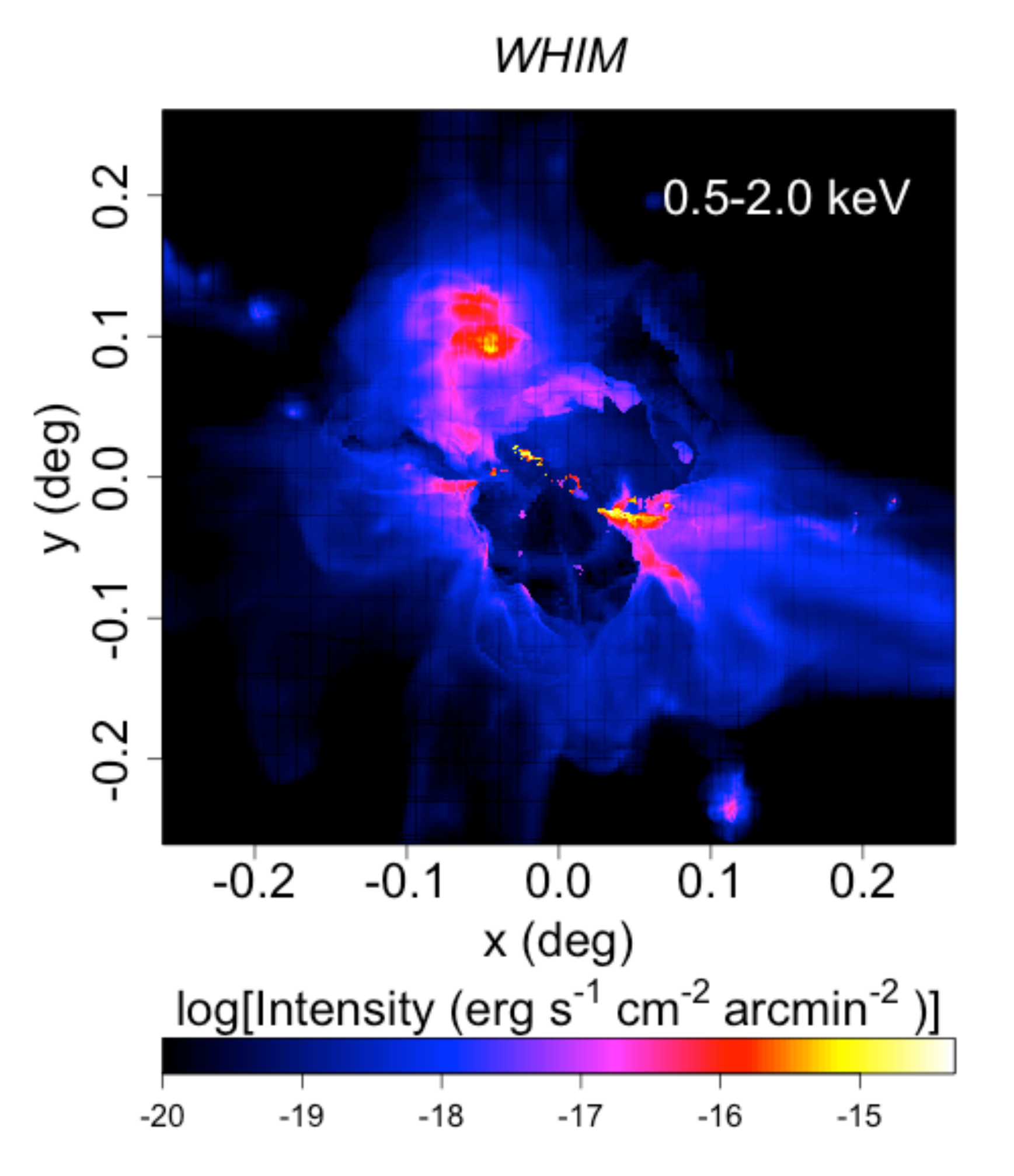}}
{\includegraphics[width=5.cm]{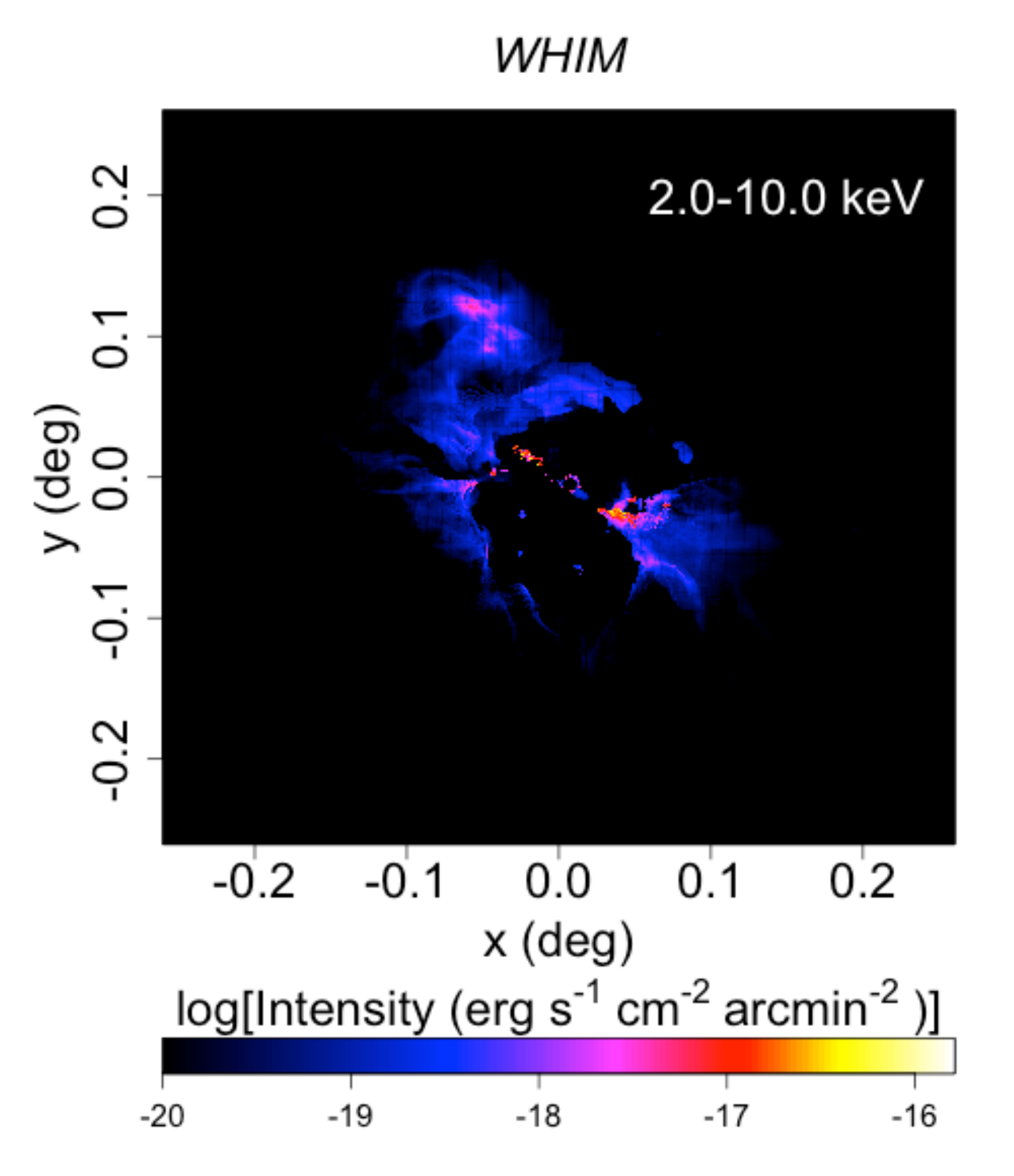}}
{\includegraphics[width=5.cm]{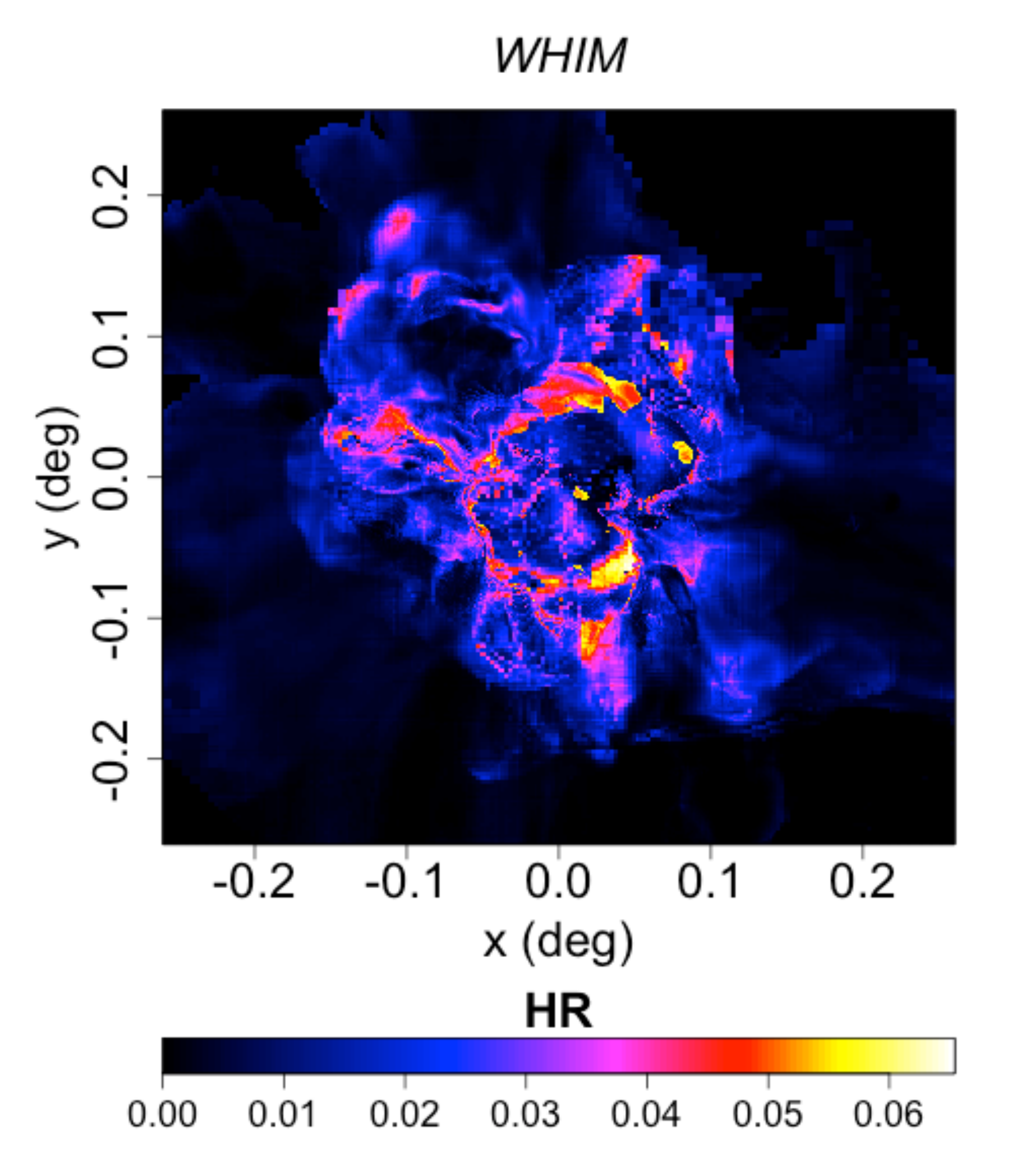}}
\caption{{\it Left and middle columns:} X-ray thermal emission maps associated to the IGM and the WHIM gas components (top and bottom rows, respectively) at the soft ($0.5-2$ keV), and hard ($2-10$ keV) energy bands. The emission, given in units of \ergscmarcmin,  is plotted in logarithmic scale according to the color code at the bottom of each panel. Each map is sampled with  $3200^2$ pixels over a region of $\sim0.6$ squared degrees around the main central halo.
 {\it Right column:} Hardness ratio (HR) maps for the IGM (top panel) and the WHIM (bottom panel) gas components, being HR the ratio of the hard over the soft X-ray emissions.}
\label{fig:th}
\end{figure*}

As explained in Section \ref{subsec:part2}, we employ {\sc SPEV} in order to compute maps of thermal X-ray emission associated to the IGM and the WHIM components in the soft  $(0.5-2$ keV) and in the hard $(2.0-10.0$ keV) X-ray energy bands.

In the following, each map has been made in order to cover a region of $\sim0.6$ squared degrees around the main central halo and is sampled with  $3200^2$ square pixels. This provides a pixel size of $\sim0.68''$, a value in line with the FWHM of \xmm\ and \chandra\ point-spread functions (PSF). The forthcoming \athena\footnote{http://www.the-athena-x-ray-observatory.eu} telescope, which has as one of its key science goals the detection and characterization of the WHIM \citep[e.g.][]{Athena_wp, Athena_2017},  will have better capabilities and will provide a detection limiting flux of $10^{-17}$ \ergscm\ in the soft band. Therefore, as for the detection limit of our synthetic maps, we decide to adopt an optimistic intensity threshold of $10^{-17}$ \ergscmarcmin.
However, we will not tune our maps with the response function or the particular capabilities of any specific observational facility.

According to this procedure, the synthetic X-ray emission maps, shown in the left and middle columns of Fig.~\ref{fig:th}, correspond to the density distributions of the IGM and the WHIM displayed in Fig.~\ref{fig:rho-Lcut}. Qualitatively, these synthetic maps are  in line with X-ray observations of the IGM. 
Indeed, as already shown in previous studies \citep[e.g.][]{Croft_2001}, we confirm that most of the IGM thermal emission is observed in the soft regime, being the hard band mainly contributed by the emission from the largest (hottest) groups and clusters. Moreover, the map obtained in the soft band highlights the existence of very small structures or density clumps with a considerable soft emission which are not visible in the hard regime. The emission from these structures is mainly contributed by metals, which tend to be located in mildly overdense warm/hot regions \citep[e.g.][]{Cen_1999a}. 
As for the emission associated to the WHIM,  there is a kind of geometrical characterization of the X-ray signals from IGM and WHIM at both energy bands: while the former is mainly connected to high-density regions and is dominant in the cluster core, the latter is mainly located in outer cluster regions around the cluster centre. Moreover, given its lower temperatures and densities, the hard emission associated to the WHIM seems fainter than the emission from the IGM \citep[e.g.][]{Croft_2001, Roncarelli_2006}. In the soft band, however, the contribution from metals would highlight the more filamentary and extended emission from the WHIM.

\begin{table}
	\begin{tabular}{ccc}
	\hline
	Energy band & IGM  & WHIM \\
	 (keV) & (\ergscmdeg)    & (\ergscmdeg)\\ \hline
	{\small $0.5-2$} & {\small $(5.09^{-0.80}_{+62.73})\times10^{-13}$}  & {\small $(0.91^{-0.44}_{+3.98})\times10^{-13}$}\\ \\
	{\small $2-10$}   &  {\small $(4.73^{-0.77}_{+68.98})\times10^{-13}$} & {\small $(0.71^{-0.44}_{+1.29})\times10^{-13}$}	\\
	\hline
	\end{tabular}
\caption{Median pixel values of the X-ray intensity maps shown in Fig.~\ref{fig:th} for the IGM and the WHIM gas components at two different energy bands.
Errors  represent the 15th and 85th percentiles of  each distribution. Median values and errors are obtained from all the pixels with intensity above 
$10^{-17}$ \ergscmarcmin, which corresponds to $3.6\times10^{-14}$ \ergscmdeg.}
\label{tbl:values}
\end{table}

Quantitatively, there is also a good match between the X-ray surface-brightness associated to our simulated cluster and recent  observations of massive galaxy clusters \citep[e.g.][]{Eckert_2015}.
We report in Table \ref{tbl:values} the median X-ray intensity values (together with the corresponding 15th and 85th percentiles) obtained for  the IGM and the WHIM gas distributions at different wavebands. The analysis of the pixel intensity maps as a function of the median instead of the mean values of the different distributions 
is less  affected by extreme pixel intensities and provides, therefore, a better idea of the `characteristic' emission.    
Only in this case, in order to facilitate the comparison with previous works, we show the brightness in units of  \ergscmdeg.
In doing such a comparison, we need to keep in mind that we only have one large galaxy cluster located in the centre of the simulated box, whereas most of previous studies were based on larger cosmological volumes  \citep[e.g.][]{Croft_2001, Phillips_2001,Roncarelli_2006, Roncarelli_2012, Ursino_2014}. Despite this issue, the maximum  brightness associated to the centre of our galaxy cluster, within $\sim~10^{-10} -10^{-9}$ \ergscmdeg, is in broad agreement with the values  associated in cosmological boxes  to  large galaxy clusters \citep[e.g.][]{Croft_2001, Roncarelli_2006}.  
Moreover, the median values reported in Table \ref{tbl:values}  also seem to enclose previous independent estimates for the IGM and the WHIM. 
As an example, in the soft band,  \citet{Croft_2001} and \citet{Roncarelli_2006} reported, respectively, mean brightness values of $\sim6.58\times10^{-12}$ \ergscmdeg\ and  
$\sim4.06\times10^{-12}$ \ergscmdeg\ for the IGM and $\sim4.15\times10^{-13}$ \ergscmdeg\ and  $\sim1.68\times10^{-12}$ \ergscmdeg\ for the WHIM component.
In our case, we obtain a mean emission in the soft band of  $6.34\times10^{-12}$ \ergscmdeg\ for the IGM and $3.18\times10^{-13}$ \ergscmdeg\ for the WHIM. Despite the different implementations, these values are in line with those previously reported by \citealt{Croft_2001} \citep[see also][]{Bryan_2001}.
In the hard band, our mean intensity values for IGM and WHIM are, respectively, $6.94\times10^{-12}$ \ergscmdeg\ and $8.68\times10^{-14}$ \ergscmdeg, slightly higher than the results obtained by \citealt{Roncarelli_2006} ($(1.01\pm1.53)\times10^{-12}$ \ergscmdeg\ and $(2.92\pm2.46)\times10^{-14}$ \ergscmdeg, respectively).
From these results we find that, on average,  in our simulation the WHIM X-ray emission corresponds to only $\sim 5$\% and $\sim1$\% of the IGM emission in the soft and hard X-ray bands, respectively. This relatively small contribution makes even more difficult the detection of the WHIM using X-rays observations.

As for the comparison with observational data, \citet{Galeazzi_2009}  employed a statistical approach on several \xmm\ observations to derive a WHIM X-ray contribution to the diffuse X-ray background of $(12\pm 5)$\%, a value which is roughly consistent with our soft band estimate.
On the other hand,  the analysis of \chandra\ deep fields by \citet{Hickox_2007} provided an upper limit for the WHIM emission in the $0.65-1$ keV energy band of $(1.0\pm0.2)\times10^{-12}$\ergscmdeg  \citep[see also observational estimates in the soft band by e.g.][]{Briel_1995, Worsley_2005}. Interestingly, and in agreement with previous numerical estimates, our predictions are well-below this value.

\begin{figure}
\begin{center}
{\includegraphics[width=6.cm]{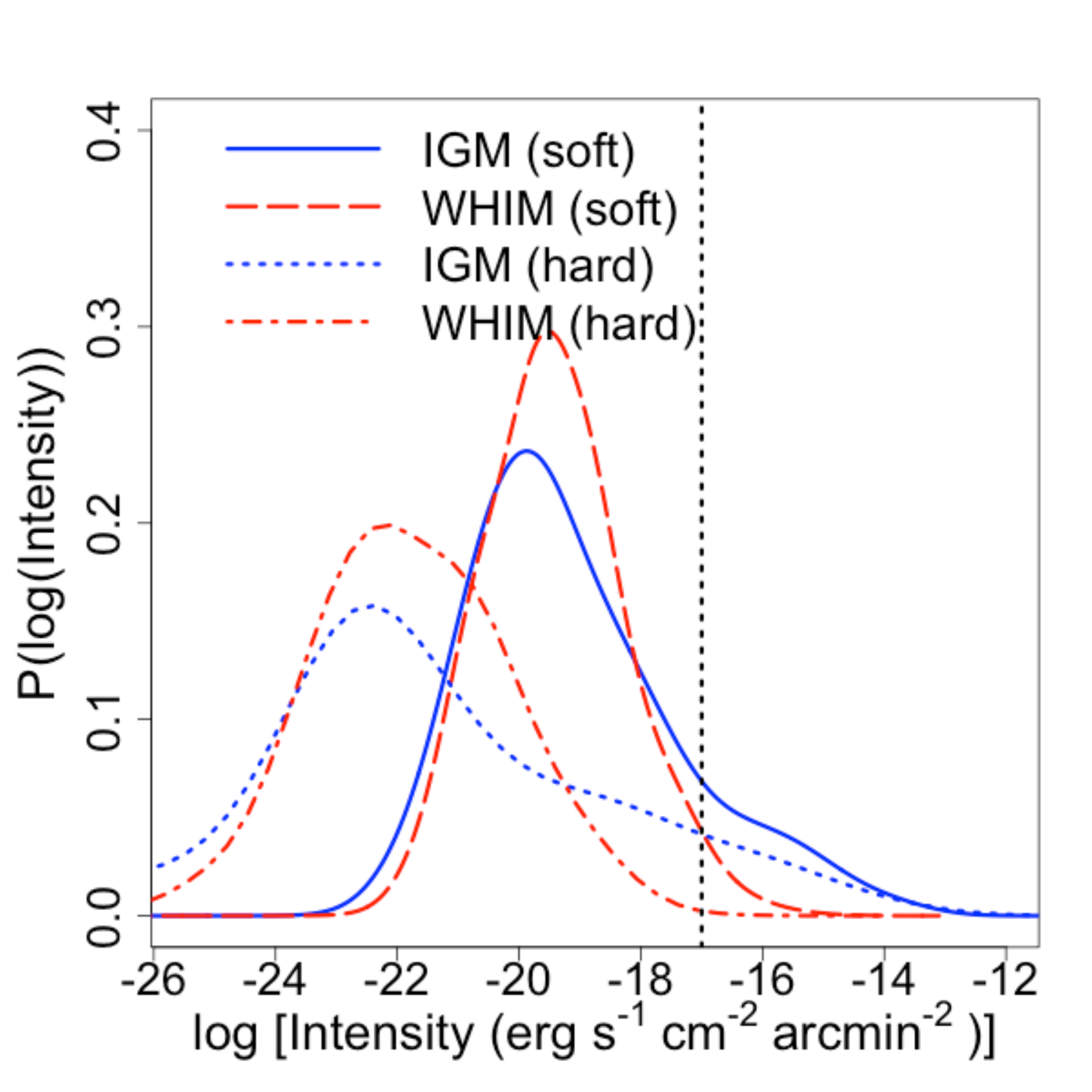}}     
\end{center}
\caption{Probability distribution function of the thermal X-ray intensity associated to the all the pixels within the maps  shown in Fig.~\ref{fig:th}. The distributions are shown for the IGM (blue lines) and the WHIM (red lines) in the soft ($0.2-0.5$ keV) and hard ($2-10$ keV) X-ray energy bands. The vertical dotted line represents a threshold detection intensity of $10^{-17}$ \ergscmarcmin.}
\label{fig:pdf}
\end{figure}

In order to highlight and compare the different emissions from the IGM and the WHIM, we show in the right column of Fig.~\ref{fig:th} the maps of the hardness ratio (HR) for each gas phase. To maximize the differences, the HR is defined as the ratio of the pixel intensities in the hard ($2-10$ keV) to the soft ($0.5-2.0$ keV) energy bands \citep[see e.g.][]{Croft_2001, Roncarelli_2006}. 
In the case of the IGM, the HR map clearly highlights the cluster centre, where the HR values lay above $\sim 0.7-0.8$, reaching the maximum ($\sim 1.18$) in the densest region. Within the cluster centre and its outskirts there is a region with lower HR values. 
Regarding the WHIM, its HR map is clearly anti-correlated with that of the IGM,  it shows a more extended distribution, and it is characterized by mean HR values smaller by a factor $\sim9$. Indeed, contrarily to the IGM, the highest HR values in the WHIM map ($\sim0.065$) are mostly detected in a thin region close to the cluster limits. In this map, however, any trace of filamentary structures has been almost completely diluted \citep[see also][]{Croft_2001, Roncarelli_2006}.  
Quantitatively, the median HR values that we obtain for the IGM ($\sim0.01$) and for the WHIM ($\sim0.006$) are slightly below the estimates from \cite{Roncarelli_2006}, who obtained $0.03$ and $0.01$, respectively.

Figure \ref{fig:pdf} shows the probability distribution functions of the thermal X-ray signal associated to  the IGM (blue lines) and to the WHIM (red lines) maps shown in Fig.~\ref{fig:th} for the two  different energy bands we analyse. For very low intensities ($\mincir10^{-20}-10^{-19}$ \ergscmarcmin) in a given energy band, associated to very low-temperature gas, IGM and WHIM distributions are very similar. Moreover, most of the pixel maps contribute to this undetectable intensity region. However, as expected, both distributions differ at the high-intensity end, since the high-temperature  gas, responsible of this emission, is included in the IGM but not in the WHIM component. Indeed, in the high-intensity region (above the detection limit marked by the vertical dotted line), the WHIM distribution tends to zero faster than that of the IGM, being almost undetectable in the hard band. Another remarkable fact is that, as already noted by other authors \citep[e.g.][]{Croft_2001, Roncarelli_2006}, as we go from the soft to the hard energy band, both distributions become wider, since the number of pixels with null flux increases.
  
To conclude, we stress that a direct comparison with previous works, either numerical or observational,  is far from being straightforward.  In this sense, we expect significant discrepancies with other numerical analyses due to not only the use of different box sizes (and, therefore, different cluster catalogues), but also due to the assumption of different cosmological parameter values, the inclusion of different physical processes, the use of different WHIM definitions, or the implementation of different methods to compute the associated X-ray emission \citep{Croft_2001, Bryan_2001, Roncarelli_2006, Tornatore_2010, Roncarelli_2012, Ursino_2014}. 

Moreover, we would like to point out that the distribution of metals is sensitive to the treatment of different feedback processes such as galactic winds from  SNIa or AGN feedback \citep[e.g.][]{Biffi_2017,Biffi_2018}. Not accounting for these effects, such as in our current simulation, could thus be a significant source of uncertainties in estimating  the X-ray emission from galaxy clusters, as already demonstrated in a number of dedicated studies \citep[e.g.][]{Bertone_2010a, Tornatore_2010, Roncarelli_2012}. In general,  the inclusion of galactic winds tends to increase the transport of metals from star-forming regions, where they are mainly produced, to outer and less dense IGM regions, producing an increase in the emission from faint lines, especially in the WHIM component \citep[][]{Roncarelli_2012}. Contrarily, \cite{Roncarelli_2012} also showed that the inclusion of AGN feedback, which tends to prevent the high-redshift collapse of dense regions, produces a reduction in line emission from both IGM and WHIM. On the other hand, as shown by \citet{Bertone_2010a}, it seems that the distribution of metals in the WHIM is quite stable to different feedback schemes, suggesting that the main point is how the distribution of the gas density is affected by different feedback processes. Therefore, given the complex interplay between different feedback processes and prescriptions, it is difficult to perform a proper comparison between different simulations. However, we would like to highlight that, despite the unavoidable uncertainties and the assumptions adopted in our numerical approach, our results seem to be in a remarkably level of agreement with current X-ray estimates obtained with more complex simulations.

From an observational point of view, we will have to wait for the launch of  \athena, by 2028, to obtain an unprecedented X-ray view of the cosmic web. In comparison with current and planned X-ray facilities, \athena\ will have a significantly improved spectral ($\sim 2.5$ eV) and angular ($\sim 5$ arcsec) resolutions, high sensitivity  and a large field of view ($\sim 40$ arcmin). With these capabilities, it will be able to reveal some of the main cosmic web properties by measuring clusters velocity fields, the group and cluster gas entropy profiles out to high redshifts, and the distribution and evolution of metals. Future observations with \athena\ will be also crucial to map the outskirts of clusters, as well as to discover the progenitors of present day galaxy clusters at $z\sim 2.5-3$. Regarding the WHIM, \athena\ will be able to detect and measure, with unprecedented precision, the distribution of gas in cosmic filaments through both absorption and emission processes, providing therefore definitive clues on the nature and evolution of the warm-hot gas component \citep[see e.g.][for further details]{Athena_wp, Athena_2017}.

\subsection{SZ observations}
\label{subsec:sz}

\begin{figure*}
{\includegraphics[width=5.cm]{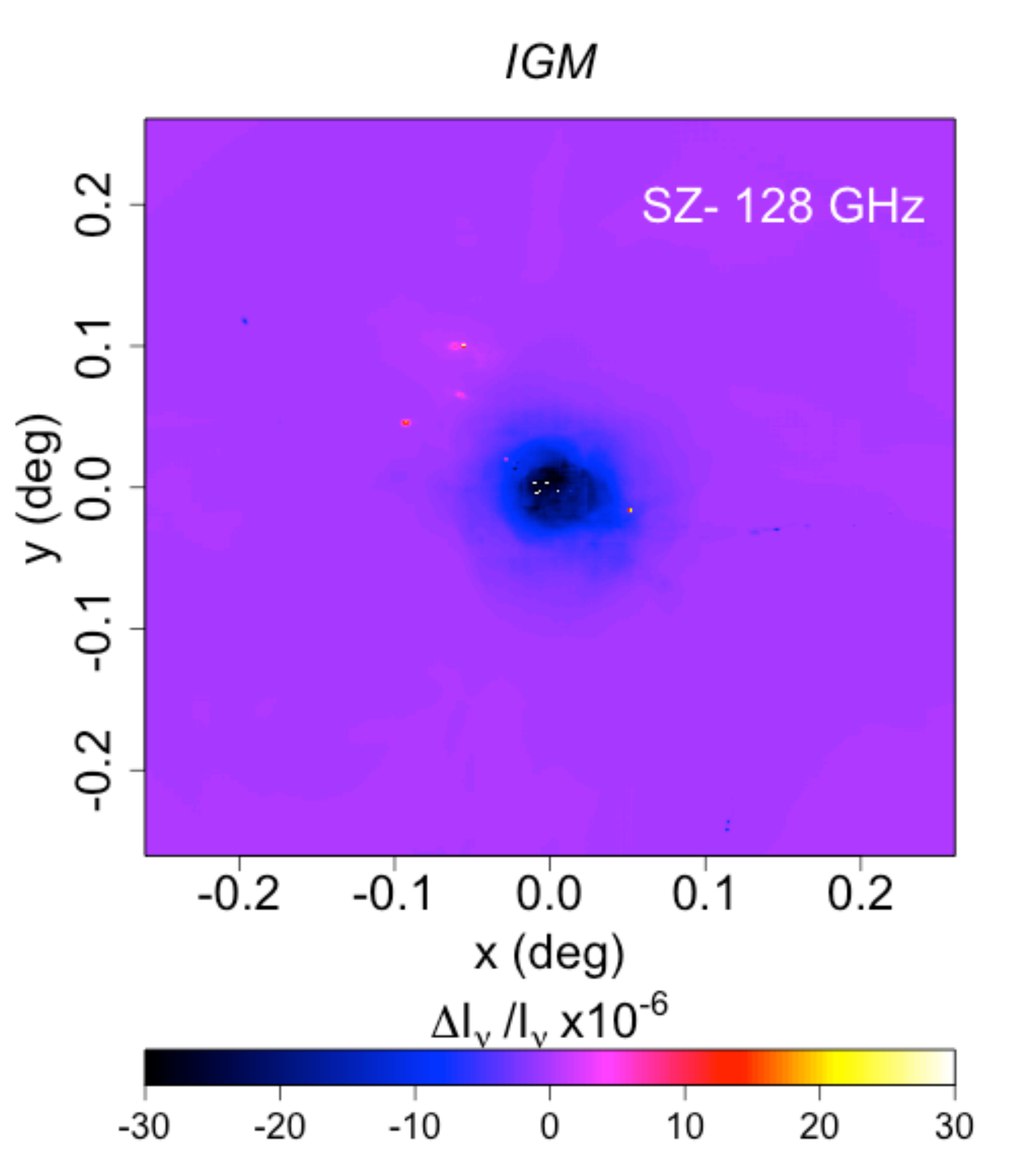}}
{\includegraphics[width=5.cm]{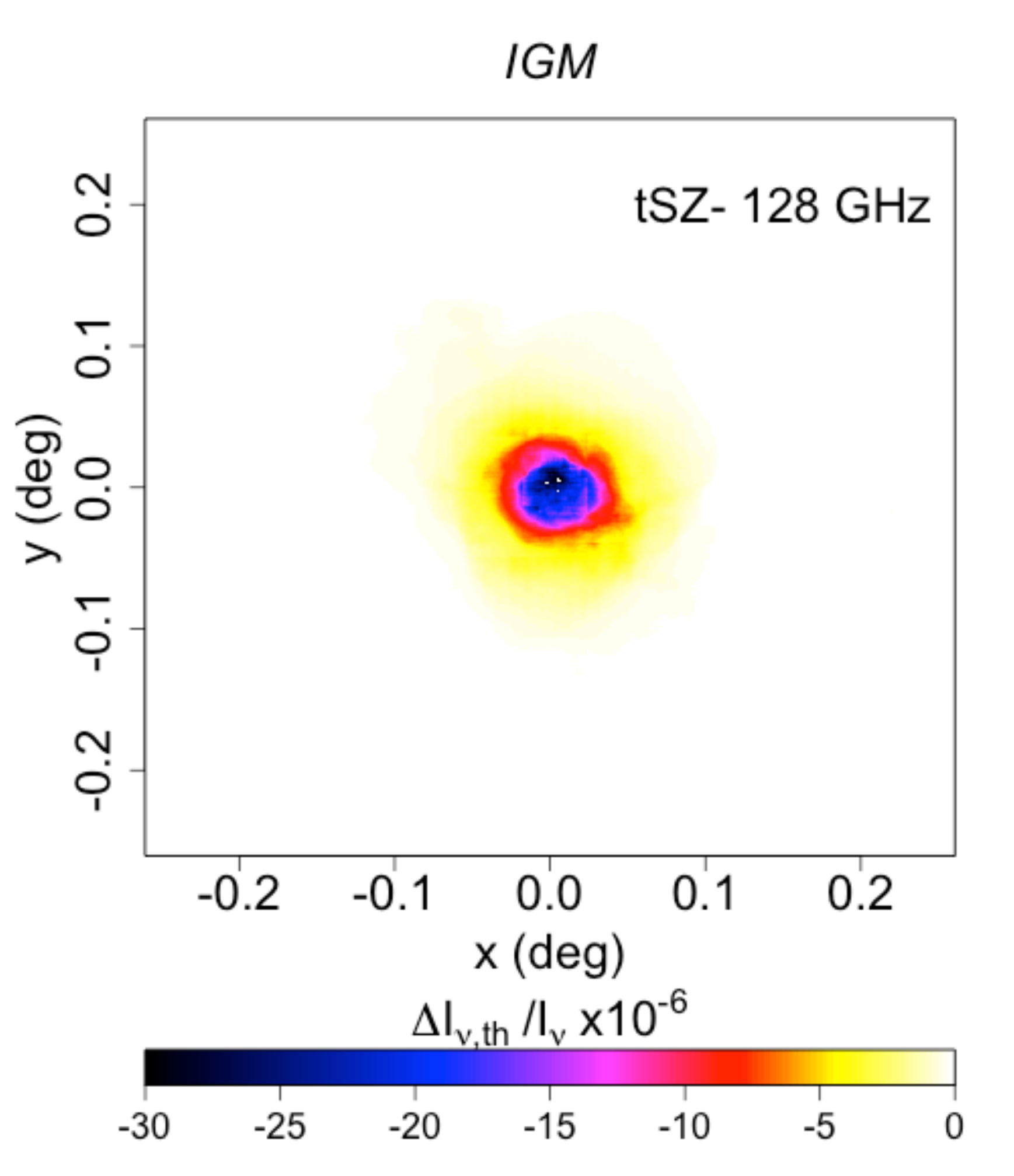}}
{\includegraphics[width=5.cm]{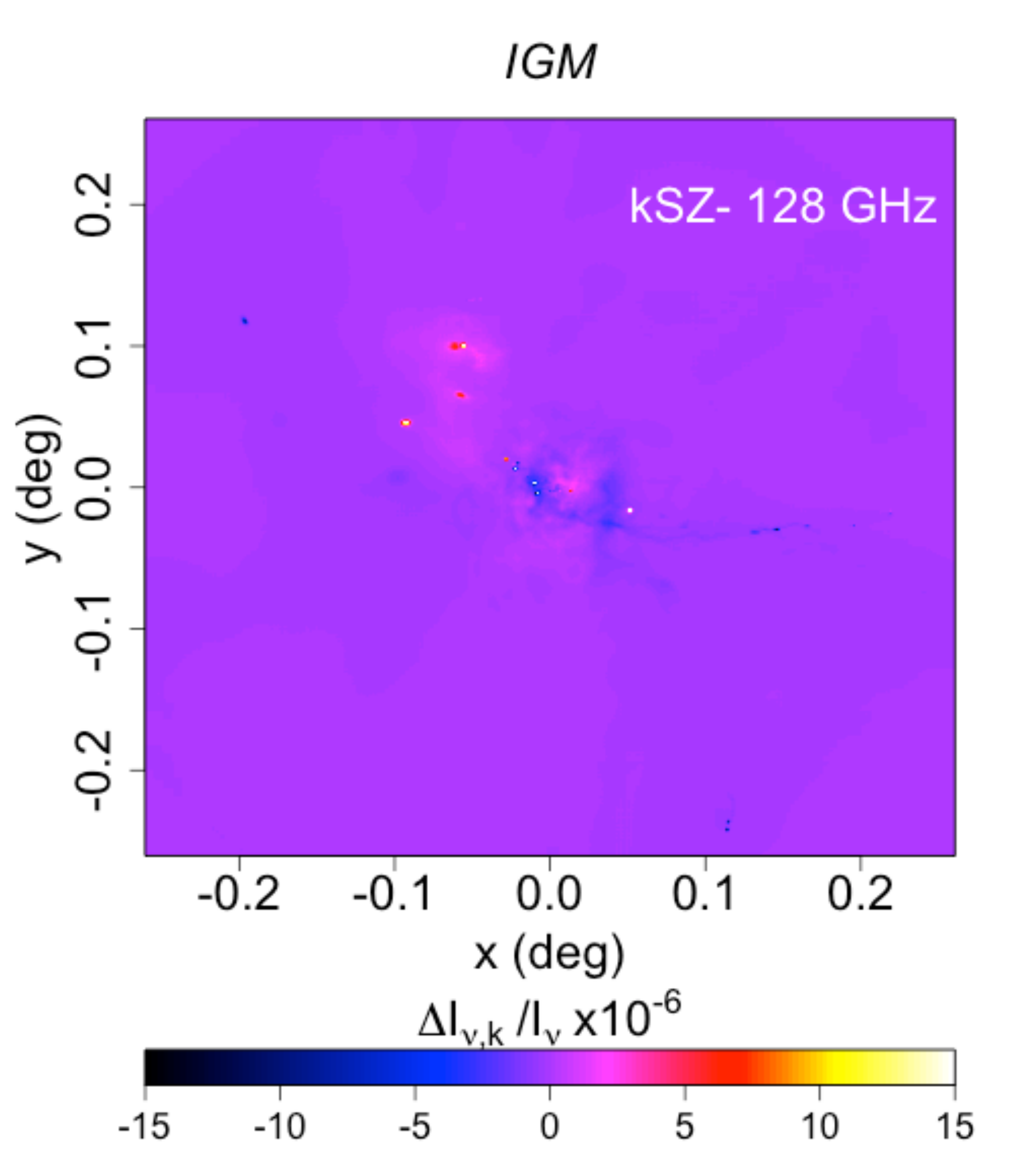}}\\
{\includegraphics[width=5.cm]{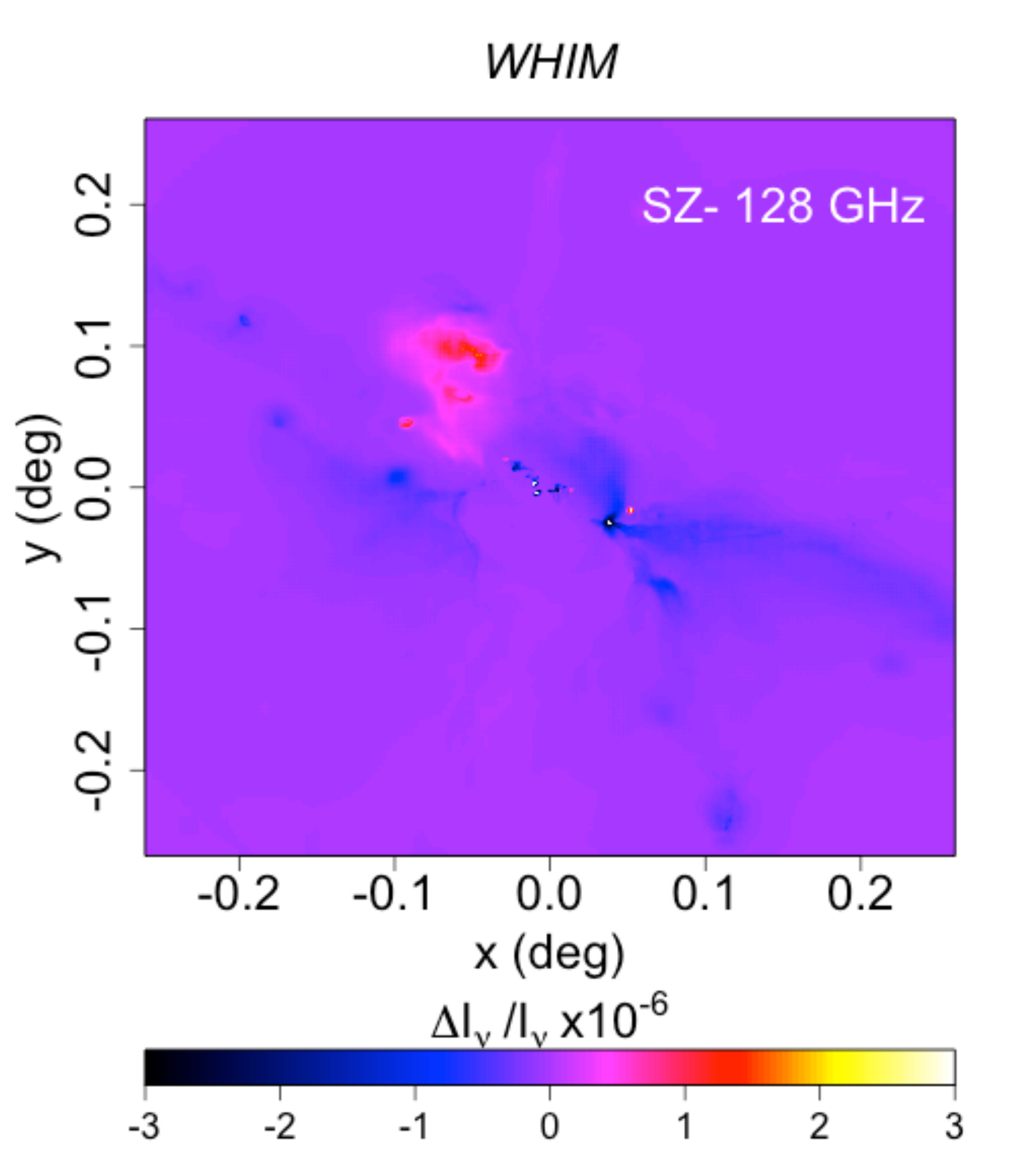}}
{\includegraphics[width=5.cm]{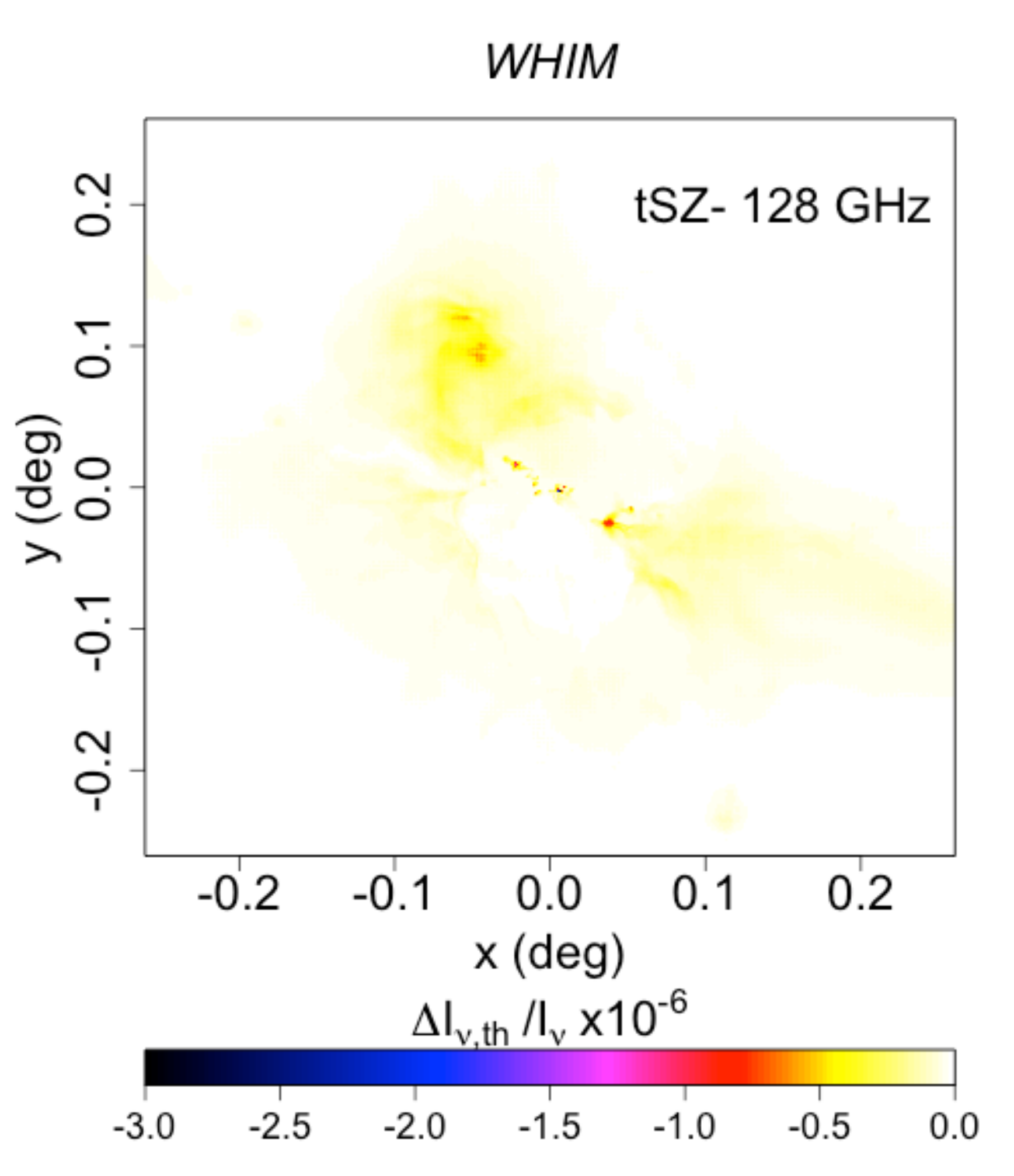}}
{\includegraphics[width=5.cm]{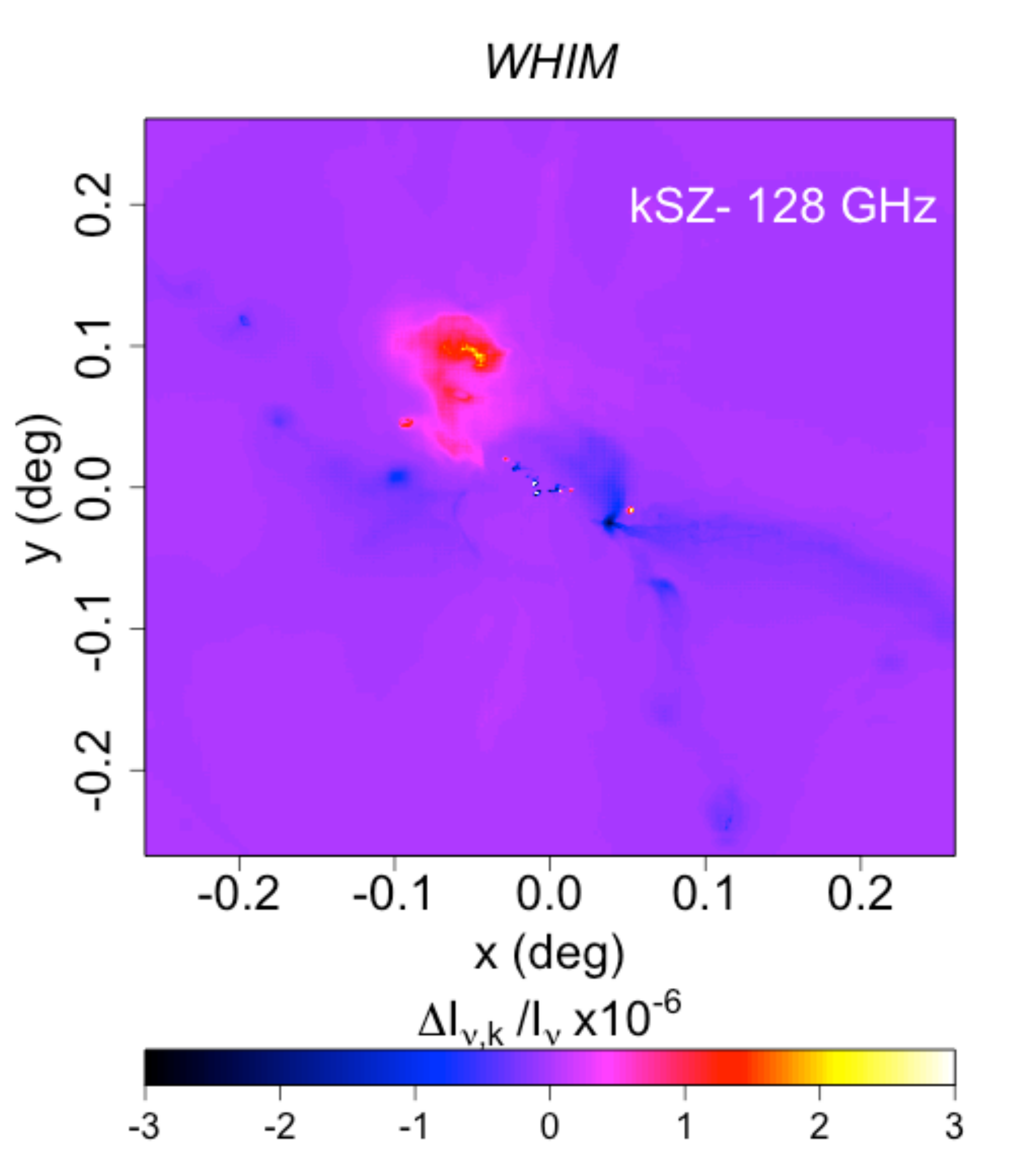}}
\caption{Intensity maps, $\Delta I/I_{\nu}$, of the total, thermal and kinematic SZ effect (panels from left to right, respectively) 
at a frequency $\nu=128$ GHz for the IGM (upper row) and WHIM (bottom row) gas components. Each map is sampled with  $1600^2$ pixels over a region of $\sim0.6$ squared degrees around the main central halo, which corresponds to a pixel size of $1.35\times1.35$ arsec$^{2}$.}
\label{fig:SZ1}
\end{figure*}

\begin{figure*}
{\includegraphics[width=5.cm]{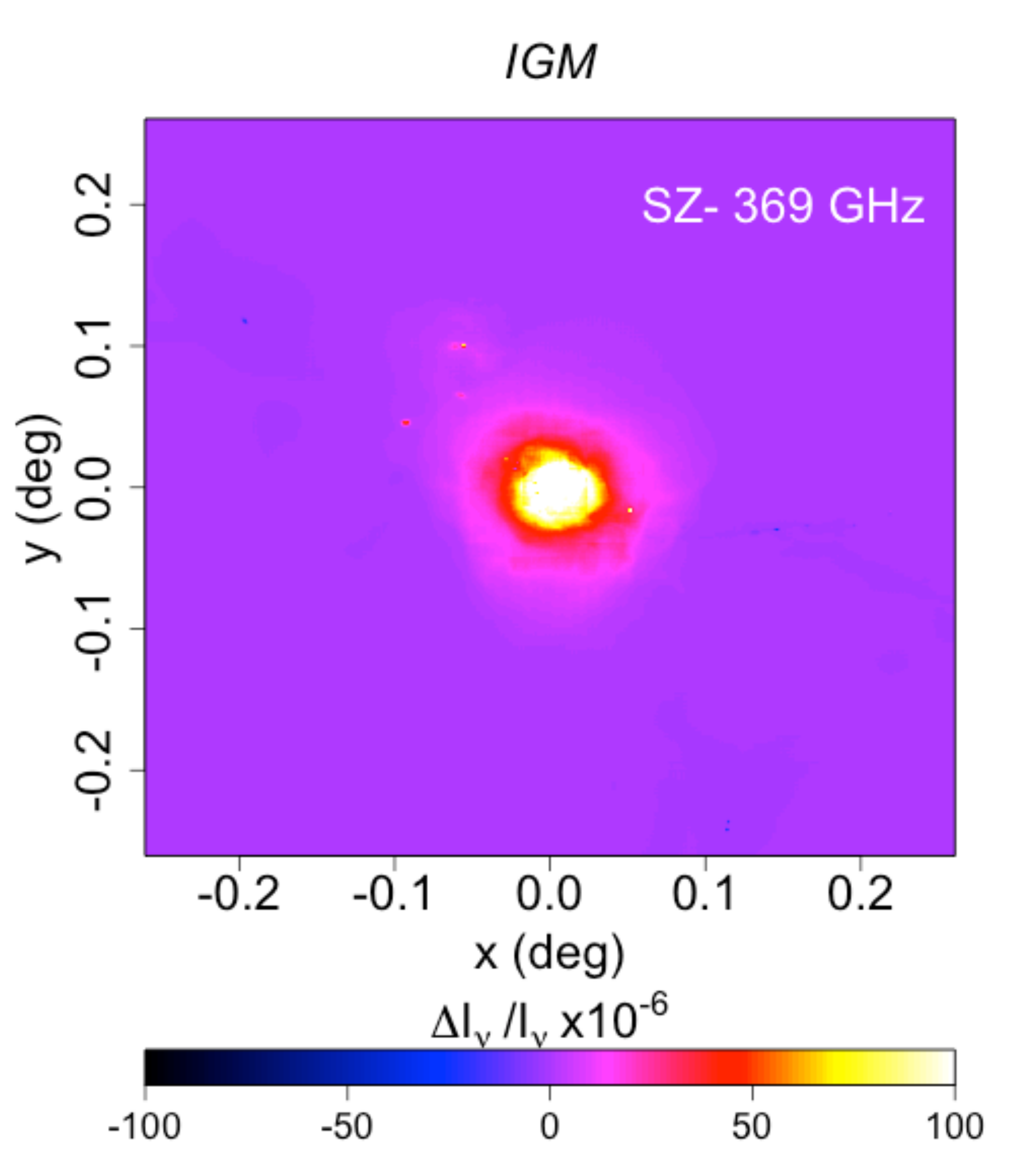}}
{\includegraphics[width=5.cm]{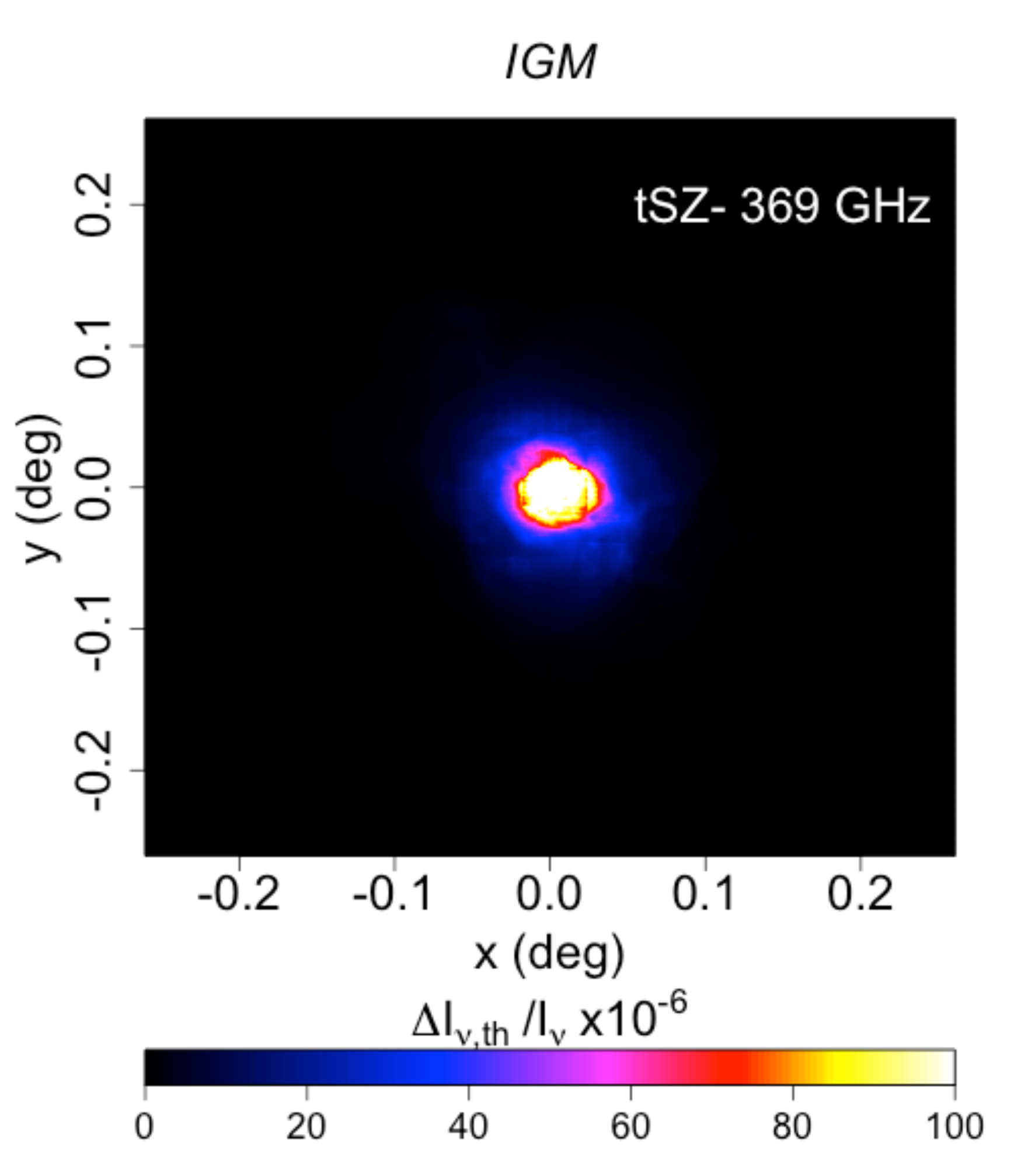}}
{\includegraphics[width=5.cm]{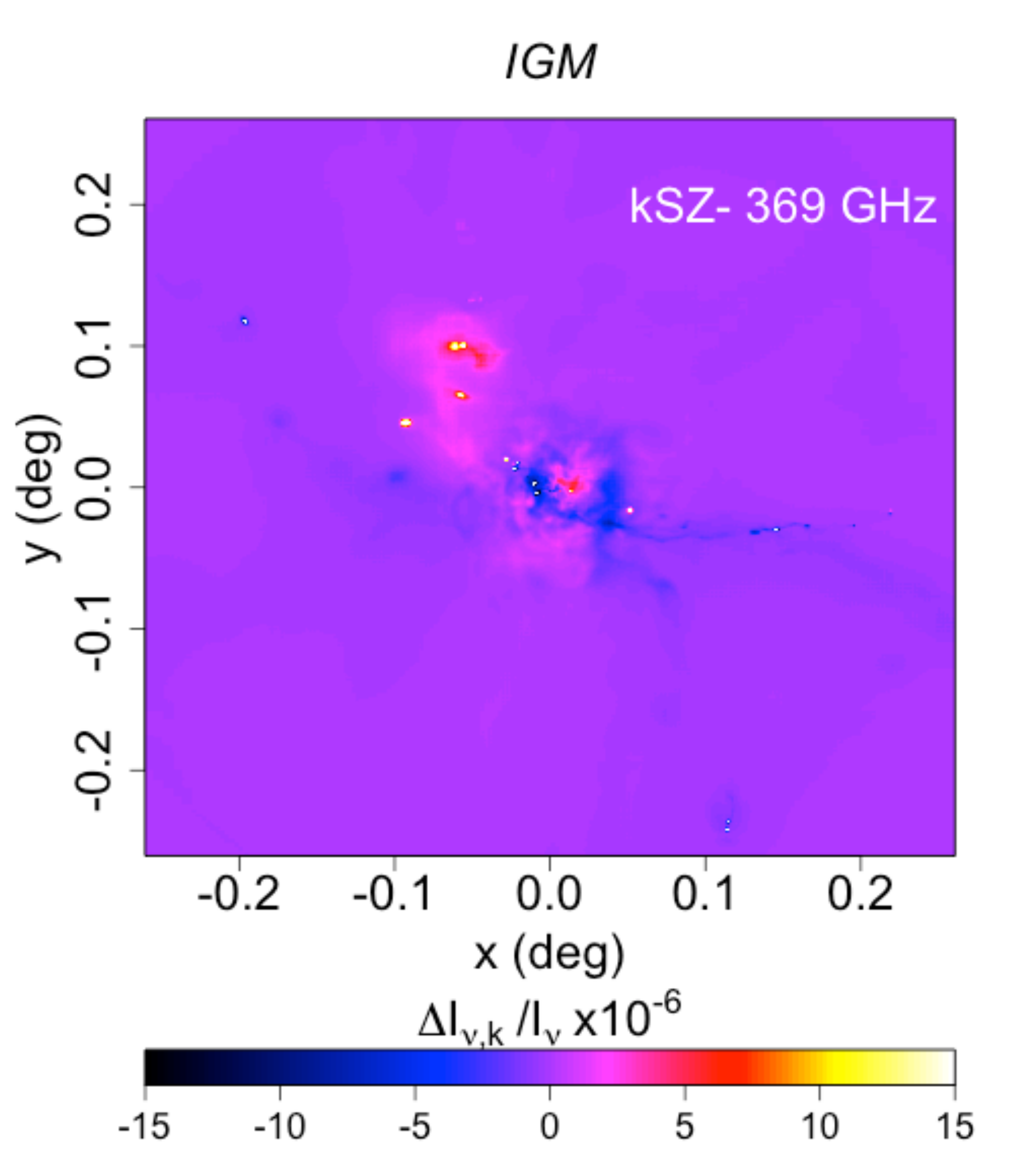}}\\
{\includegraphics[width=5.cm]{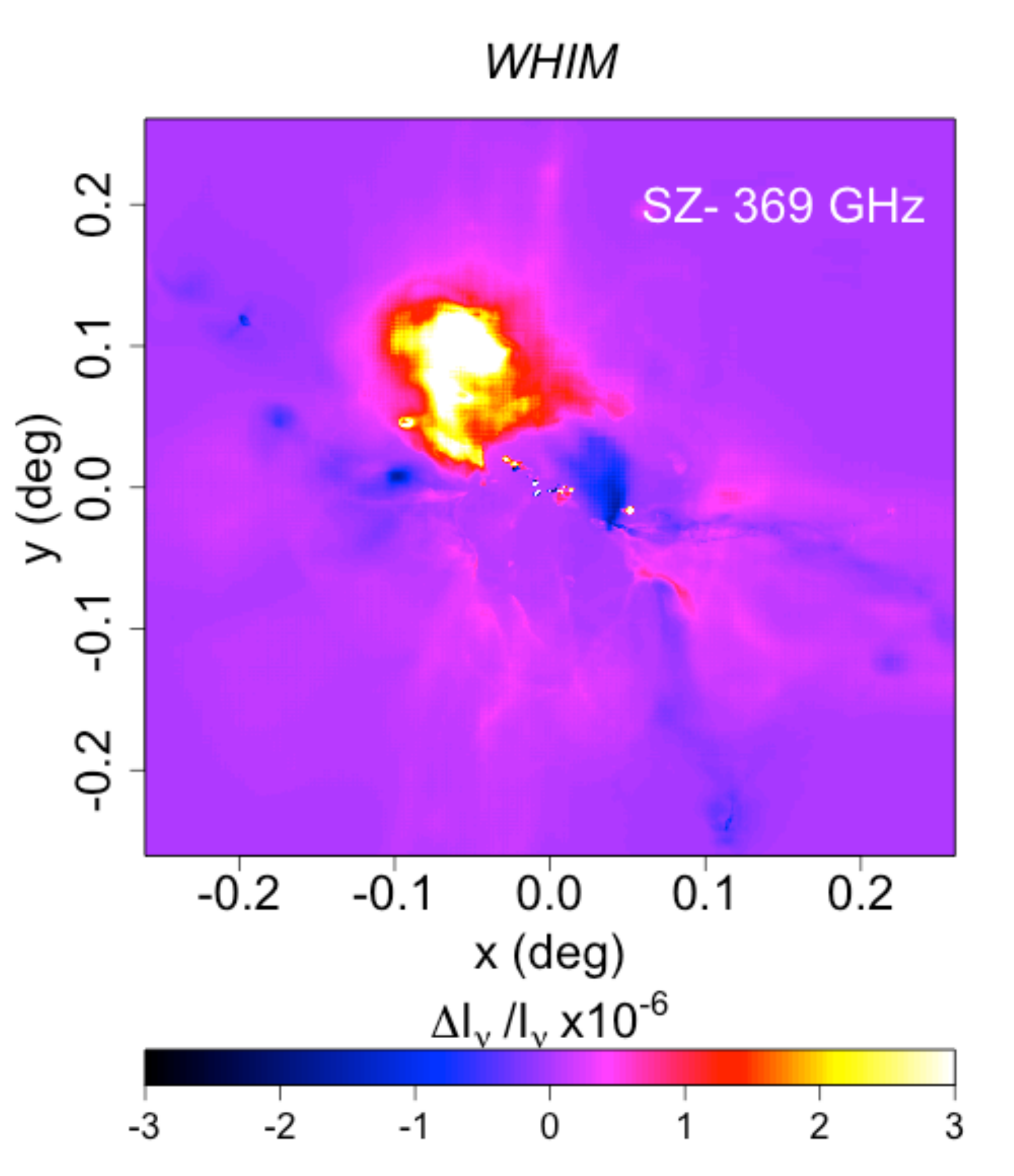}}
{\includegraphics[width=5.cm]{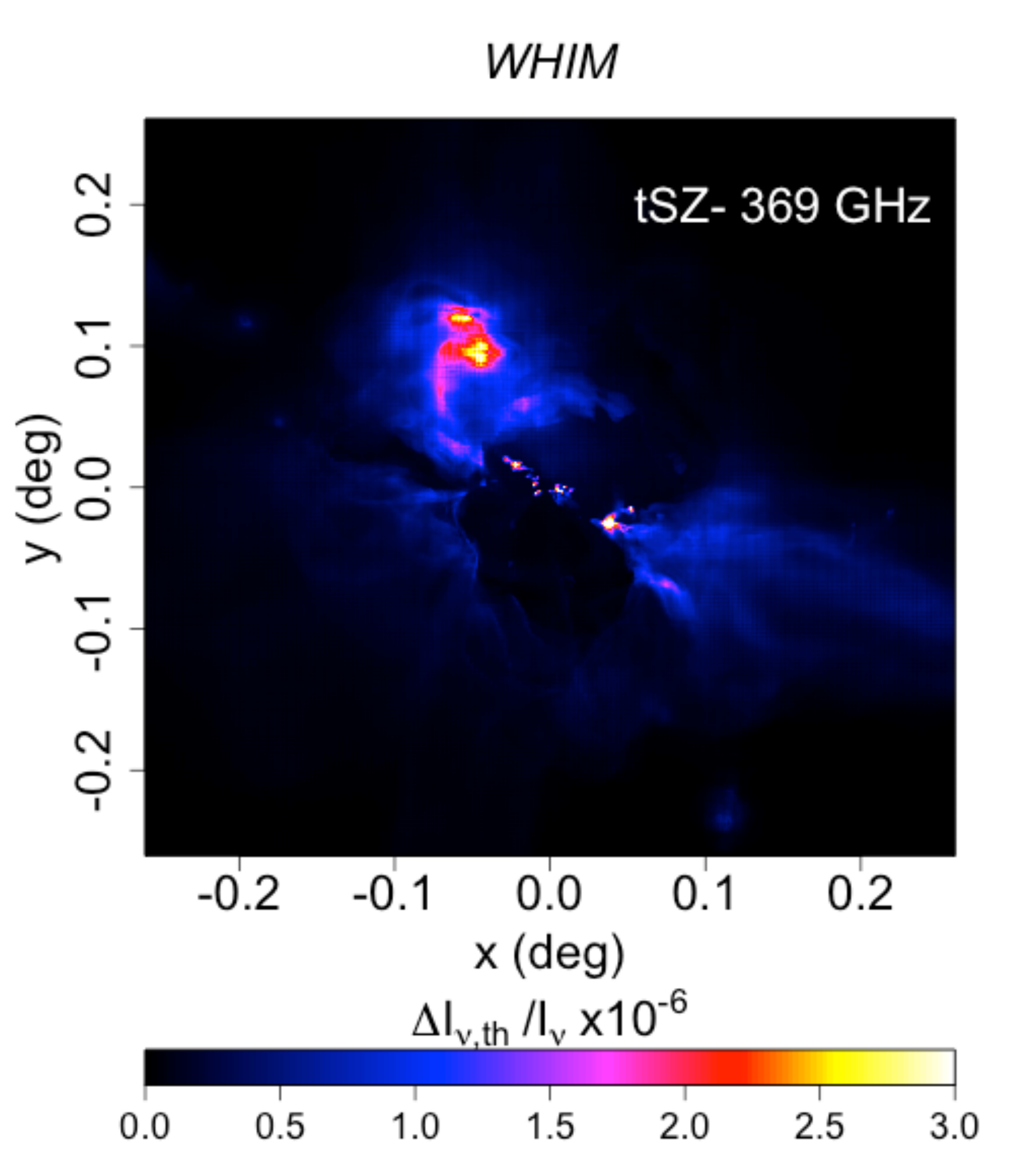}}
{\includegraphics[width=5.cm]{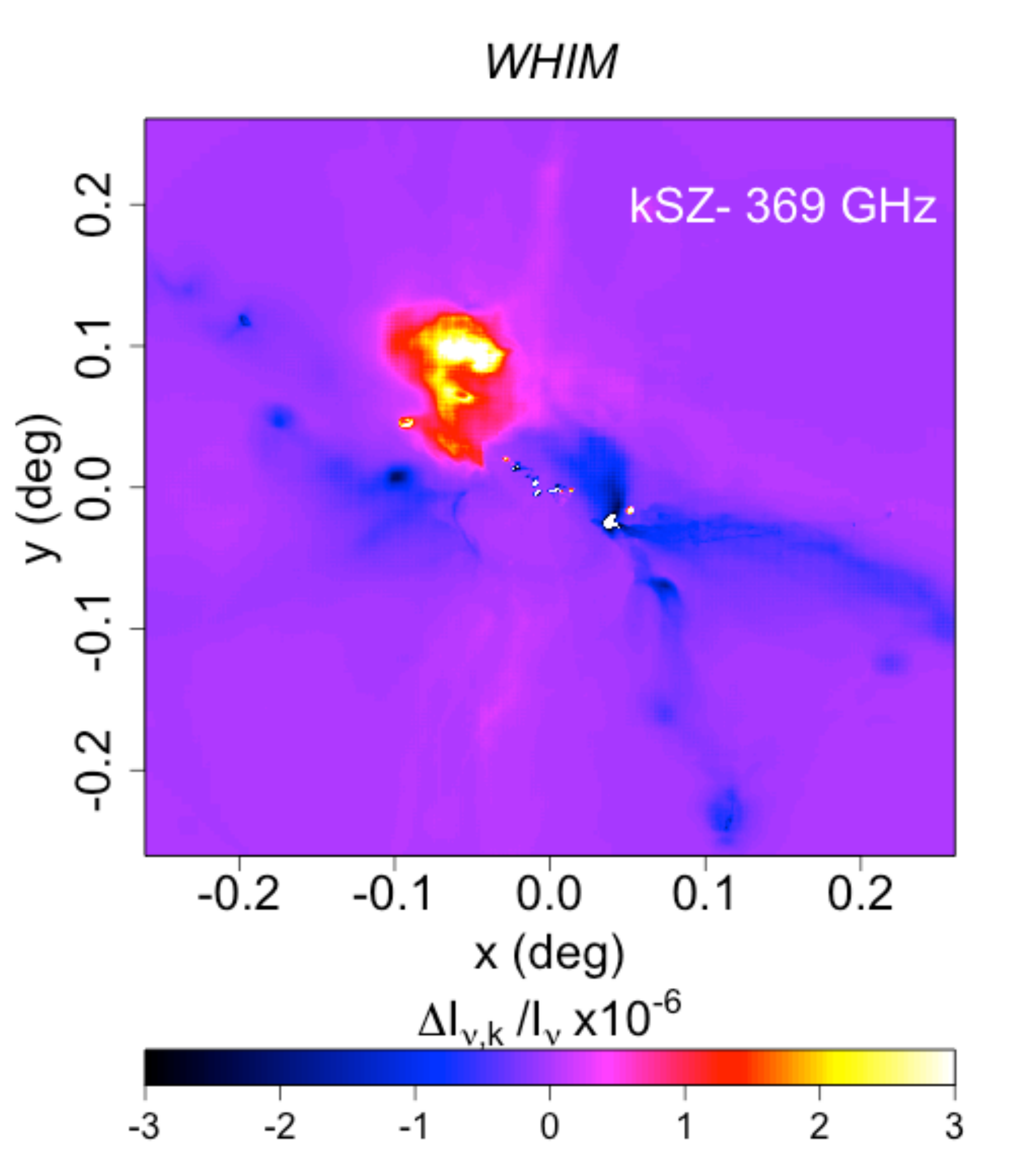}}
\caption{Intensity maps, $\Delta I/I_{\nu}$, of the total, thermal and kinematic SZ effect (panels from left to right, respectively) 
at a frequency $\nu=369$ GHz for the IGM (upper row) and WHIM (bottom row) gas components. These maps have the same resolution than those displayed in Fig.~\ref{fig:SZ1}.}
\label{fig:SZ3}
\end{figure*}

\begin{table*}
	\begin{tabular}{cccc}
	\hline
	 \small{$\nu$ (GHz)} & \small{128} & \small{217} &  \small{369} \\
	\hline
	 \small{IGM-tSZ} & \small{$(-2.87^{-9.19}_{+1.38})\times10^{-6}$} & \small{$(-1.30^{-18.83}_{+0.18})\times10^{-10}$} &  \small{$(4.63^{-1.42}_{+23.40})\times10^{-6}$}\\ \\
	 \small{IGM-kSZ} & \small{$(-1.04^{-1.76}_{+1.66})\times10^{-6}$} & \small{$(-0.19^{-697.22}_{+354.02})\times10^{-10}$} &  \small{$(1.19^{-2.00}_{+2.09})\times10^{-6}$}\\ \\
	 \small{WHIM-tSZ} & \small{$(-1.58^{-2.63}_{+1.11})\times10^{-6}$}  & \small{$(-1.19^{-4.15}_{+0.19})\times10^{-10}$} &  \small{$(1.33^{-1.07}_{+1.99})\times10^{-6}$}\\ \\
	 \small{WHIM-kSZ} & \small{$(1.19^{-1.07}_{+1.52})\times10^{-6}$}  & \small{$(-66.58^{-62.35}_{+26.78})\times10^{-9}$} &  \small{$(1.45^{-1.11}_{+2.34})\times10^{-6}$}\\
	\hline
	\end{tabular}
\caption{Median intensity variation values, $\Delta I/I_{\nu}$, of the tSZ  and kSZ effects  as obtained from the IGM and the WHIM gas components at three different frequencies. Errors  represent the 15th and 85th percentiles of  each distribution. We have assumed a frequency-independent threshold intensity variation of $\Delta I/I_{\nu}=10^{-6}$.}
\label{tbl:szvalues}
\end{table*}

In this Section we discuss how our cluster is observed  via the thermal and kinematic SZ effects.
Specifically, as explained in Section \ref{sec:SZ}, we compute the total, the  thermal and the kinematic SZ contributions to the CMB spectrum at three particular frequencies:  $\nu\sim 128$ GHz, where the tSZ shows its minimum value; $\nu\sim 217$ GHz, where the tSZ is almost negligible and the spectrum is dominated by the kSZ signal; and $\nu\sim369$ GHz, where the contribution from the tSZ effect to the total spectrum is maximal \citep[e.g.][]{Birkinshaw_1999}. Numerical \citep[e.g.][]{Kay_2008, Prokhorov_2010, Prokhorov_2011, Morandi_2013} and observational  \citep[e.g.][]{Colafrancesco_2010} studies have shown that this multi-frequency analysis of the SZ signal represents an extremely useful method to derive the ICM temperature distribution. 

Figure \ref{fig:SZ1} shows the intensity maps, $\Delta I/I_{\nu}$, of synthetic SZ observations of our cluster at $\nu =128$ GHz for both the IGM and the WHIM gas components (top and bottom rows, respectively). As expected, at this frequency the tSZ map (middle panels) shows always negative values since it produces a reduction in the temperature of the CMB photons with respect to the CMB temperature. On the contrary, the kSZ effect (right-hand panels) represents a positive or a negative change in the intensity depending on whether the gas is approaching or going away from the observer. For the sake of completeness, we also show the intensity map associated to the total SZ effect (left-hand panels) computed according to Eq.~\ref{eq:totalSZ}.
From the inspection of the tSZ map obtained for the IGM, it is clear that the signal is dominated  by the central galaxy cluster, which can reach intensities as high as  $\abs{\Delta I_{\nu, th}/I_{\nu}}\sim 3 \times 10^{-5}$ in its very central region. Overall, the tSZ signal is broadly distributed in shells of decreasing intensity from the cluster core out to its outskirts.  
As for the WHIM gas component, most part of the tSZ signal associated to the cluster core is removed, leaving an outer and much fainter signal around the cluster dominated by a more filamentary structure across the central region. 
If we compute the mean signal from all the pixels contributing to the maps, we obtain $\abs{\avg{{\Delta I_{\nu, th}/I_{\nu}}}}\sim 4\times 10^{-7}$ for the IGM and $\abs{\avg{{\Delta I_{\nu, th}/I_{\nu}}}}\sim 3\times 10^{-8}$ for the WHIM component, indicating that most part of the tSZ signal is contributed by hot and/or massive structures. Indeed, the tSZ maps obtained for the IGM and the WHIM at this frequency can be easily correlated with the X-ray thermal emission shown in Fig.~\ref{fig:th}, especially at the soft X-ray band. 
Specifically, the brightest regions in X-rays show the largest negative signals in the SZ maps \citep[e.g.][]{Ursino_2014}.
 
On the other hand, when galaxy clusters show bulk motions with respect to the CMB rest frame,  SZ observations can be significantly biased by the existence of a significant kSZ signal.  In our case, the comparison of the kSZ maps associated to the IGM and to the WHIM  (right-hand panels of Fig.~\ref{fig:SZ1}) provides a more similar spatial distribution of their intensities than in the case of the thermal maps. In this case, considering all the pixels contributing to the maps, we obtain mean values of $\avg{\abs{\Delta I_{\nu, k}/I_{\nu}}}\sim 10^{-9}$  for both the IGM and the WHIM. In general, whereas the gas producing the tSZ signal is also the main contributor to the X-ray emission, the peaks of intensity in the kSZ maps do not always have a clear counterpart in the tSZ maps, since they are mainly associated to small objects with a significant bulk motion \citep[e.g.][]{Roncarelli_2007}.  Indeed, in this case, most of the kSZ high-intensity regions can be correlated with the smaller subhaloes, the filamentary structure and the mid-overdense regions shown in Fig.~\ref{fig:rho-Lcut}. Moreover, in comparison with Fig.~\ref{fig:th}, it is also easy to find their X-ray counterpart in the soft energy band, where the smaller overdensities are highlighted, but not in the hard energy range, where only the highest-density regions are observed. 
Given the complementarity of the tSZ and kSZ maps, the total SZ signal shows a more accurate description of both contributions since it highlights both, the signal from high-density and hot central gas concentrations and the one coming from lower-density and colder small substructures with a non-negligible motion.

As shown in Fig.~\ref{fig:SZ3}, the qualitative distributions of the different SZ signals at  $\nu=369$ GHz is very similar to those obtained at  $\nu=128$ GHz  (Fig.~\ref{fig:SZ1}). The main difference between these two figures is, as expected, in the values of the tSZ maps since, at this higher frequency, the tSZ effect produces a positive deviation from the CMB spectrum. Moreover, the magnitude of the SZ signal at this high frequency is larger than at lower frequencies. According to the analysis presented in Appendix \ref{app:SZeffect}, we must take these results with caution since, at  $\nu=369$ GHz, the fact of  avoiding the relativistic corrections to compute the SZ effect might have a significant contribution in the final result (see Appendix \ref{app:SZeffect} for further details).

For the sake of completeness, we provide in Table \ref{tbl:szvalues} the median values of the intensity variations, $\Delta I/I_{\nu}$, as obtained from the tSZ  and kSZ maps  associated to the IGM and the WHIM gas components at the three characteristic frequencies we look at. To compute these values we have assumed a  frequency-independent threshold intensity variation of $\Delta I/I_{\nu}=10^{-6}$, which is in line with  the detection sensitivity of  current microwave facilities. 
In general, as confirmed in previous numerical analyses and observations, the kSZ effect associated to the IGM shows an  intensity variation slightly below its corresponding tSZ signal; only at $\nu= 217$ GHz, where the tSZ effect has a tiny contribution (several orders below the assumed threshold intensity), both signals are comparable. Moreover, as expected, the WHIM component  shows, at all frequencies, a smaller tSZ intensity variation than the one associated to the IGM. However, it is interesting to note that the median kSZ signal associated to the WHIM  is always above the median kSZ values associated to the IGM. This indicates that, in our selected region, whereas the WHIM (a warm-hot, mildly-overdense intergalactic gas distributed in small substructures and filaments)  only has a minor contribution to the total tSZ signal, it dominates the kSZ effect. As far as we are aware of, this is the first comparison between the kSZ signals associated to the IGM and the WHIM. Nevertheless, \cite{Nagai_2003} already showed the spatial distribution of the IGM kSZ signal and  how this effect could be used to determine cluster peculiar velocities.
On average, we find that, at all frequencies, the tSZ signal associated to the WHIM corresponds to only  $\sim 6$\% of the tSZ signal coming from the IGM,  a value that is roughly consistent with the estimation provided by \cite{Hallman_2007}.  
In contrast to X-ray observations that show a significant surface-brightness dimming with redshift, SZ observations are redshift-independent. Therefore, despite the weakness of the tSZ signal associated to the WHIM, the detection of this warm-hot gas through the SZ effect provides a valuable complementary tool to X-ray observations, especially at high redshift  \citep[e.g.][]{Ursino_2014}.
 
In this regard, current multifrequency SZ facilities, such as {\it Planck}\footnote{http://www.esa.int/Our\_Activities/Space\_Science/Planck} \citep[e.g.][]{Planck_2011} or {\it SPT}\footnote{https://pole.uchicago.edu} \citep[e.g.][]{Bleem_2015},  are detecting a significant number of high-z clusters. Concurrently, a number of additional SZ instruments, such as {\it CCAT-prime}\footnote{http://www.ccatobservatory.org} \citep[e.g.][]{Mittal_2017}, with higher resolution, sensitivity and spectral coverage, are already planned and will play a crucial role in providing a deeper picture of the ICM and WHIM properties by means of SZ observations alone. Indeed, while there are measurements of the tSZ effect for a large number of clusters \citep[e.g.][]{Bleem_2015}, measuring the kinematic \citep[e.g.][]{Sayers_2013} or the relativistic SZ effects \citep[e.g.][]{Zemcov_2012} for individual systems is much more challenging. In this regard, {\it CCAT-prime} will provide a multi-band detection of the SZ effect with improved capabilities.  In comparison with \planck, {\it CCAT-prime} is expected to have a much better sensitivity (by an order of magnitude in the $95-405$ GHz range) and a much better angular resolution  \citep[by a factor $\sim 6$; see][for further details]{Erler_2017}. Moreover, at higher frequency, it will be more efficient in removing point sources. With these capabilities, it will be able to determine individual cluster temperatures and peculiar velocities with an unprecedented precision \citep[see also][for an analysis of the ICM detection with future {\it CCAT-prime} observations]{Morandi_2013}.

\subsection{Radio emission}
\label{subsec:radio}

\begin{figure*}
\center
{\includegraphics[width=5.cm]{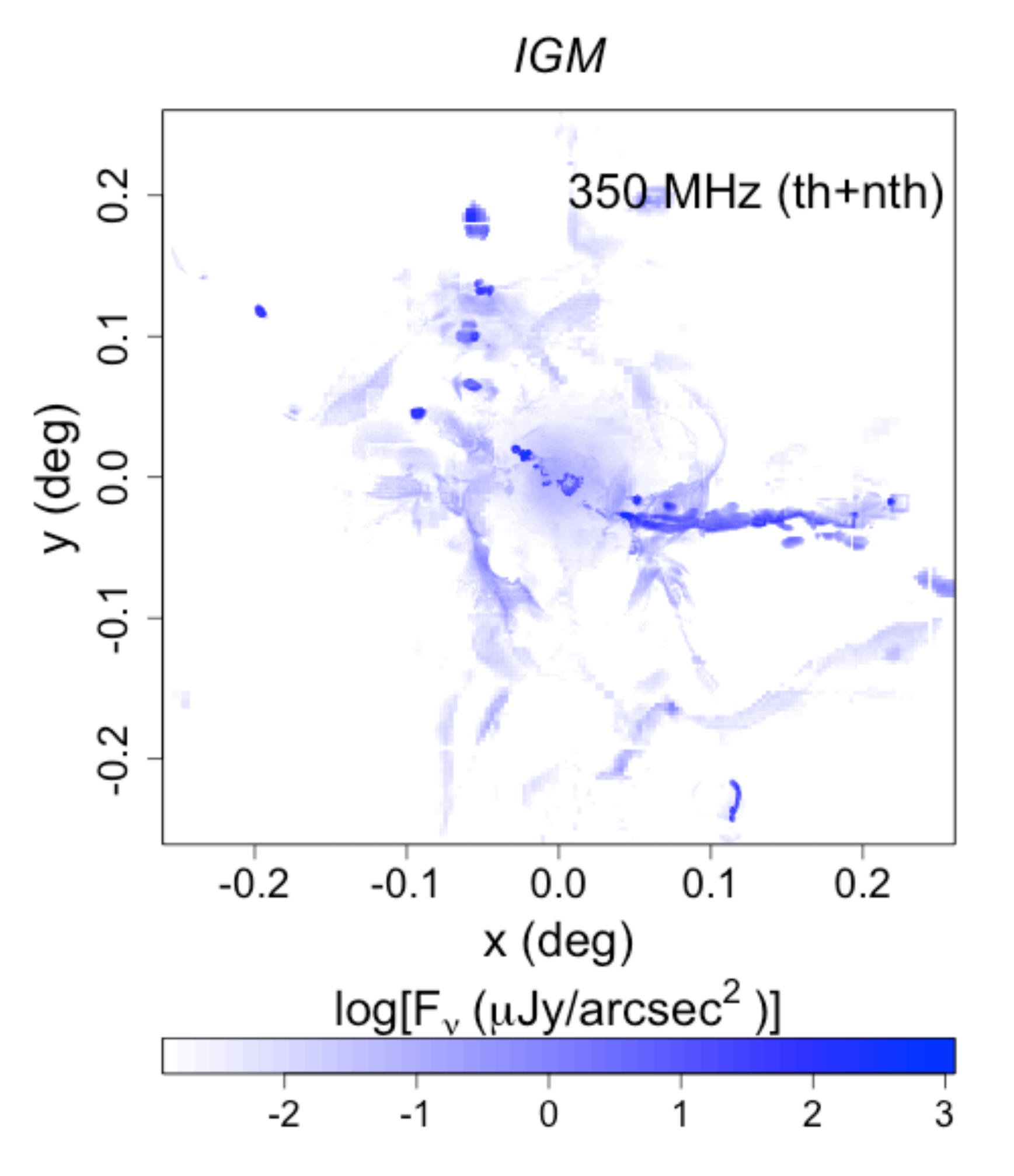}}
{\includegraphics[width=5.cm]{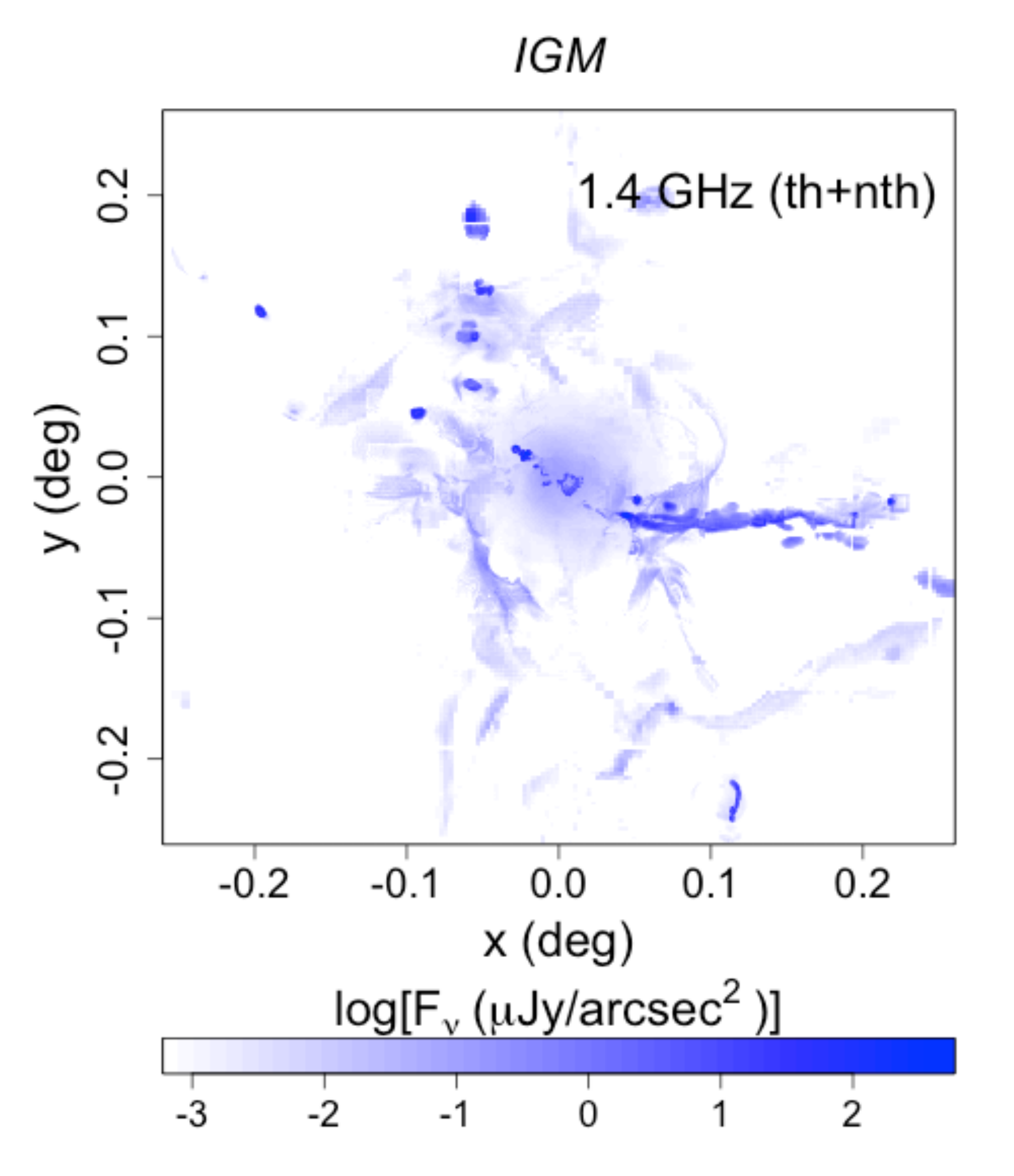}}
{\includegraphics[width=5.cm]{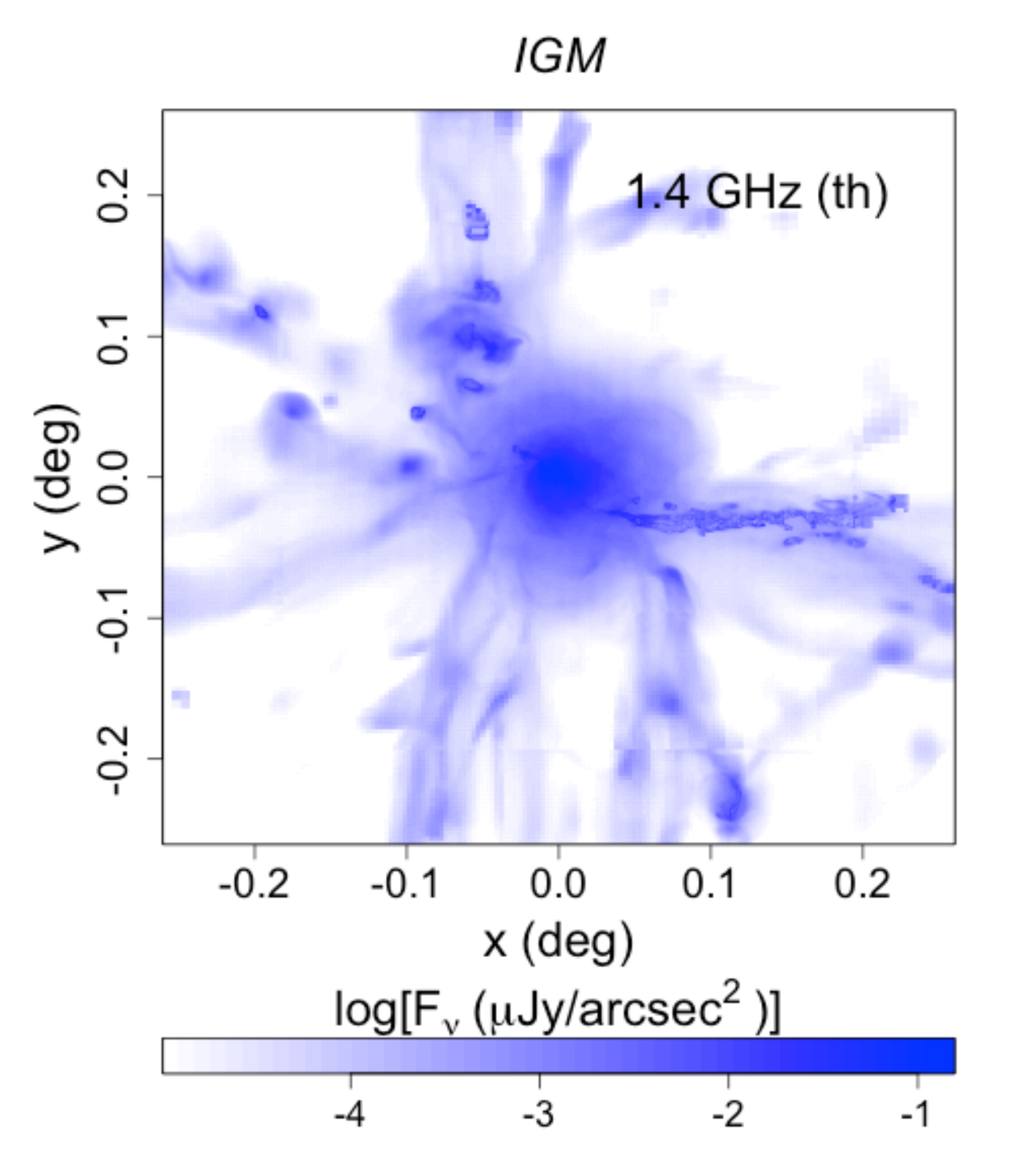}}\\
{\includegraphics[width=5.cm]{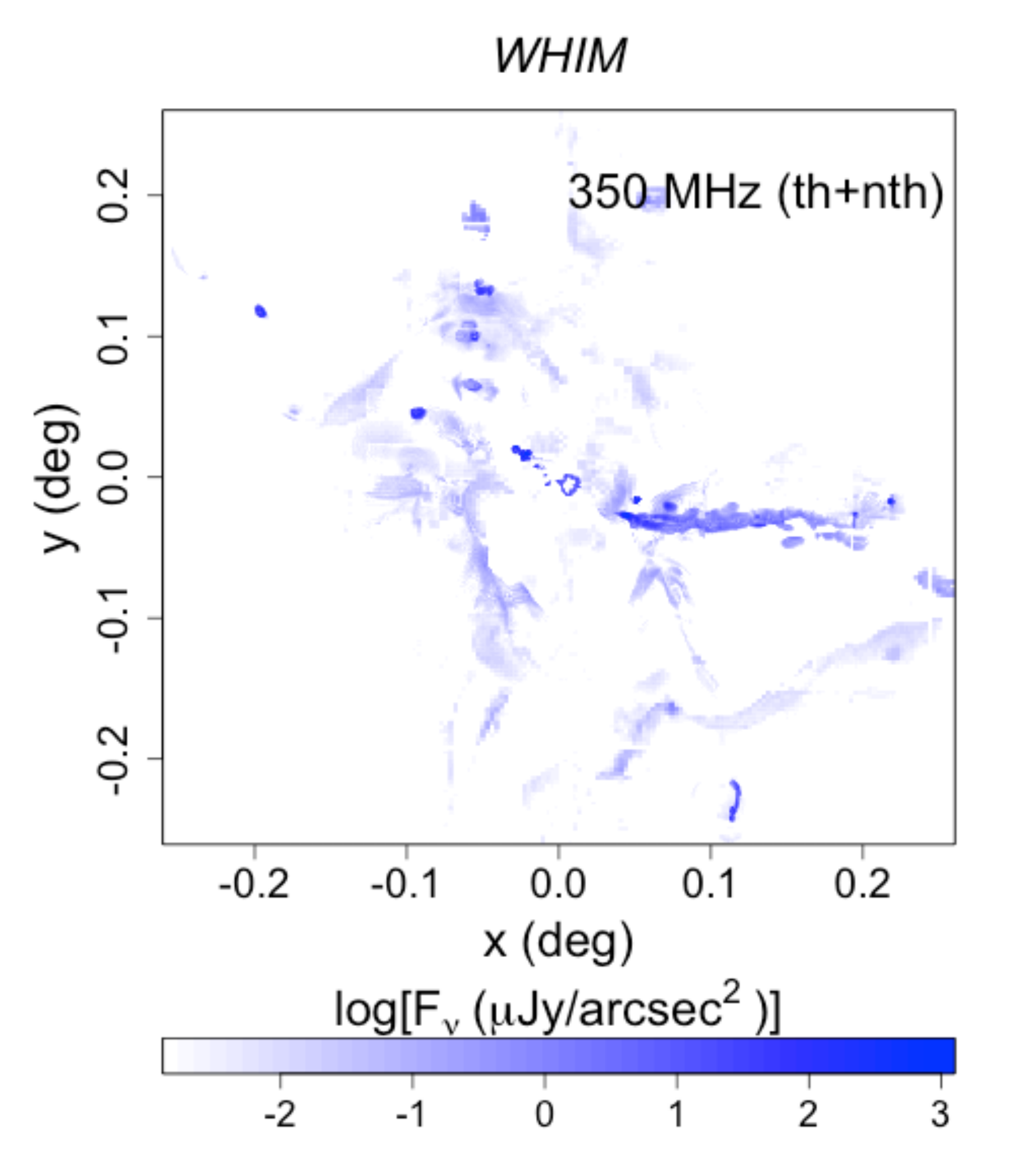}}
{\includegraphics[width=5.cm]{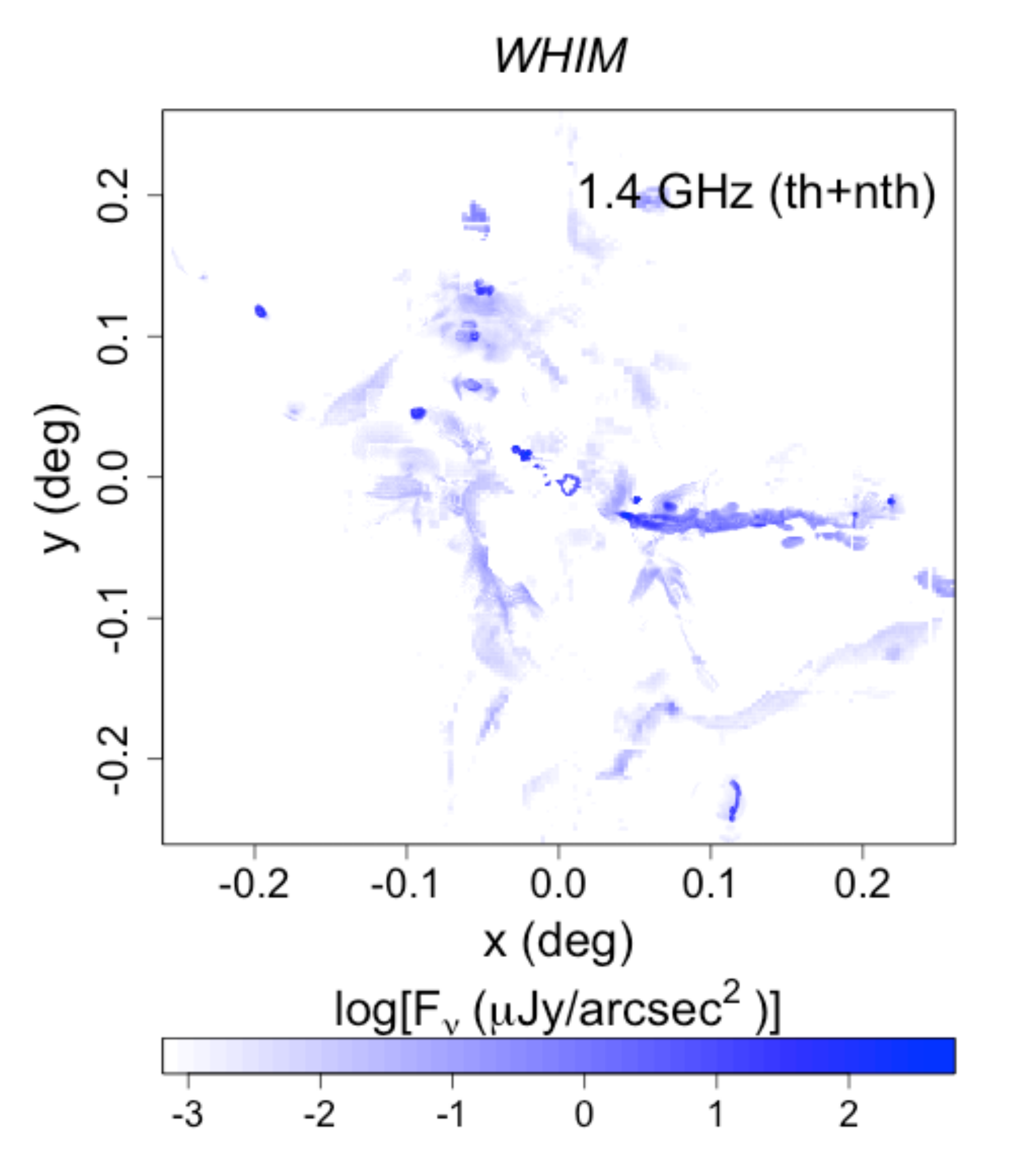}}
{\includegraphics[width=5.cm]{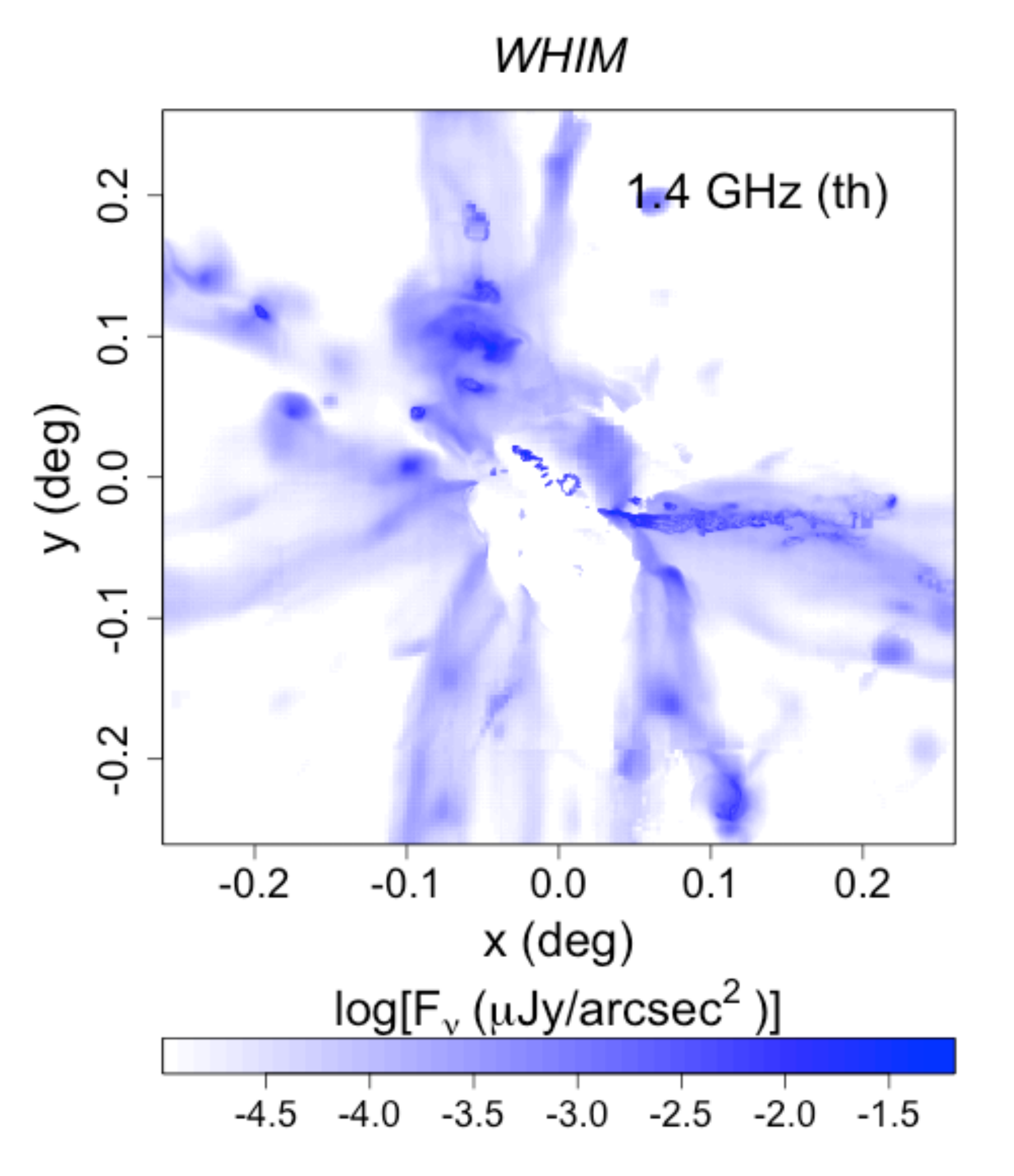}}\\
\caption{Mock radio observations of the IGM and the WHIM components 
(top and bottom rows, respectively) at $350$ MHz  and $1.4$ GHz for a 
beam size of $1\times1$ arcsec$^2$. Left and middle panels show the 
contribution from the thermal plus non-thermal components ($th+nth$) to 
the radio maps, whereas the right column shows the subdominant contribution 
from the thermal component alone.}
\label{fig:radio}
\end{figure*}

\begin{table}
\center
	\begin{tabular}{ccc}
	\hline
	 $\nu$ & IGM & WHIM\\
	  (MHz)   &   {($\mu$Jy~arsec$^{-2}$)} & {($\mu$Jy~arsec$^{-2}$)} \\ 
	\hline
	\small{$350$} (th$+$nth) & {$(1.50^{-0.47}_{+12.13})\times 10^{-3}$}  & {$(1.07^{-0.42}_{+7.62})\times 10^{-3}$}  \\ \\
	\small{$1400$} (th$+$nth)  & {$(1.33^{-0.43}_{+10.22})\times 10^{-3}$}  & {$(0.96^{-0.39}_{+6.22})\times 10^{-3}$}  \\ \\
	\small{$1400$ (th)}  & {$(1.14^{-0.42}_{+5.84})\times 10^{-3}$}  & {$(0.65^{-0.36}_{+2.57})\times 10^{-3}$}  \\
	\hline
	\end{tabular}
\caption{Median pixel values of the synthetic radio maps shown in Fig.~\ref{fig:radio} for the IGM and the WHIM gas components at $350$ MHz  and $1.4$ GHz for a beam size of $1\times1$ arcsec$^2$. Errors  represent the 15th and 85th percentiles of  each distribution. These values are obtained assuming an hypothetical threshold detection intensity of $\sim3\times10^{-4}\mu$Jy~arcsec$^{-2}$.}
\label{tbl:valuesradio}
\end{table}

\begin{figure*}
\center
{\includegraphics[width=5.cm]{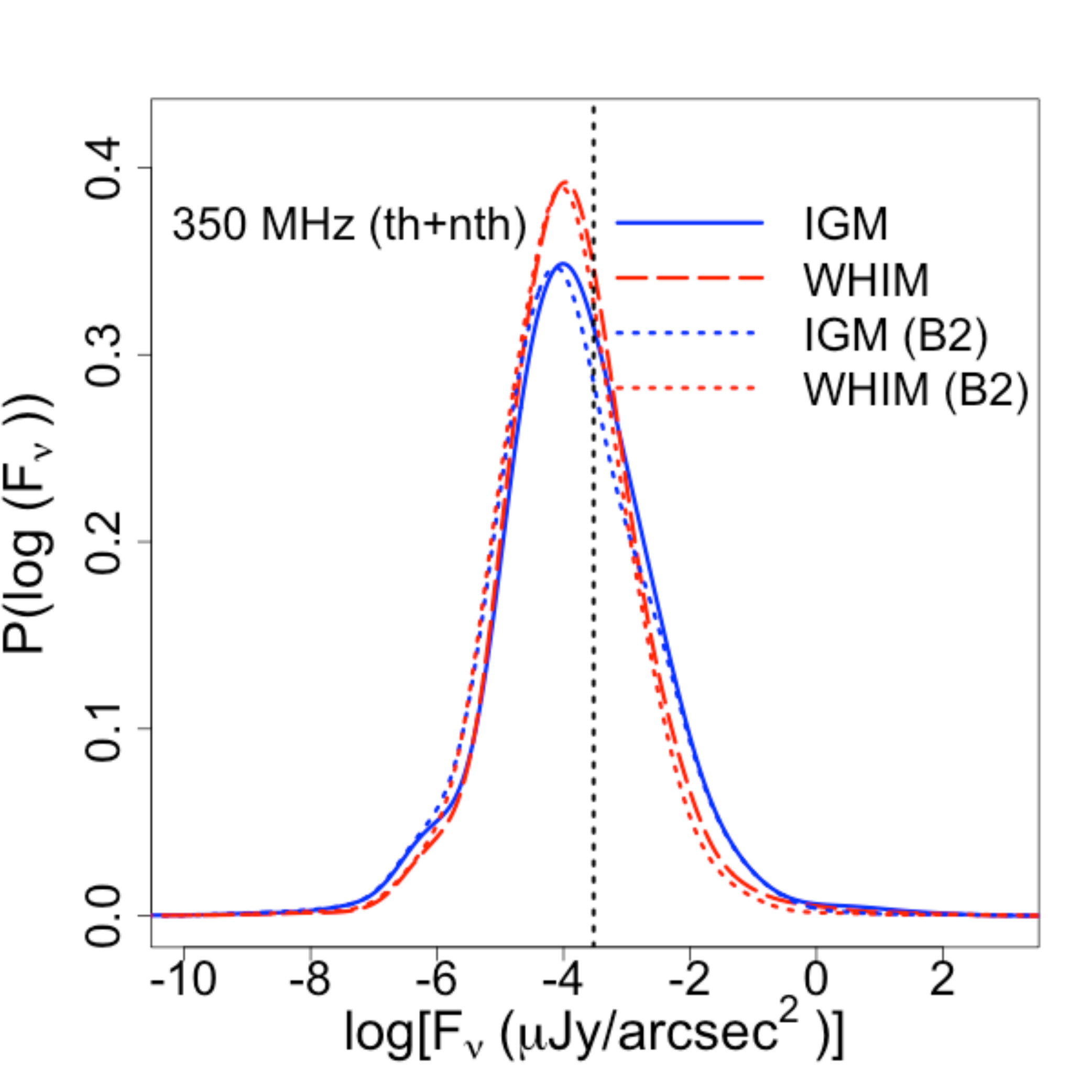}}
{\includegraphics[width=5.cm]{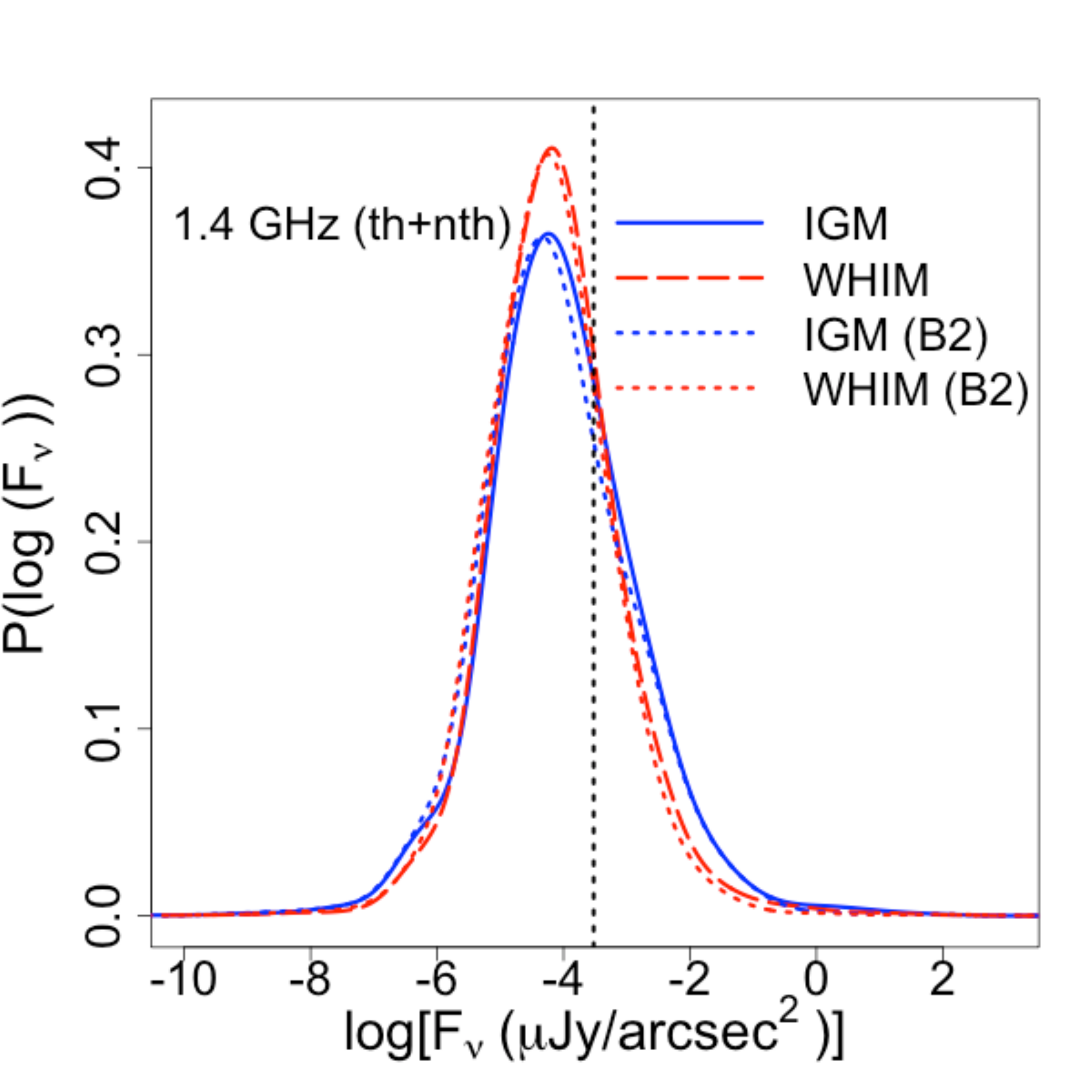}}
{\includegraphics[width=5.cm]{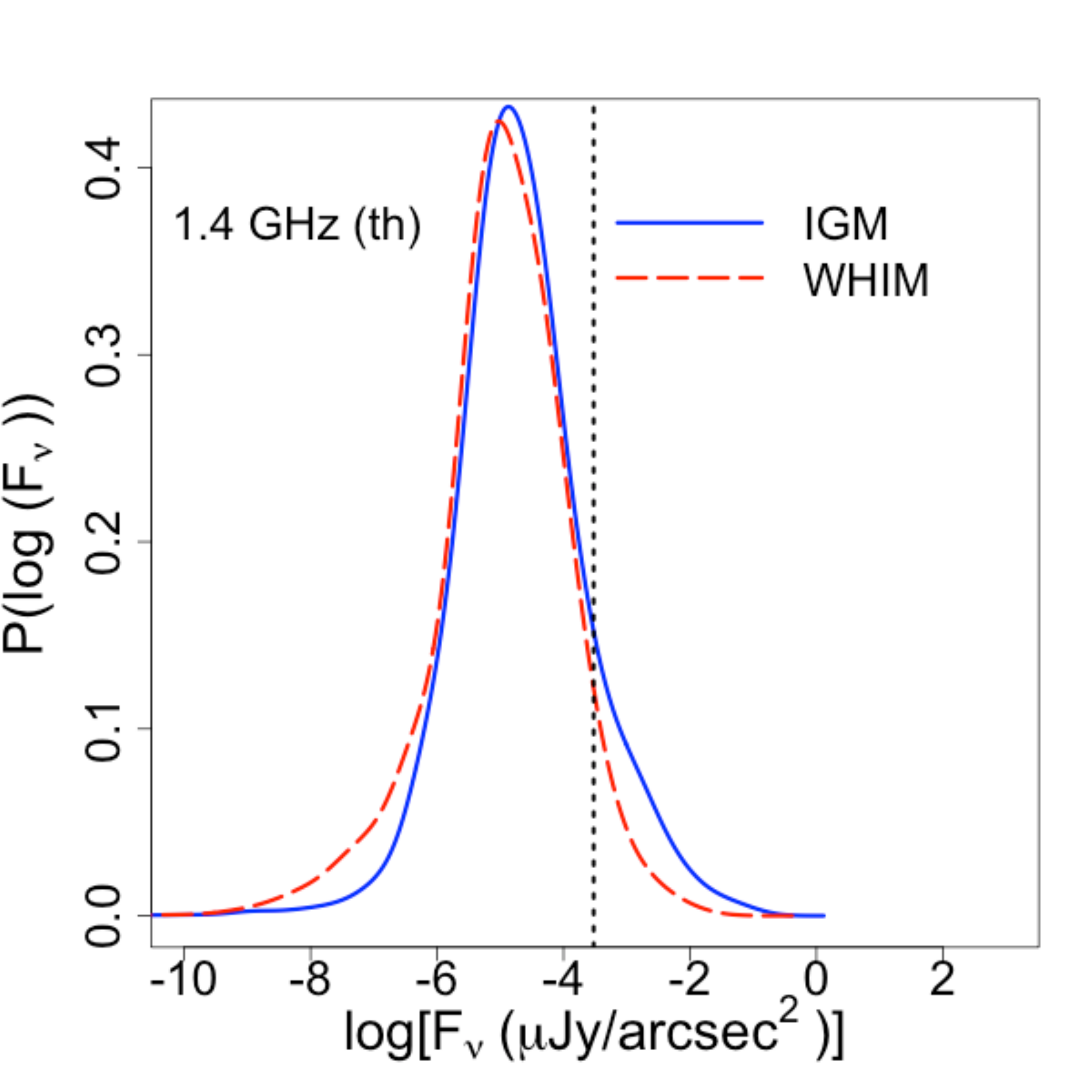}}
\caption{Probability distribution function of the radio signal associated to the pixels composing the maps shown in  Fig.~\ref{fig:radio}. The distributions are shown for the IGM (solid blue line) and the WHIM (dashed red line) as contributed by the thermal plus non-thermal components at $350$ MHz  and $1.4$ GHz for a beam size of $1\times1$ arcsec$^2$ (left and middle panels, respectively) and by the thermal component alone at $1.4$ GHz (right panel).
The vertical dotted line in each panel represents a threshold detection intensity of $1~\mu$Jy~arcmin$^{-2}$, which corresponds to $\sim3\times10^{-4}\mu$Jy~arcsec$^{-2}$.
The distribution of pixel intensities obtained with the simpler magnetic field model provided by Eq.~\ref{eq:Bfield2} (labelled as model B2) are also shown for comparison. }
\label{fig:psfradio} 
\end{figure*}

\begin{figure}
\center
{\includegraphics[width=9.cm]{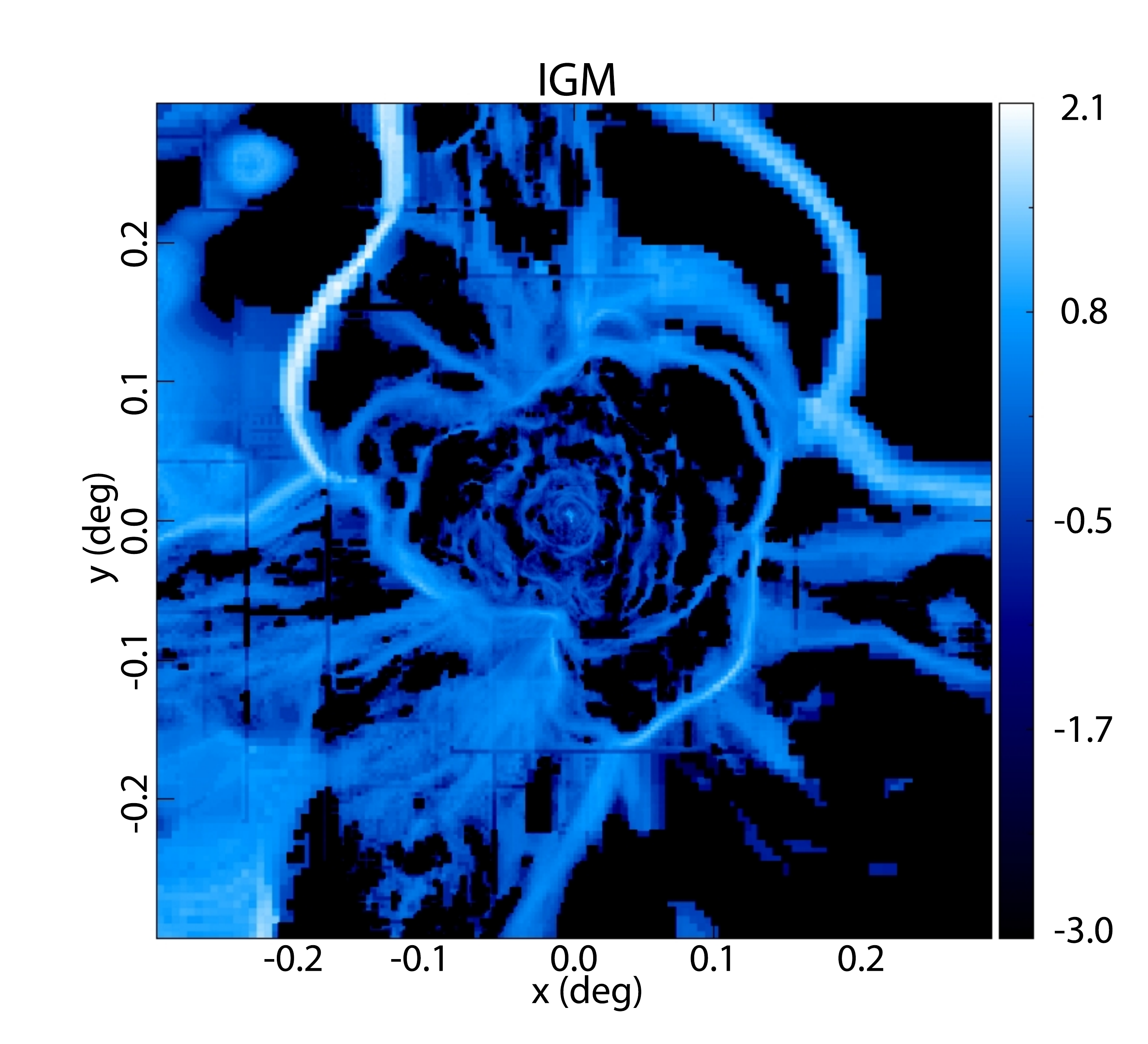}}
\caption{Distribution of shock waves, represented by their shock Mach numbers, within the same region as shown in previous figures around our massive cluster. The map represents a projection along the line of sight for a slice of thickness $0.1$ Mpc. The color bar shows the Mach number in log scale.}
\label{fig:mach}
\end{figure}

Radio emission from the cosmic web, connected to merger and accretion structure formation shocks, can also provide independent and complementary information on the existence and evolution of the WHIM \citep[e.g.][]{Brown_2011, ArayaMelo_2012}.  

As explained in Section \ref{subsec:part2}, we have implemented a simplified phenomenological model for shock acceleration in {\sc SPEV} in order to compute the synchrotron emission in our simulation. We have built synthetic radio maps for the same region around our simulated galaxy cluster as shown in previous sections. We convolve the images with a Gaussian beam whose full width half maximum is $1\times 1$ arcsec$^2$. 
In the following, in order to quantify the emission from different bands, we will consider a threshold detection intensity of $1~\mu$Jy~arcmin$^{-2}$, which corresponds to $\sim3\times10^{-4}\mu$Jy~arcsec$^{-2}$. However, we will not adapt our radio maps to the technical properties of any observational instrument.

In particular, the left and middle panels of Fig.~\ref{fig:radio}
show, respectively, the radio emission from the IGM and the WHIM
computed at $350$ MHz and at $1.4$ GHz.  These images account for the
contribution to the radio emission from the thermal and non-thermal
gas components ($th+nth$). As expected, we find that only a small
fraction of the cluster emits significantly in radio. Moreover, the
IGM radio emission is more clearly correlated with high-density and
high-temperature regions. Thus, given the features of our emission
model, since the density (and, therefore, the magnetic field) and the
gas temperature decrease at the cluster periphery, we obtain a very
little (if any) radio emission at the very outer cluster regions
(around and beyond $\sim4\times R_{vir}$). A similar result is also
obtained for the WHIM although in this case the emission is
distributed in a more filamentary structure, reducing therefore its
extension.  As obtained in previous simulations, this result is in
line with the strong dependence of the radio emission with the cluster
temperature \citep[e.g.][]{Hoeft_2008}.  However, although the
radio emission associated to the WHIM is less extended, at $350$
MHz, the mean radio signals associated to the WHIM
($\sim0.55~\mu$Jy~arcsec$^{-2}$) and to the whole IGM
($\sim0.63~\mu$Jy~arcsec$^{-2}$) are very similar. This is obviously
due to the fact that, when computing the mean pixel intensity, the
high-intensity pixels have more weight than the low-intensity ones
in the WHIM maps. If we look instead to the median pixel values
reported in Table \ref{tbl:valuesradio}, the radio signal associated
to the IGM is larger than the one associated to the WHIM by a factor
$\sim 1.4$. Moreover, when we go from $350$ MHz to $1.4$ GHz, we
obtain a $\sim 60$\% fainter, but a slightly more spatially
distributed, mean radio emission from both the IGM and the WHIM
components. Overall, at $1.4$ GHz, we obtain for IGM and WHIM a
maximum radio signal of half mJy arcmin$^{-2}$, whereas at $350$ MHz
the maximum emission is of the order of $\sim$1$-$2 mJy
arcmin$^{-2}$.

We can compare these results, obtained with our magnetic field reference model given by Eq.~\ref{eq:Bfield}, with those obtained with the simpler B2 model provided by  Eq.~\ref{eq:Bfield2}. In this case we obtain, at both considered frequencies, an increase of a factor $\sim2$ and $\sim3$ for the mean emissions from the IGM and the WHIM, respectively. Moreover, the maximum emissions associated to the IGM and to the WHIM are also increased by a factor  $\sim3$ and $\sim5$, respectively. Given the characteristics of our simulation and the approximations adopted in our post-processing, we should take with caution our results. Nevertheless, we can accept as an upper limit the results on the radio emission obtained with the magnetic field model B2 provided by Eq.~\ref{eq:Bfield2}.

For the sake of completeness, the right column of Fig.~\ref{fig:radio} shows the contribution of the thermal component alone to the radio maps  at $1.4$ GHz. As expected, the thermal contribution is highly correlated with the density maps shown in Fig.~\ref{fig:rho-Lcut} and its global contribution to the radio signal is significantly smaller (by a factor $\sim4$) than the non-thermal contribution.  

Despite the clear difference in the emission level, Fig.~\ref{fig:psfradio} shows that the probability distribution functions of the radio signal associated to the pixels composing the radio maps of Fig.~\ref{fig:radio} display a similar gaussian-like distribution, with the IGM  dominating over the WHIM  radio emission at the high-flux end of the distributions at both frequencies. 
A similar conclusion is obtained if we compare the pixel intensity distributions resulting from the reference and the B2 magnetic field models. As expected, these two models only affect the non-thermal emission and, therefore, they  generate an identical thermal distribution at $1.4$ GHz (right panel of Fig.~\ref{fig:psfradio}).
 
To complement these radio maps, Fig.~\ref{fig:mach} shows the distribution of shock waves within our region of interest for a thin slice of $\sim 0.1$ Mpc along the line of sight. In particular, shock waves are represented by their Mach numbers, which give an idea of the shock strength.
From this distribution, it is easy to distinguish internal, low-Mach number shocks ($\mathcal{M}< 10$) from more external and stronger shocks  ($\mathcal{M}>10$) mainly wrapping the central region.  Although Fig.~\ref{fig:radio} and Fig.~\ref{fig:mach} are not directly comparable, they suggest that, in our case,  most of the cluster radio emission is associated to non-thermal internal shock events within the cluster, whereas radio emission connected to external accretion shocks is negligible \citep[see also][]{Hoeft_2008, ArayaMelo_2012}.
It has been suggested that the radio emission associated to large-scale accretion shocks could be detected by the next generation of radio telescopes, such as the {\it Low Frequency Array}\footnote{https://www.lofar.org/} \citep[LOFAR; e.g.][]{Rottgering_2003} or the {\it Square-Kilometre Array}\footnote{https://www.skatelescope.org/} \citep[SKA; e.g.][]{Keshet_2004, Acosta-Pulido2015}. 
However, the results of our analysis on the radio band suggest that such a low emission from the accretion region ($\sim4\times R_{vir}$) would be undetectable even with future radio facilities.

Observationally, the detection of the WHIM radio emission is extremely difficult. Previous simulations \citep[e.g.][]{Pfrommer_2008, Battaglia_2009},  including a complex treatment for the physics of the non-thermal electron population, predict a maximum level of radio emission in the outskirts of massive clusters of only a few $\sim\mu$Jy~arcmin$^{-2}$ at $1.4$ GHz. Moreover, radio observations at high frequencies also suffer contamination from large spatial scales and confusion sources \citep[e.g.][]{Brown_2011, Clarke_2014}. Indeed, above $1$ GHz, the detection of shock structures through future polarization observations seems to be the most hopeful method to detect the radio emission associated to the WHIM. In addition, later analyses have also suggested that the next generation of radio telescopes will be able to detect the cosmic web at lower frequencies \citep[$< 200$ MHz; e.g.][]{Vazza_2015, Vazza_2015b,Vazza_2016}. 
In this regard, the \ska\ will allow observations of the cosmic web with unprecedented resolution and sensitivity over a large range of frequencies \citep[$\sim 0.05-20$ GHz;][]{Kale_2016} and beyond $z=0.4$. It will be able to detect lower surface brightness (by a factor $10-50$ than current facilities) and extended cluster radio emission, thus allowing the detection of a significant number of clusters (as an example, it is expected to discover thousands of new relics, radio haloes and mini haloes with a typical emission of  $\sim 0.2-1 \mu$Jy~arcsec$^{-2}$ at $1.4$ GHz). Regarding the WHIM, as predicted by \cite{ArayaMelo_2012}, if electrons are accelerated by accretion shocks,  radio fluxes of $\sim0.12~\mu$Jy at $150$ MHz  from filaments at $z=0.15$ will be detected. This will also allow to put constraints on the amount of magnetic fields and cosmic rays in filaments.

\section{Summary and conclusions}
\label{sec:conclu}

The combination of multi-wavelength observations of the cosmic web is crucial to deepen our knowledge of the Universe in a number of aspects, such as the physics of baryons, the magnetization of the IGM, the pressure of the hot ICM gas or the formation and evolution of cosmic shock waves. Therefore, in view of improved and  upcoming observational facilities, such as \athena, the \ska\ or the \ccat,  it is crucial to analyse multi-wavelength synthetic observations performed with full cosmological simulations.

In this paper, we analyse a massive galaxy cluster formed and evolved in an Eulerian-AMR cosmological simulation in order to investigate the spatial distribution and the emission associated to the WHIM gas component, defined as the gas with a temperature in the range $10^5-10^7$ K. A set of  multi-wavelength synthetic observations around the largest galaxy cluster ($M_{vir}\sim 3.2\times 10^{14} M_\odot$ and $R_{vir}\sim 1.7$ Mpc) developed in our simulation  is built.
To do so, we present a novel numerical methodology in which we apply  the relativistic full-radiative transfer code {\sc SPEV} \citep{Mimica_2009, CMC_2015, Mimica_2016} on the outcomes of the simulation in order to compute the emission associated to both the whole IGM and the WHIM gas components. 
The upgrade version of {\sc SPEV} computes the intensity along each line of sight by integrating the transfer equation along the null-geodesic of photons.  
The emission in the three studied bands, namely, X-ray, SZ, and radio,  is computed using exactly the same numerical scheme and,  therefore, all bands are consistently treated. In order to study the different spatial distributions of both gas components, we consider a region of $5\times R_{vir}$ around our central system. This choice allows us to produce detailed synthetic emission maps of both gas phases, from the cluster core out to a very external region covering an angular extension of $\sim0.6$ squared degrees.

Besides the presentation of our numerical procedure, the main results obtained in the current analysis on the emission properties of the IGM and the WHIM can be summarised as follows:
\begin{itemize}

\item  The z-evolution of different IGM gas components within the whole  ($40$ Mpc) simulated box shows that most of the volume ($\sim90\%$) is dominated by the warm ($T< 10^5$ K) gas phase, whereas the WHIM ($10^5$ K $<$ T $<10^7$ K) and the  hot ($T> 10^7$ K) gas components occupy, respectively,  $\sim10\%$ and a few per cent of the domain. 

\item As evolution proceeds and the largest cosmic structures begin to be formed, the amount of gas in the hot phase increases from $~10\%$ at $z=1$ up to $20\%$ at $z=0$. Concurrently, while the amount of gas in the warm component slightly decreases down to $~30\%$ at $z=0$, the amount of WHIM mildly augments from $50\%$ at  $z=1$ up to a value of $\sim55\%$ at the present epoch \citep[in line with previous numerical analyses; e.g.][]{Cen_1999}.   

\item The density maps obtained for the whole IGM and for the WHIM reveal two different distributions: while in the former, high-density regions are clearly connected  with the largest haloes and subhaloes in the simulation,  the WHIM tends to reside around the main central halo with a more filamentary structure that extends out to more external regions. 

\item On average, we obtain that the WHIM X-ray emission corresponds to $\sim 5$\% and  $1$\% of the IGM emission in the  soft ($0.5-2$ keV) and hard ($2-10$ keV) X-ray bands, respectively. This makes the detection of the WHIM a challenge for current observational instruments.

\item The maximum values of the X-ray surface brightness  in the centre of our galaxy cluster ($\sim10^{-11}-10^{-9}$  \ergscmdeg) are in broad agreement with previous estimates \citep[e.g.][]{Croft_2001, Roncarelli_2006}. 

\item The tSZ maps obtained for the IGM and for the WHIM at $\nu=128$ GHz  can be easily correlated with the X-ray thermal emission of both components, especially at the soft X-ray band. Specifically, the brightest regions in X-rays show the largest negative signals in the SZ maps \citep[e.g.][]{Ursino_2014}. We obtain intensities as high as  $\abs{\Delta I_{\nu, th}/I_{\nu}}\sim 3 \times 10^{-5}$ in the very central cluster region.

\item The kSZ maps associated to the IGM and to the WHIM  show a more similar spatial distribution of their intensities than in the case of the thermal maps. In this case, we obtain mean values of $\avg{\abs{\Delta I_{\nu, k}/I_{\nu}}}\sim 10^{-9}$  for both gas components \citep[see also][]{Nagai_2003}. 

\item Qualitatively, the distribution of the different SZ signals at  $\nu=369$ GHz is very similar to those obtained at  $\nu=128$ GHz. However, our results at  this high frequency, obtained in the non-relativistic approximation, must be taken with caution. Indeed, through a  comparison with the SZ signal obtained with {\sc SZpack} \citep{SZPack_2012, SZPack_2013}, we have confirmed that avoiding the relativistic corrections at  $\nu=369$ GHz can significantly affect our estimations, especially for the IGM.  

\item On average, we find that, at all frequencies, the tSZ signal associated to the WHIM corresponds to $\sim 6$\% of the tSZ signal from the IGM \citep[roughly consistent with the estimate by][]{Hallman_2007}. 
 
\item We find that only a small fraction of the cluster emits in radio. In addition,  whereas the IGM radio emission is quite correlated with high-density and high-temperature regions, the radio emission associated to the WHIM is distributed in a more filamentary structure.  

\item On average, at $1.4$ GHz we obtain a radio signal  $\sim 60$\% fainter than at $350$ MHz for  both the IGM and the WHIM. 
As a mean upper limit for the radio signal at $1.4$ GHz ($350$ MHz) we get a value of  half mJy~arcmin$^{-2}$ ($1-2$ mJy~arcmin$^{-2}$) for the IGM and the WHIM.

\item At both radio frequencies, $350$ MHz  and $1.4$ GHZ,  the cluster radio emission seems to be mainly associated to non-thermal internal shock events within the cluster.  Contrarily, we obtain a negligible radio emission connected to external accretion shocks \citep[see also][]{Hoeft_2008, ArayaMelo_2012}.

\end{itemize}

We would like to point out that the results presented in this paper should be taken as our first step towards designing a technique that allows us to treat, consistently, the IGM emission at any frequency. To fulfill this objective, we use a full-radiative transfer code that implies no approximations nor simplifications, as the transfer equation is solved along the null-geodesic of photons. 
Nevertheless, despite this ambitious intention, we are aware of the current limitations of our simulation and of our numerical procedure. In this regard, a self-consistent treatment of the metal distribution, a proper modeling  of fundamental feedback processes, such as AGN feedback or galactic winds, or the inclusion of the relativistic corrections to estimate the SZ effect represent significant improvements to be accounted for.
Having in mind these limitations, we chose as a first target for our study a very high-resolution galaxy cluster ($\sim 610$ pc), in contrast with previous numerical approaches which were focused on larger but lower resolution cluster samples. The idea behind this choice was to show the capability of {\sc SPEV} to compute multi-wavelength synthetic observations and to produce well-resolved mock maps in order to facilitate the comparison with the imminent  observational data arriving from forthcoming facilities. Now that the method has been fully tested and is fully functional, we are working on a new set of simulations that, while keeping a very high resolution, will produce a large catalogue of clusters.  
Undoubtedly, a major step forward in our understanding of the nature and the role of gas in cosmic structures will come from the combined analysis of multi-wavelength observations produced by the next generation of telescopes, together with  mock observations derived from high-resolution cosmological simulations including all relevant physical processes.
In this regard, the unprecedented observational capabilities of upcoming X-ray, radio and submillimeter observatories, such as \athena, \ska\ or \ccat, are expected to provide the most valuable conclusions on the main properties of the IGM and the WHIM gas components.

\section*{ACKNOWLEDGEMENTS} 
 
Authors would like to thank Jens Chluba for valuable discussions and his help in running the {\sc SZpack} code as well as the anonymous 
referee for his/her constructive criticism.
We acknowledge support by the {\it Spanish Ministerio de Econom{\'i}a y 
Competitividad} (MINECO, grant AYA2016-77237-C3-3-P) and the Generalitat Valenciana (grant GVACOMP2015-227).
SP is ``Juan de la Cierva'' fellow (ref. IJCI-2015-26656) of the Spanish MINECO.
PM acknowledges the support from the European Research Council (grant CAMAP-259276).
CC-M acknowledges the support of ACIF/2013/278 fellowship. We acknowledge the use of publicly available 
\verb+R+ packages \verb+tidyr+ and \verb+dplyr+ \footnote{http://https://www.tidyverse.org/} for data manipulation 
and \verb+ggplot2+\footnote{https://www.ggplot2.org/} for producing graphics as well as the VisIt visualisation software \citep{visit}.

\bibliographystyle{mnbst}
\bibliography{WHIM_paper}

\appendix
\section{Computing thermal X-ray emission using {\sc Cloudy}}
\label{app:Cloudy}

We use the publicly available code {\sc Cloudy} \citep[version 17.00;][]{Cloudy_2017} to construct a large interpolation table of the X-ray emission in the soft and hard bands. In this section we give details about the computation, as well as comparisons with thermal Bremsstrahlung and the interpolation errors.

\subsection{Interpolation table}

To compute the X-ray images in this work we need to be able to model the plasma emission in a large dynamic range of parameters. To achieve this we used {\sc Cloudy} to compute a grid of models with the following axes:
\begin{itemize}
\item temperature $T$: $51$ logarithmically spaced nodes between $10^{3.6}$ K and $10^{8.6}$ K,
\item hydrogen density $n_H$: $131$ logarithmically spaced nodes between $10^{-7.7}$ cm$^{-3}$ and $10^{5.3}$ cm$^{-3}$,
\item logarithm of metallicity relative to solar $log\left(Z/Z_\odot\right)$: $17$ points uniformly spaced between $-2.2$ and $-0.6$,
\item redshift $z$: since we know which simulation snapshots can possibly contribute to the emission (see Section~\ref{subsec:part3}), to reduce the table computing time we use their redshifts instead of a denser grid; the $6$ redshifts are $0.3516$, $0.3821$, $0.4126$, $0.4429$, $0.4732$ and $0.5051$,
\item ${\cal D}_Z$, not used by {\sc Cloudy}, but used to construct the final table: a combination of the Doppler factor and the redshift (${\cal D}_Z:=1/(1+z)/\Gamma/(1-v_\parallel/c)$, where $\Gamma:=(1-v/c)^{1/2}$), $20$ logarithmically spaced nodes between $0.5$ and $1.1$.
\end{itemize}
We determined these parameter ranges by analyzing the cosmological simulation data.

The first four parameters generate $681462$ models that are computed by {\sc Cloudy}. For each model we assume a thermal equilibrium, use the \verb+CMB+ and \verb+HM05+ {\sc Cloudy} commands to generate the CMB and UV background radiation fields \citep[][]{Haardt_1996}, and save the diffuse continuum. We then integrate the continuum in X-ray bands for each value of ${\cal D}_Z$. To be more specific, the soft X-ray emission is integrated between $0.5/{\cal D}_Z$ keV and $2/{\cal D}_Z$ keV, while the hard band is between $2/{\cal D}_Z$ keV and $10/{\cal D}_Z$ keV (the division by ${\cal D}_Z$ simultaneously takes care of the redshift and the proper motion of emitting cells). This gives us $1.362924\times 10^7$ final models organized in a 5D array. In general we use a five-way linear interpolation of the emissivity logarithm to predict the emission for arbitrary parameter combinations. However, in this particular case we do not interpolate in redshift, because the redshifts of our snapshots are the nodes of the table.

\subsection{Interpolation accuracy}

To assess the accuracy of the interpolation table, we generated $10^4$ models using random combinations of parameters in the range mentioned in previous section. We then run {\sc Cloudy} to compute the spectrum of each model in the soft band, and then integrate it to obtain the correct emission in that band. Finally, we compare these values with those predicted by the table for the $10^4$ models.

\begin{figure*}
  \center
  \includegraphics[width=8cm]{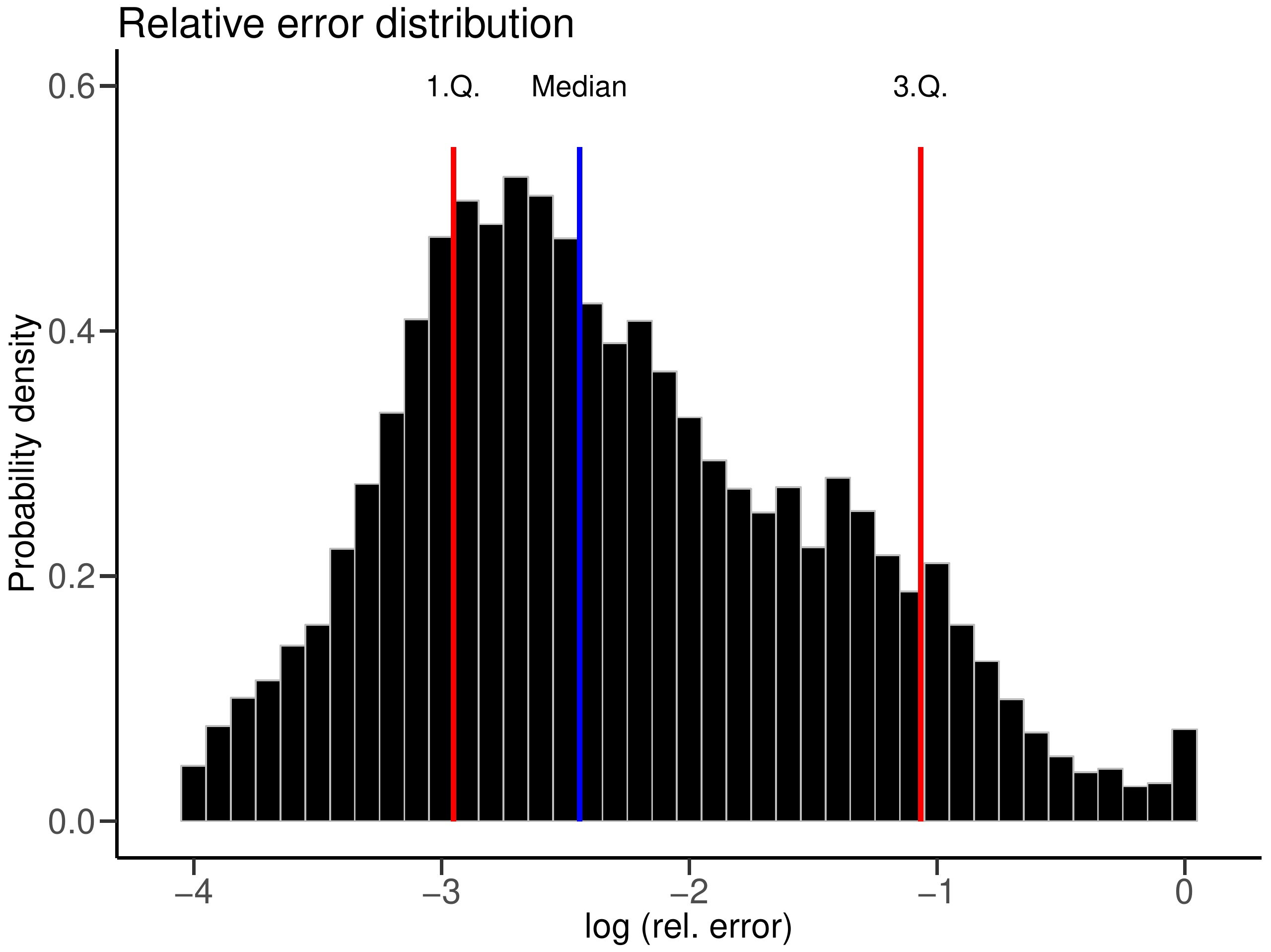}
  \includegraphics[width=8cm]{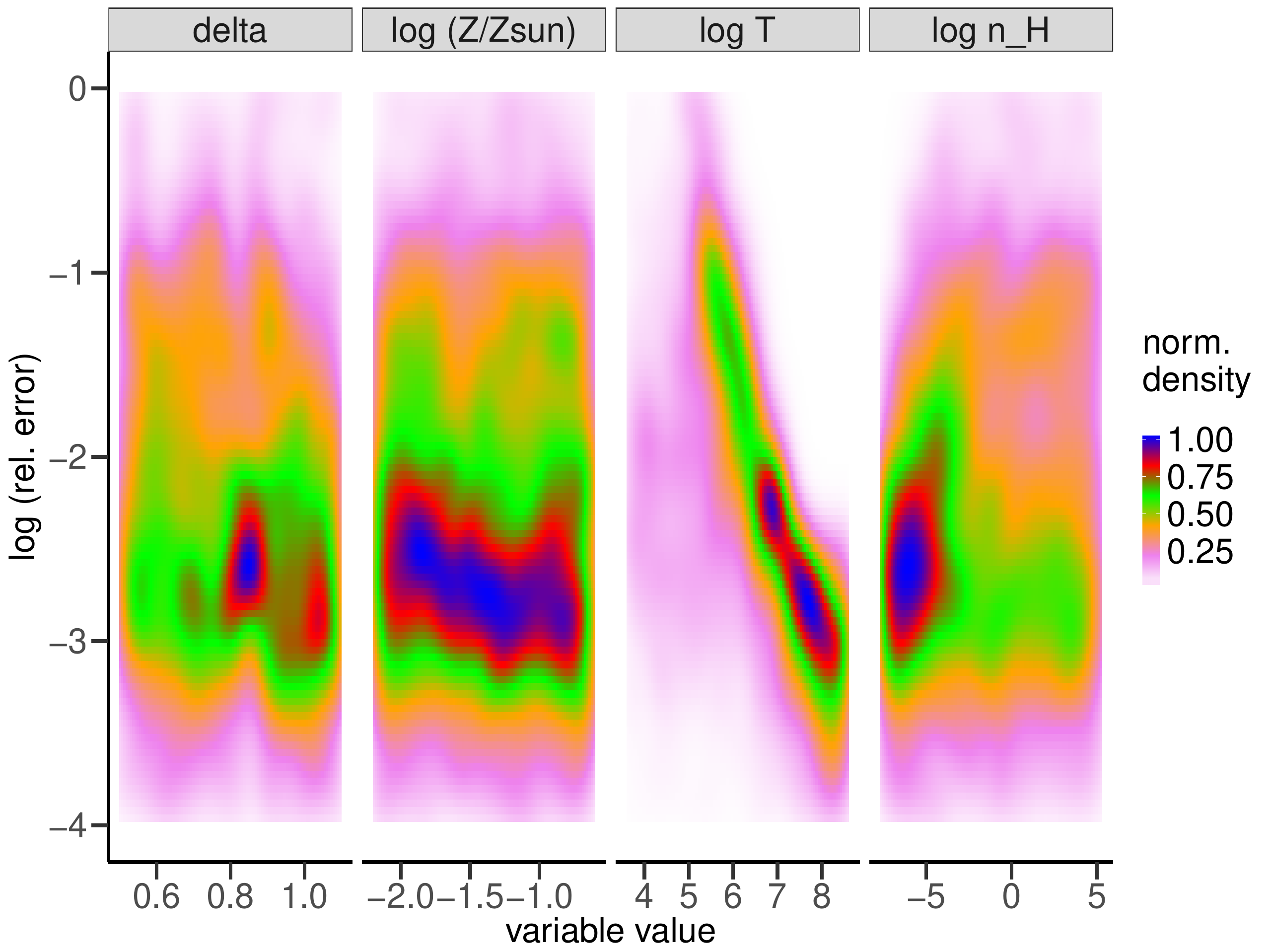}
  \caption{Relative errors of the soft X-ray emission predicted by the interpolation table and computed directly by {\sc Cloudy} for $10^4$ models whose parameters have been randomly drawn. {\it Left:} Relative error distribution histogram. The vertical lines denote the first quartile, the median and the third quartile of the distribution. {\it Right:} Dependence of relative errors on each of the four interpolation variables (note that in this test we fix the redshift $z$). The colors show the relative errors density distribution (separately normalized in each panel).}
\label{cloudy:errors}
\end{figure*}

The left panel in Fig.~\ref{cloudy:errors} shows a histogram of relative errors $|I_{\rm table} - I_{\rm Cloudy}| / I_{\rm Cloudy}$, where $I_{\rm table}$ and $I_{\rm Cloudy}$ are the intensities in $0.5-2$ keV band predicted by the table and computed by {\sc Cloudy}, respectively. The vertical lines denote the first quartile, median and the third quartile of the distribution. As can be seen, half of the errors are smaller than $0.35$\% and three-quarters of the errors are smaller than $8.6$\%. We note that this is a very conservative estimate in that we used random values for each parameter independently. In our calculation the errors are expected to be much smaller. The right panel of Fig.~\ref{cloudy:errors} shows the distribution of errors when the interpolation parameters are varied. As can be seen, for temperatures above $10^6$ K the errors decrease significantly, while for the rest of the parameters there is no clear trend. Nevertheless, the accumulation around $0.35$\% is evident. We conclude that this interpolation table gives an acceptable tradeoff between the computational, memory and storage cost of running a large number of {\sc Cloudy} models and the interpolation accuracy.

\begin{figure}
  \center
  \includegraphics[width=8cm]{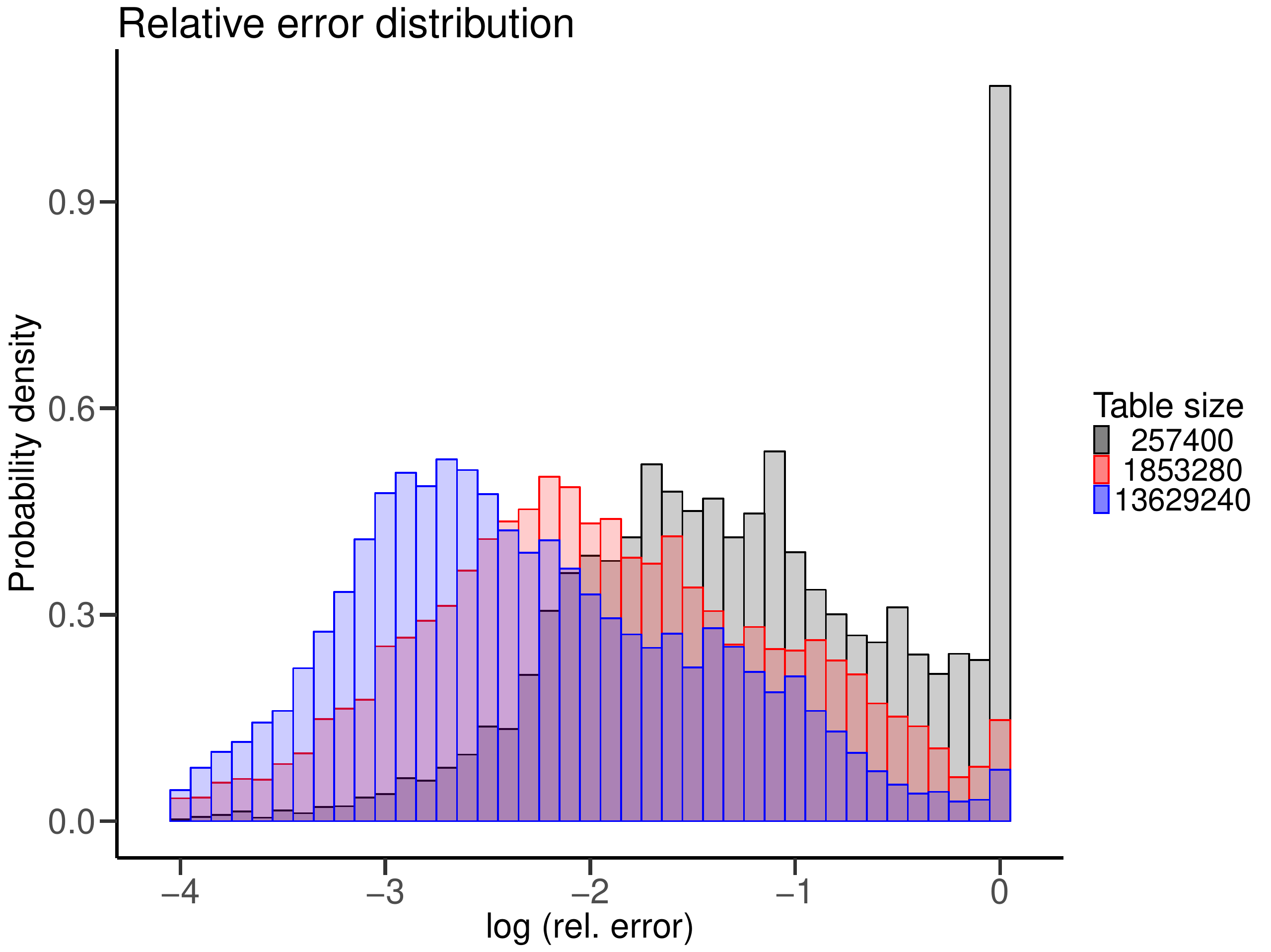}
  \caption{Relative errors of the soft X-ray emission predicted by different interpolation table sizes using $10^4$ test models. Gray, red and blue histograms show the error distribution for $2.574\times 10^5$, $1.85328\times 10^6$ and $1.362924\times 10^7$ points, respectively. All three tables cover the same parameter space.}
\label{cloudy:tables}
\end{figure}

Fig.~\ref{cloudy:tables} shows the dependence of table accuracy on the number of points. The gray line shows the results for a small interpolation table ($\sim 250$ thousand points) that has very large errors. The medium-sized table ($\sim 2$ million points) behaves better, but the median of its distribution is around $\sim 1$\%. We decided that this is still a relatively large error, so we computed an even larger table ($\sim 14$ million points) so that the median decreases to $\sim 0.35$\%, which we deem acceptable for this application.

\subsection{Comparison with thermal Bremsstrahlung}

We run a low-resolution image to compare the intensity distribution between the default {\sc SPEV} thermal emission method \citep{CMC_2015} and the new interpolation table based on {\sc Cloudy}. 

\begin{figure}
\center
\includegraphics[width=8cm]{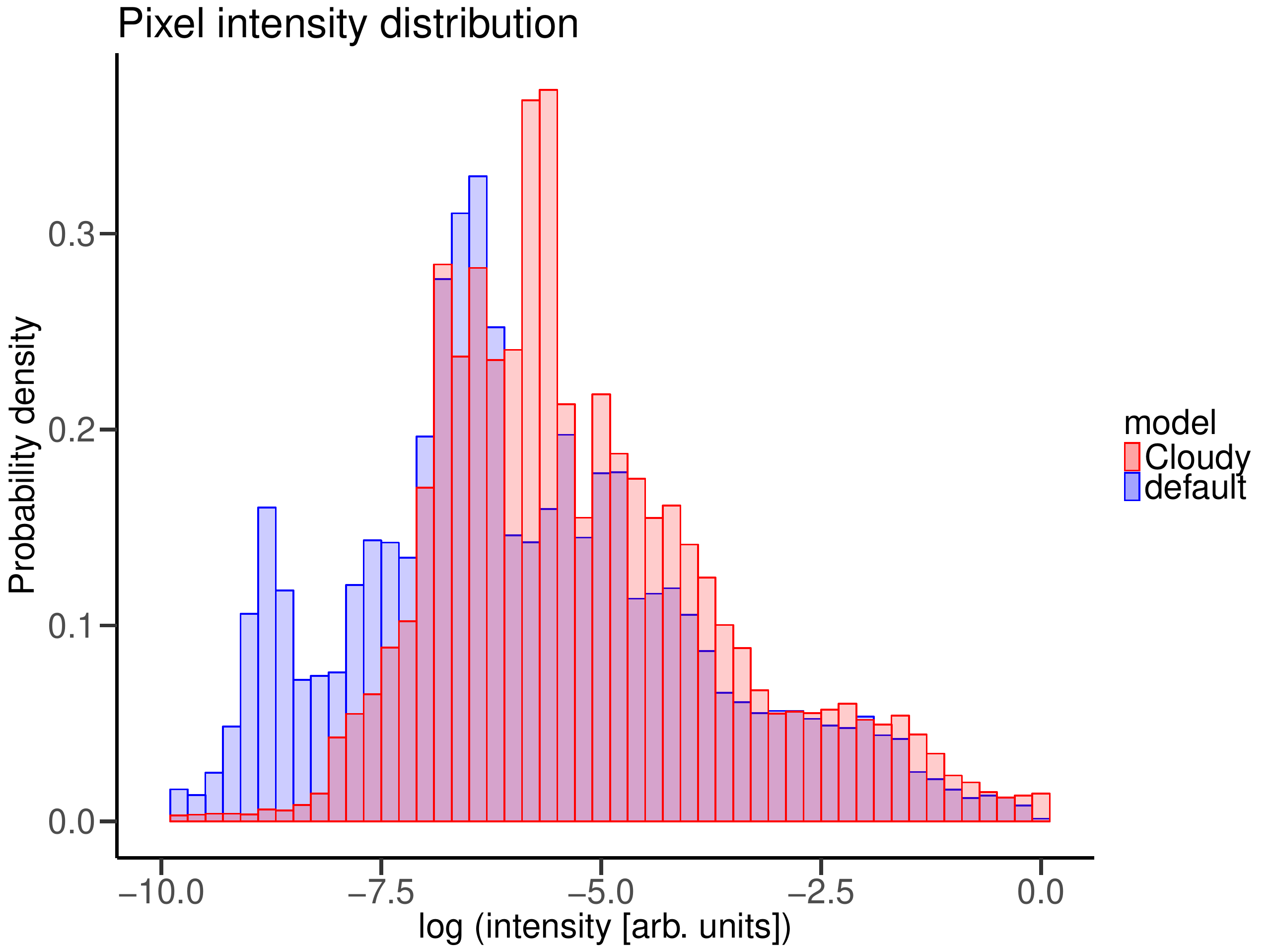}
\caption{Distribution of unconvolved image pixel intensities (in arbitrary units) in the soft band ($0.5-2.0$ keV) computed using the thermal emission as estimated by the method of \citealt[][(``default'', see Section \ref{subsubsec:th} for details)]{CMC_2015} and by an interpolation table based on {\sc Cloudy} (``Cloudy'', see Section \ref{subsubsec:th} for details).}
\label{cloudy:comparison}
\end{figure}

Fig.~\ref{cloudy:comparison} shows the unconvolved image pixel intensity distribution in soft X-rays. As can be seen, the distributions are quite similar for higher intensities (corresponding to higher temperatures), but the metal emission lines taken into account by {\sc Cloudy} produce consistently more intense emission at lower densities and temperatures than does the \citealt{CMC_2015} model (which was not meant to be used in that regime). From this we conclude that the use of {\sc Cloudy} was warranted in this application, as it allowed us to reliably predict the emission from low-density and low-temperature media.

\section{Computing the SZ effect. Comparison between {\sc SPEV} and {\sc SZpack}}
\label{app:SZeffect}

\begin{figure*}
\center
{\includegraphics[width=8cm]{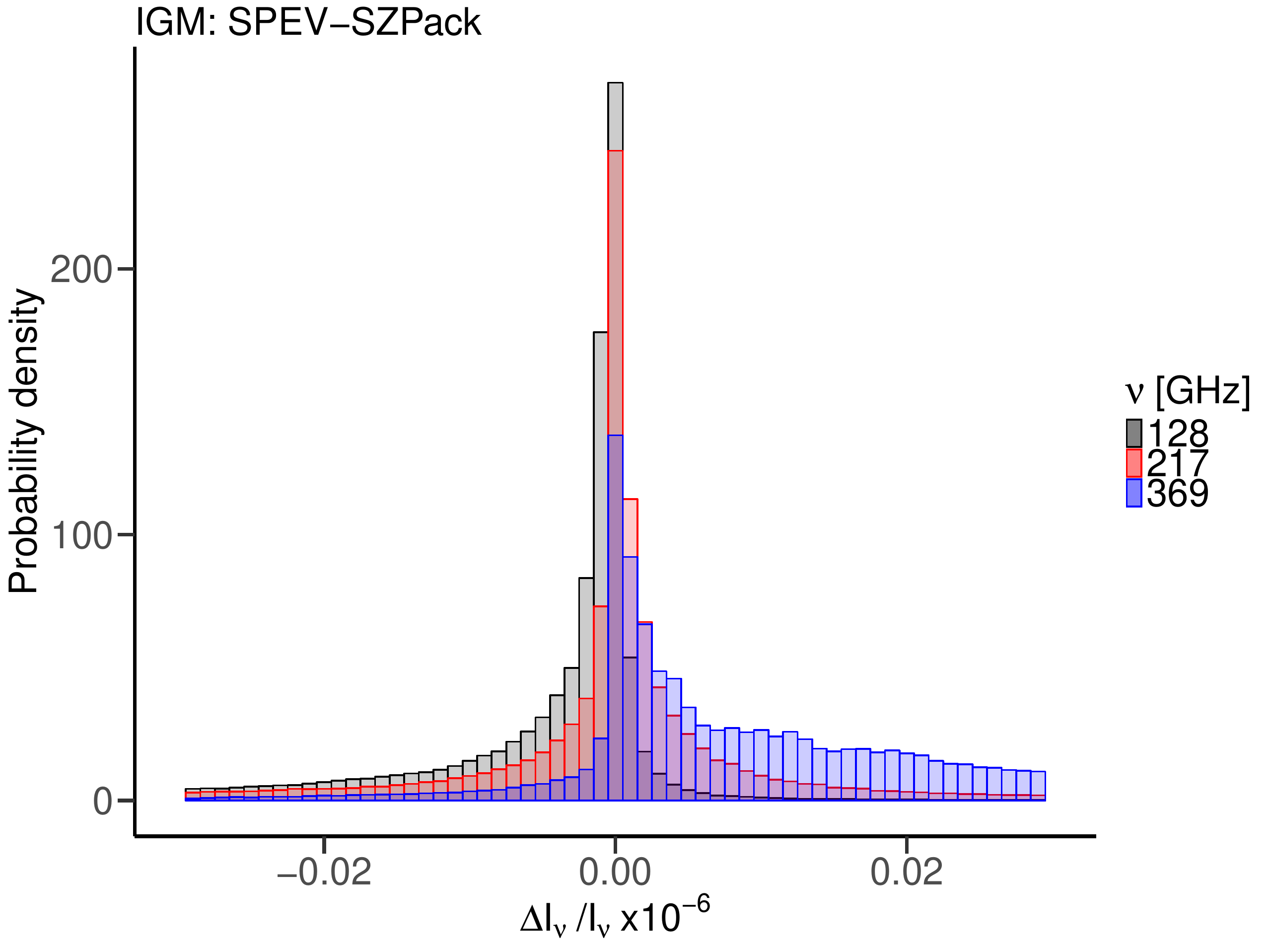}}
{\includegraphics[width=8cm]{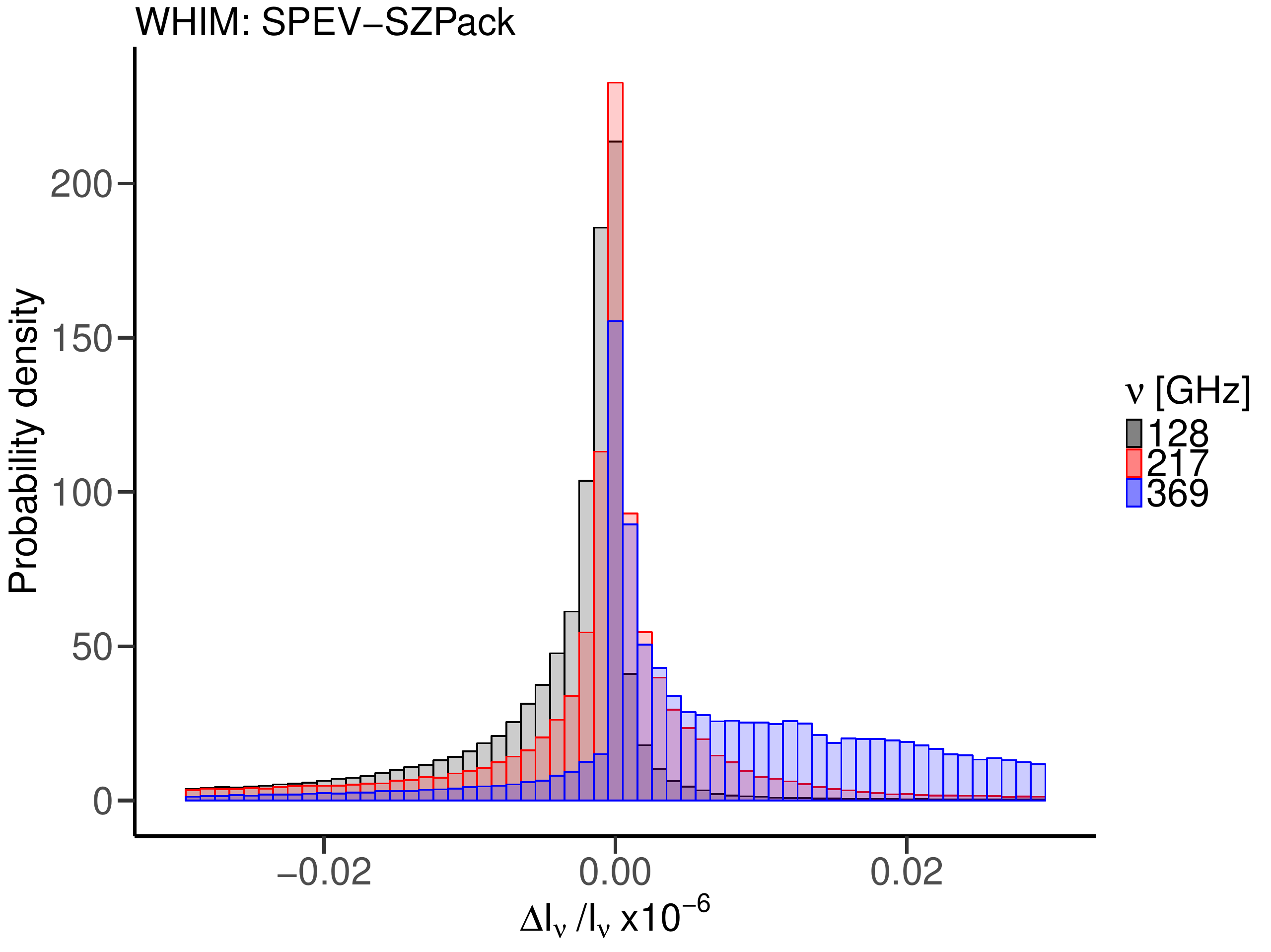}}
\caption{Absolute difference between the total SZ effect computed using {\sc SPEV} and 
{\sc SZpack} at $\nu=128, 217$ and $369$ GHz for the IGM  and the WHIM 
(left and right panels, respectively).}
\label{fig:szpack}
\end{figure*}

In Section  \ref{subsec:sz} we employed {\sc SPEV}  to compute the total, the thermal and the kinematic SZ signals associated to our selected cluster at different observational frequencies.
However, when  obtaining these synthetic observations, we avoided to account for relativistic corrections, which can be particularly relevant for high ICM temperatures ($T_e>10$ keV), large cluster peculiar velocities and high frequencies \citep[$x>10$; e.g.][]{Nozawa_2005}. In order to get an impression of the relevance of this simplification, we compare here our results with those obtained with the publicly available library {\sc SZpack} \citep{SZPack_2012, SZPack_2013}.  Taking into account relativistic corrections, {\sc SZpack} allows for a precise estimation of the SZ effects  in massive galaxy clusters for electron temperatures of up to $kT_e\sim25$\,keV and cluster peculiar velocities as large as $v/c\sim0.01$.

To perform such a comparison we have considered the region of interest around our central galaxy cluster ($\sim0.6$ squared degrees) and we have computed the total SZ effect with {\sc SPEV} and with  {\sc SZpack} at different frequencies. Now, to speed up the comparison, each map has been sampled with $600^2$ pixels, which provides a pixel size of $3.6\times3.6$ arsec$^{2}$, a resolution $\sim 4$ times lower than the one employed in  Section \ref{subsec:sz}.

Figure \ref{fig:szpack} shows the absolute difference between the total SZ signal obtained with {\sc SPEV} and with  {\sc SZpack} at $\nu=128, 217$ and $369$ GHz for the IGM (left panel) and the WHIM (right panel). In general, the distributions obtained for both gas phases are quite similar. As a function of the observation frequency, {\sc SPEV} tends to  overestimate positively the SZ effect at $369$ GHz, whereas at lower frequencies the resulting signal tends to be negative. However, all distributions show a dominant peak around zero and are nearly symmetric. On average,  95\%  of the distributions have an absolute error for IGM (WHIM) below  $0.16$\% ($0.3$\%), 4\% (2\%), and 325\% (14\%) at  $\nu=128, 217$ and $369$ GHz, respectively. These results are quite encouraging, since they confirm that the synthetic total SZ maps generated with {\sc SPEV} in the non-relativistic approximation are in a remarkably good agreement with the results obtained with  {\sc SZpack}, at least at $\nu=128$ and $217$ GHz. However, at  the highest considered frequency the discrepancies generated between {\sc SPEV} and {\sc SZpack} are more significant, especially for the IGM. 
According to Figs.~\ref{fig:SZ1} and \ref{fig:SZ3}, the total SZ signal obtained with {\sc SPEV} at $\nu=369$ GHz is higher than at lower frequencies. In our selected volume, considering all the redshifts of interest, the average number of gas elements (cells) with a temperature above $10$ keV is only $\sim0.1$\%. Moreover, our central cluster is moving with a velocity of $v/c\sim0.003$. These arguments  suggest that, in our case, increasing the frequency of observation could make it necessary to account for relativistic corrections. Indeed, at  $\nu=369$ GHz ($x\sim6.5$) it seems already necessary to include these corrections in order to reproduce the results obtained with {\sc SZpack}.

\begin{figure}
\center
\includegraphics[width=8cm]{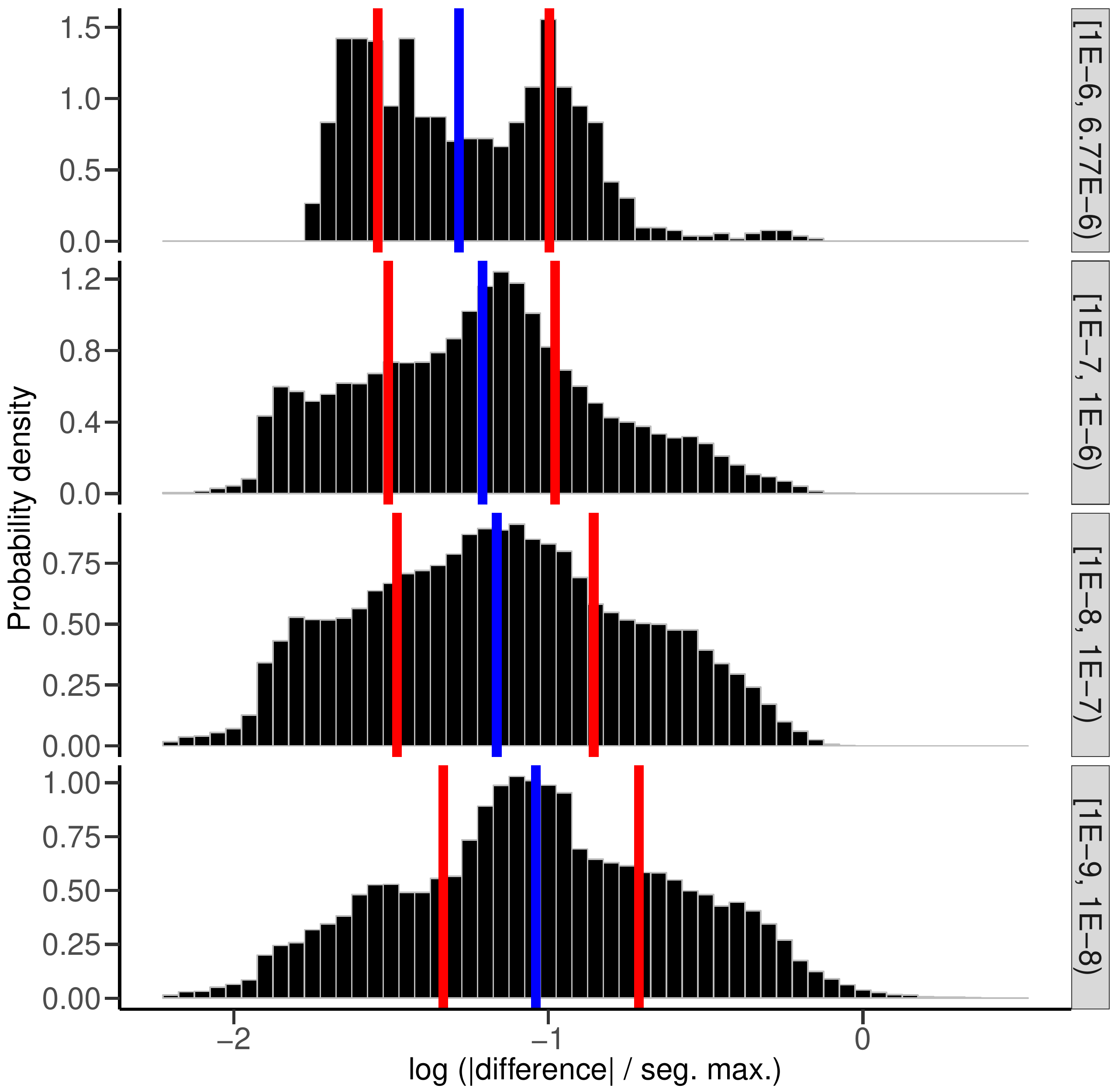}
\caption{Normalized dfiferences between {\sc SPEV} and {\sc SZpack} at $217$ GHz. The error distribution has been partitioned in four segments according to the size of the total SZ effect computed by {\sc SPEV}. From top to bottom the segments are $[10^{-6}, 6.77\times 10^{-6})$,  $[10^{-7}, 10^{-6})$,  $[10^{-8}, 10^{-7})$ and  $[10^{-9}, 10^{-8})$. In each segment the absolute difference between {\sc SPEV} and {\sc SZpack} has been divided by the upper limit of that segment. We show the distribution of the logarithm of that quantity. In addition, the blue line shows the median of each segment, while the red lines show the first and third quartiles.}
\label{fig:szpack-segs}
\end{figure}

Figure \ref{fig:szpack-segs} shows the analysis of the differences between {\sc SPEV} and {\sc SZpack} in four different SZ strength segments at $217$ GHz. In each segment the difference has been normalized to the upper limit of that segment. While not being equivalent to relative differences (which could not be reliably computed since {\sc SPEV} and {\sc SZpack} may give values of different signs or values close to zero when the SZ effect is small), this representation gives us a clue about the typical relative error we make by not taking into account the relativistic effects. The blue lines show the median in each segment, which is always below $10$\%. The third quartiles are below $20$\%. We note that the differences are smaller for pixels with stronger SZ effect.

With these tests in mind, we choose to keep using {\sc SPEV} because most of the errors are not significant and because it allows us to optimize the calculation ({\sc SPEV} runs in parallel). In a future work we plan to extend our analysis including the relativistic approach in {\sc SPEV} and estimating the SZ effect for a larger sample of clusters.

\end{document}